\documentclass[emulateapj]{emulateapj}
\usepackage{natbib}
\usepackage{amsmath}
\begin{document}
\title{
%Spitzer's View of Terrestrial and Icy Planet Formation:\\
A Deep Spitzer Survey of Circumstellar Disks in the Young Double Cluster, h and $\chi$ Persei
% and a Global Analysis of Disk Evolution From 5 Myr to 50 Myr}
}
\author{Ryan Cloutier\altaffilmark{1}, Thayne Currie\altaffilmark{1}
George H. Rieke \altaffilmark{2}, 
Scott J. Kenyon\altaffilmark{3}, Zoltan Balog \altaffilmark{4},
Ray Jayawardhana\altaffilmark{1,5}
}
\altaffiltext{1}{University of Toronto., 50 St. George St., Toronto, ON}
\altaffiltext{2}{Steward Observatory, University of Arizona}
\altaffiltext{3}{Harvard-Smithsonian Center for Astrophysics, 60 Garden St. Cambridge, MA 02140}
%\altaffiltext{2}{Department of Astronomy, University of Michigan}
%\altaffiltext{3}{Institute of Astronomy, University of Cambridge}
\altaffiltext{4}{Max Planck Institute for Astrophysics-Heidelberg}
\altaffiltext{5}{Department of Physics and Astronomy, York University}
\email{cloutier@cita.utoronto.ca, currie@astro.utoronto.ca, grieke@as.arizona.edu, skenyon@cfa.harvard.edu}
\begin{abstract}
We analyze very deep IRAC and MIPS photometry of $\sim 12,500$ members of
the 14 Myr old Double Cluster, h and $\chi$ Persei, building upon on our earlier, shallower
 Spitzer Cycle 1 studies (Currie et al. 2007, Currie et al. 2008).
   Numerous likely members show infrared (IR)
excesses at 8 $\mu$m and 24 $\mu$m indicative of circumstellar dust.
The frequency of stars
with 8 $\mu$m excess is at least 2\% for our entire sample, slightly lower
(higher) for B/A stars (later type, lower-mass stars).   
Optical spectroscopy also identifies gas in about 2 \% of systems but with 
no clear trend between the presence of dust and gas.  Spectral energy
distribution (SED) modeling of 18 sources with detections at optical wavelengths through 
MIPS 24 $\mu m$ reveals a diverse set of disk evolutionary states, including a high fraction of transitional disks, 
although similar data for all disk-bearing members would provide constraints.   Using Monte
Carlo simulations, we combine our results with those for other young
clusters to study the global evolution of dust/gas disks.  For nominal
cluster ages, the e-folding times ($\tau_0$) for the frequency of warm dust
and gas are 2.75 Myr and 1.75 Myr respectively.  Assuming a revised set of ages for some
clusters \citep[e.g.][]{Bell2013}, %(e.g. \cite{Bell2013})
these timescales increase to 5.75 and
3.75 Myr, respectively, implying a significantly longer typical
protoplanetary disk lifetime than previously thought.   In both cases the transitional disk duration, averaged over 
multiple evolutionary pathways, is $\approx$ 1 $Myr$.
Finally, 24 $\mu m$ excess frequencies for 4--6 $M_{\odot}$ stars appear lower than for 1--2.5 $M_{\odot}$ stars 
in other 10--30 $Myr$ old clusters.
\end{abstract}
\section{Introduction}

Planet formation around other stars provides a reference point for
understanding the formation and early evolution of the solar system.
Nearly all stars are born surrounded by optically thick disks of gas and
dust %\citep[e.g.][]{adamslada1987,Strom1989,
%Hernandez2007,Gutermuth2009}
\citep{adamslada1987,Strom1989,Hernandez2007,Gutermuth2009}, which comprise the building
blocks of planets.  Over time, dust grains in these protoplanetary disks
settle towards the midplane and coagulate into larger, lower-opacity
bodies \citep{Dullemond2005}.  Protoplanetary disk gas (and dust
entrained in it) is drained by accretion onto the star 
\citep{Hartmann1998}.   Hence, the frequency of detectable gas-rich dusty
protoplanetary disks decreases with time \citep{Hernandez2007,
CurrieSiciliaAguilar2011}.

By 10-20 Myr, the frequency of gas-rich dusty disks has dropped
effectively to zero.   Few stars show infrared broadband excess consistent
with warm, optically thick protoplanetary disk dust \citep{Silverstone2006,
Currie2007a}, cold gas from outer disk
regions \citep{Pascucci2006,Dent2013}, or accreting gas from
inner disk regions \citep{Jayawardhana2006,Fedele2010}.   Optically-thin, gas poor debris
disks have second-generation dust originating from planetesimal collisions 
 \citep{Wyatt2008,KenyonBromley2008}.  They comprise most of the disk
population at ages beyond 10 $Myr$ \citep{Rieke2005,Su2006,Carpenter2009a} 
%\citep[e.g.][]{Rieke2005,Su2006,Carpenter2009a}
and first emerge around stars nominally
age-dated to 3-5 Myr \citep{Carpenter2006,CurrieLada2009,CurrieKenyon2009}.

Stars with ages of several Myr to 10-20 Myr, therefore, clarify the
timescale for and probe the structures of dissipating protoplanetary disks
and the transition from these disks to debris disks.  Most recent studies
suggest an e-folding timescale of $\sim 3$ Myr for protoplanetary disk dust and
gas \citep{Hernandez2007,Fedele2010,CurrieSiciliaAguilar2011}.  
Fewer than $\sim$ 20\% of such disks around stars of
solar mass or greater remain at $\sim$ 5 Myr \citep{CurrieLada2009, CurrieKenyon2009}.

 The morphologies of protoplanetary disks at these intermediate ages also show evidence for 
 significant evolution.  Fewer disks have optically thick
at all IR wavelengths, characteristic of nearly all $Myr$-old ``primordial" disks.  
More disk exhibit reduced IR emission, consistent
with inner disk clearing. As shown by SED modeling, many of these transitional
protoplanetary disks have
%suggests two pathways for disk clearing: those with
an optically thin \textit{inner hole/cavity} that retain an optically thick outer
disk \citep{Calvet2005,Espaillat2007}.  Many other disks 
exhibit 
%reveal 
%and those that have
a reduced luminosity at IR to submm wavelengths and likely represent a second evolutionary pathway,
\textit{homologously depleted} transitional disks \citep{CurrieLada2009}, although this population 
is challenging to distinguish from some primordial disks with dust settling 
\citep{Luhman2010,CurrieSiciliaAguilar2011}.  Nevertheless, the relatively high frequency
of the combined transitional disk population suggests that this phase, 
averaged over all pathways, comprises an appreciable fraction (e.g. 30\%) of the total disk
lifetime \citep{CurrieLada2009,CurrieSiciliaAguilar2011}, although the
exact sizes of the transitional disk populations and relative lifetimes for different morphologies 
are still being debated \citep{Muzerolle2010,
Luhman2010,CurrieSiciliaAguilar2011,Espaillat2014}.

However, our understanding of protoplanetary disk and early debris disk
evolution hinges on key, recently challenged assumptions about the samples
comprising the above mentioned studies.  In particular, while the relative 
ages of young clusters can be reliably determined, absolute ages are highly 
uncertain \citep[see discussion in ][]{Soderblom2014}.
\cite{Bell2013} and \cite{Pecaut2012} find that stars in many clusters nominally
age-dated to 2-5 Myr and used to derive disk evolution timescales are
about twice their nominal age.  
%Reviewing the recent literature, 
%\cite{Soderblom2014} cast doubt on our ability to constrain the ages of stars to
%better than a factor of two.  
However, studies deriving the
protoplanetary disk timescale and the transitional disk duration generally assume
nominal, systematically younger ages \citep{Hernandez2007,CurrieSiciliaAguilar2011}.

Moreover, our knowledge about disk evolution is limited by poor
constraints on the properties of disks around some kinds of stars at 10-20
Myr. Few protoplanetary disks remain around any stars age-dated at $\sim$ 10-20
Myr. Ensembles of such disks much larger than the populations of nearby
clusters are needed to investigate the morphologies and relative frequency of the
longest-lived protoplanetary disks.  While
planets appear around stars as massive as 4.5 M$_{\odot}$ \citep{Hatzes2005}, 
studies focused on nearby clusters have only explored the disk
populations around $M$ $\lesssim$ 2-3 M$_{\odot}$ (e.g. late B/early A type) stars.
For any reasonable initial mass function \citep[e.g.][]{MillerScalo1979}, 
these massive stars are far less frequent than the AFGKM stars normally analyzed 
in disk evolution surveys. Therefore 
extremely populous clusters at 10-20 Myr are required to 
investigate statistically the last vestiges of the protoplanetary disk phase and
complete our understanding of the frequencies of debris disks around all
types of planet-bearing stars.

The massive 14 Myr old Double Cluster, h and $\chi$ Persei, provides us with a valuable 
laboratory for studying disk evolution at 10-20 Myr. The large population of stars 
($\sim 12,500$) in this region at a wide range of evolutionary states 
allows us to constrain the frequency and morphologies of the longest-lived 
protoplanetary disks. Additionally, h and $\chi$ Persei contains a 
large number of very early type, massive stars (SpT $\lesssim$ B6, M $\geq 4$M$_{\odot}$) 
relative to other nearby 
clusters, allowing us to investigate the disk populations of the most massive stars.
 Lastly, compared to most other clusters, the age of h and 
$\chi$ Persei is well constrained. 
A number of independent diagnostics such as pre-main-sequence 
isochrones, the main-sequence turnoff, and the luminosity of M supergiants 
\citep{Slesnick2002,Currie2010a,Bell2013} independently point to an age of 14 $Myr$. Because the derived disk lifetimes are 
dependent on cluster age, precise knowledge 
of the age of h and $\chi$ Persei helps to clarify the protoplanetary disk 
lifetime as well as the relative frequency and morphology of long-lasting
protoplanetary disks.

Previous \textit{Spitzer Space Telescope} data for h and $\chi$ Per provide a first but very limited probe of the Double Cluster's disk population \citep{Currie2007a,Currie2007b,Currie2008a}.
 IRAC 3.6--8 $\mu m$ data revealed many stars on the field with red colors consistent with disk emission but was too shallow to detect the photospheres of stars less massive than about 
 1.3 $M_{\odot}$.   MIPS-24 $\mu m$ data identified many large-excess sources consistent with a plateau/peak in debris emission at 10--20 $Myr$ for (some) stars more massive than the Sun due to icy debris disk evolution, which was verified by analyzing MIPS excesses in nearby regions also found in other studies \citep[][]{Currie2008a,Hernandez2007,Chen2011}.  However, these data detected just 17 stars less than 3 $M_{\odot}$ with the most luminous debris emission and could not probe the photospheres of the most massive planet-bearing stars.  
 Finally, at the time of these publications, membership lists for the Double Cluster were limited to post-main sequence stars, Be stars, and the most massive main sequence stars \citep[e.g.][]{Slesnick2002}.  The candidate member list from \citet{Currie2010a} extending h and $\chi$ Per's census to subsolar-mass stars was not yet available: warm dust and gas frequencies derived from \citet{Currie2007a,Currie2007c} are then prone to uncertainties due to contamination from non-members.
 
%In order to provide a robust disk evolution timescale we require deep observations of a large number of stars across a wide range of spectral types to obtain accurate measurements of the frequency of disks at various evolutionary states. The massive cluster of h and $\chi$ Persei provides us with an invaluable sample of thousands ($\approx 13, 000$) of stars with varying spectral type (from B0.4-M6.3, for members with a known spectral type) and at a range of stellar evolutionary states. At a well-defined age of 14 Myr \citep{something}, stellar members of h and $\chi$ Persei occupy the interval in a disk's lifetime between long-lived protoplanetary disks and optically thin debris disks; the disk evolution timeframe of interest. In summary h and $\chi$ Persei provides us with a statistically significant sample at a time of peak terrestrial and icy planet formation with which to constrain the evolution of gas and dust.

%Early photometric infrared studies of h and $\chi$ Persei with the \emph{Spitzer Space Telescope} (\emph{SST}) and numerous ground-based observations showed that *

In this study we present a deep photometric, infrared study of h and $\chi$ Persei members using new $IRAC$ and MIPS data from Spitzer and the 
membership list from \citet{Currie2010a}. 
%In the bandpasses of particular interest to our study we detect cluster members at $\sim 1.5$ magnitudes dimmer in both [8] and [24] compared to \cite{Currie2007a} (IRAC) and \cite{Currie2008a} (MIPS) (cf. Figs. \ref{numbercounts}, \ref{mag24error}).
%6635 cluster members at 8 $\mu$m and 413 at 24 $\mu$m which is a factor of $\sim *$ greater than the previous cluster survey(s) *. 
The paper is organized as follows: in \S \ref{sectdata} we describe our data reduction techniques 
and quantify our completeness limits for both the IRAC and MIPS data. In \S \ref{sectdust} we 
identify and quantify the frequency of stars surrounded by warm and slightly cooler 
circumstellar dust while \S \ref{sectgas} details a similar analysis of 
gas-rich disks. To further investigate the morphologies of disks around h and $\chi$ Persei 
stars, we then compare the SEDs of members with 8 $\mu$m and 24 $\mu$m excess to fiducial models 
representing a range of disk evolutionary states (\S \ref{sectseds}). In \S \ref{sectdiskevol}, 
we combine our results with those for other clusters to constrain the lifetime of protoplanetary 
disks and the duration of the transitional disk phase. We conclude with a summary of our results, 
contrast them with other recent studies, and identify future 
work that can further constrain the h and $\chi$ Persei disk population and clarify key aspects 
of protoplanetary disk evolution in \S \ref{sectdiscussion}.

\section{Data} \label{sectdata}
\subsection{IRAC 3.6--8 $\mu$m Data}
The Infrared Array Camera (IRAC; \citealt{Fazio2004}) observed h 
and $\chi$ Persei on October 30, 2008 (AOR IDs 2182740, 21828608, 
21828096, 21828864, 21828352, and 2182912).  
Solar activity was normal to below-average.  Zodical emission ranged between $\sim$ 0.02 and 2 MJy sr$^{-1}$ from 
3.6 $\mu$m to 8 $\mu$m.
Image processing and photometry were performed separately for the short exposure
and long exposure frames.  

\subsubsection{Image Processing and Photometry for Short Exposure Data}
The short-exposure (0.6 $s$) \textit{corrected BCD} (Basic Calibration Data; CBCD) frames
 have very low source density, no diffuse mid-IR emission 
(e.g. nebulosity), and extremely few residual cosmic ray hits.  We 
processed these data using \textit{MOPEX} \citep{Makovoz2005}, applying 
an overlap correction and mosaicing the images.  Photometry 
was performed using \textit{PhotVis} \citep{Gutermuth2005}, a slightly-modified 
GUI-based adaption of the standard IDLPHOT aperture photometry package.  
We used a 2-pixel (2\farcs{}4) radius aperture, a sky annulus 
ranging between 2 and 6 pixels in radius, and the aperture corrections 
listed in the IRAC data handbook.

\subsubsection{Image Processing and Photometry for Long Exposure Data}
The source density for long exposure data is significantly higher; 
the long exposure data identify far more stars, including 
numerous AFGKM cluster members that are too faint (m[IRAC] $\gtrsim$ 13) 
to be detected by the 0.6s exposures \citep[see][]{Currie2007a,Currie2008phd}.  
To optimize our photometry and thus better insulate our results against 
photometric uncertainties, we adopt a more detailed approach 
to image processing and photometry.  %Our approach adopts the method of 
%\citet{CurrieLada2009} as a starting point but improves upon their 
%mosaicing and photometry.  

%Appendix A details and justifies each step of our image processing and photometry.
Briefly, we first applied the \textit{Array-dependent Pixel Response} correction 
\citep{Quijada2004} to each CBCD frame and then mosaic the CBCD frames 
using MOPEX\footnote{We also tested the artifact mitigation routines 
described in \citet{Cpk2008} to the "uncorrected" BCD frames.  These routines also 
removed image artifacts (e.g. column pulldown, \textit{muxstriping}), though they did not improve 
upon corrections from pipeline scripts.  To make our results easier to reproduce, 
we simply used the CBCD frames.}. 
We applied an overlap correction to each frame, masking pixels 
containing bright point sources and calculating each frame's zero-point 
offset using iterative 3$\sigma$ clipping to determine the frame's background 
level.  

To determine the distribution of pixel values for each position on the final 
mosaic, we interpolated values from individual frames using the \textit{drizzle} 
algorithm with a "drizzle factor" of 0.1.  Though very computationally 
expensive, the \textit{drizzle} algorithm produced the most sharply defined 
point sources and did not introduce low-frequency pattern noise like that 
produced by bicubic interpolation.  Outlier rejection 
was performed using the \textrm{mosaic outlier} and \textit{rmask mosaic} 
modules.  
%For each pixel stack, pixels with deviations from the median 
%value exceeding 3$\sigma_{stack}$ were flagged as outliers and removed as 
%were deviant pixels identified by the \textit{rmask} mosaic.  
Final flux 
densities for each pixel were derived from the average value of the 
sigma-clipped stack of pixel values.

For photometry, we use a slightly modified version of the IDLPHOT aperture 
photometry package, improving the sky background determination, 
centroiding, and photometric error determination.
  To identify point sources, we set the detection threshold in 
each filter in the \textrm{find.pro} routine equal to three times 
the measured dispersion in the sky background ($\sim$ 3$\sigma_{rms, avg}$).  
We apply pixel masks on the mosaic to remove the central pixels of all stars 
and the wings of bright stars from sky background determination.  The 
adopted sky background is set to the median value, not the 
"mode".  As with the short-exposure data, we adopt a two-pixel aperture, a 
2--6 pixel sky annulus, and the aperture corrections from the IRAC data handbook.  
From the uncertainty in the sky background, photon noise, and read noise, 
we determine the photometric uncertainty, following \citet{Cpk2008}.

To select a final list of IRAC detections, we compared lists from  
the long and short exposure data.  For sources brighter than magnitude 10.5 
in the long exposure data, we replaced the entries with those from the short exposure data 
due to saturation \citep[see also][]{Cpk2008}.  By matching our catalog to the 2MASS 
All Sky Survey catalog, we fine-tuned the IRAC data astrometry and 
produce a combined list of IRAC sources with and without 2MASS counterparts.
In all band merging between the IRAC channels and IRAC to MIPS/2MASS etc. 
we adopted a matching radius of 1\arcsec{}.

\begin{figure}
\centering
\plotone{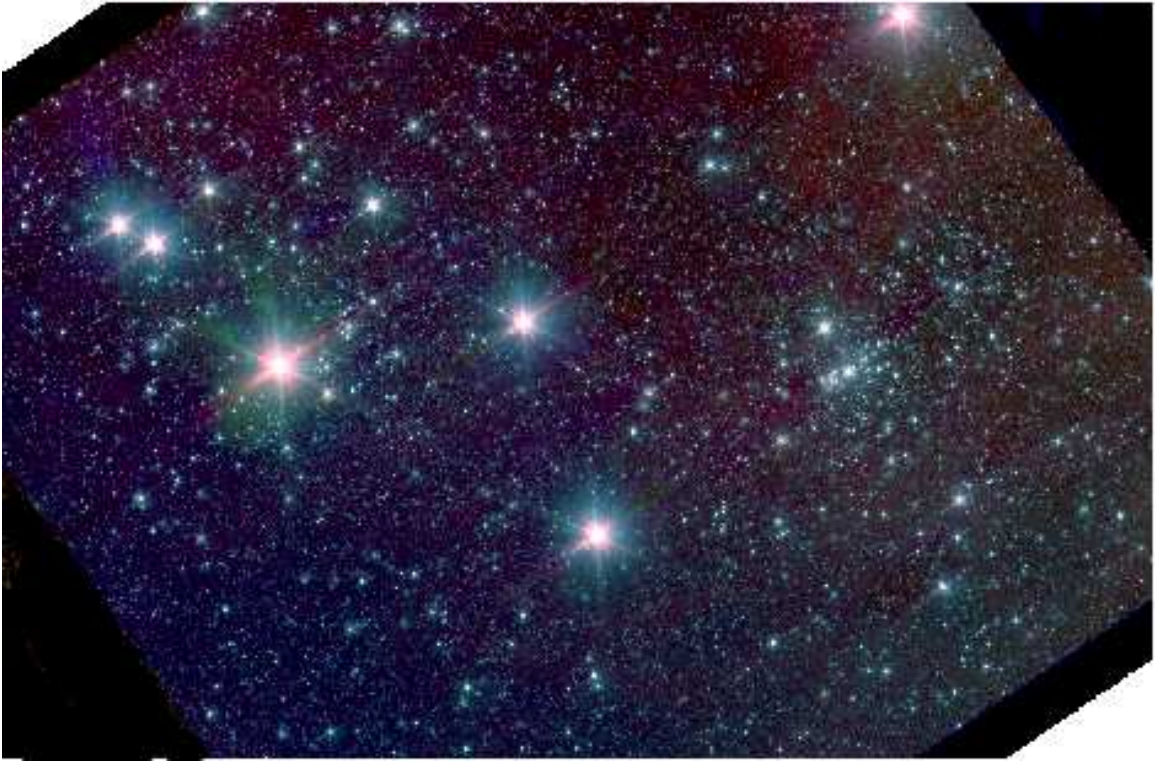}
\caption{Multicolor mosaic image centered on h and $\chi$ Persei.  Blue identifies emission 
from the 3.6 $\mu$m filter, green from the 4.5 $\mu$m filter, and red from the 8 $\mu$m filter.  
The core of h Persei is centered on the overdense region of stars at the middle-right; the $\chi$ Persei core is 
identified by an M supergiant at the middle-left.  }
\label{rgbimage}
\end{figure}

Figure \ref{rgbimage} shows a color mosaic image produced from the IRAC [3.6], [4.5] and [8] filters (blue, green, and red).

%\subsubsection{Completeness and Magnitudes vs. Errors in the IRAC bands}
\subsubsection{IRAC Source Counts and Detection Limits}
\begin{figure}
\plottwo{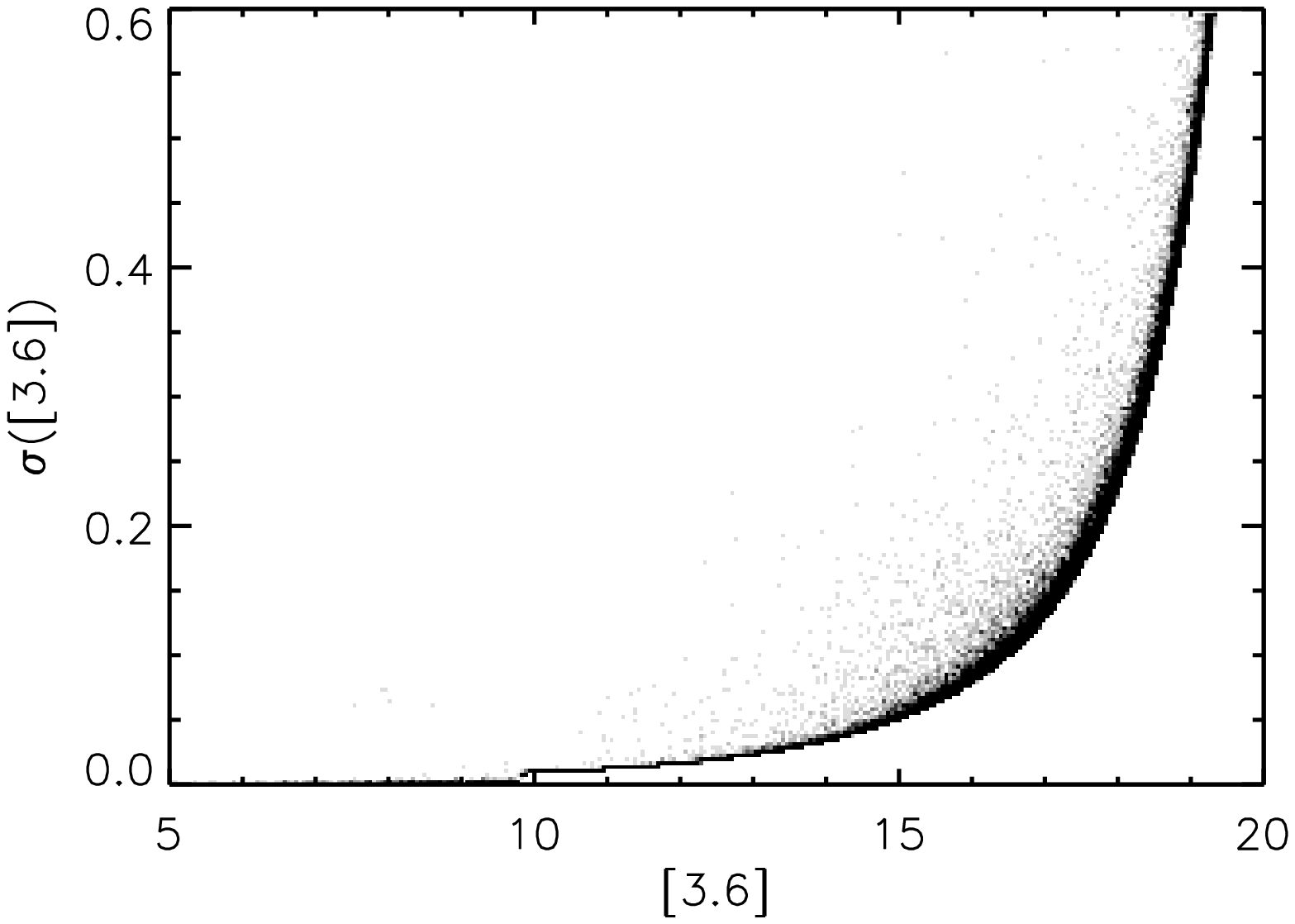}{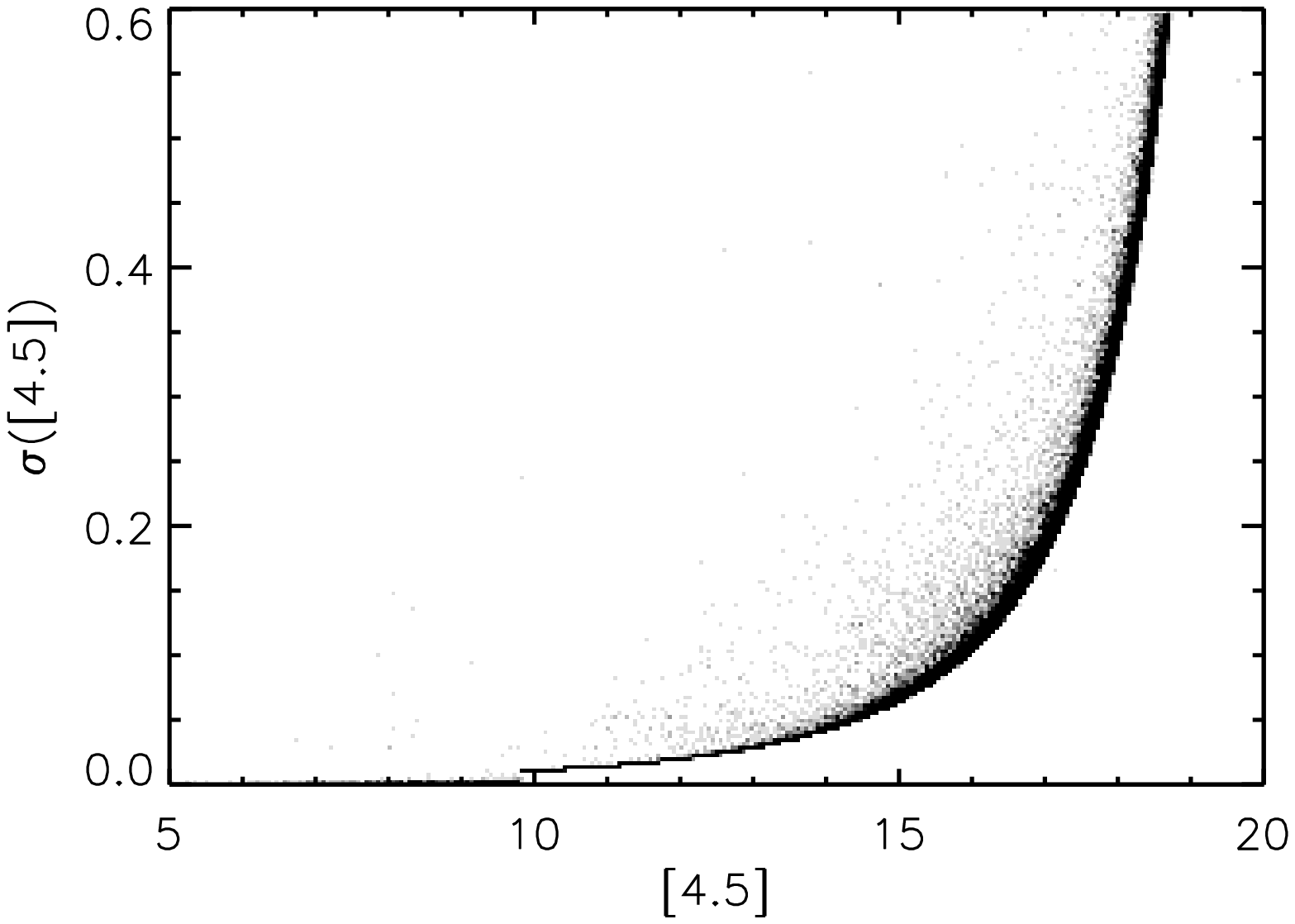}
\plottwo{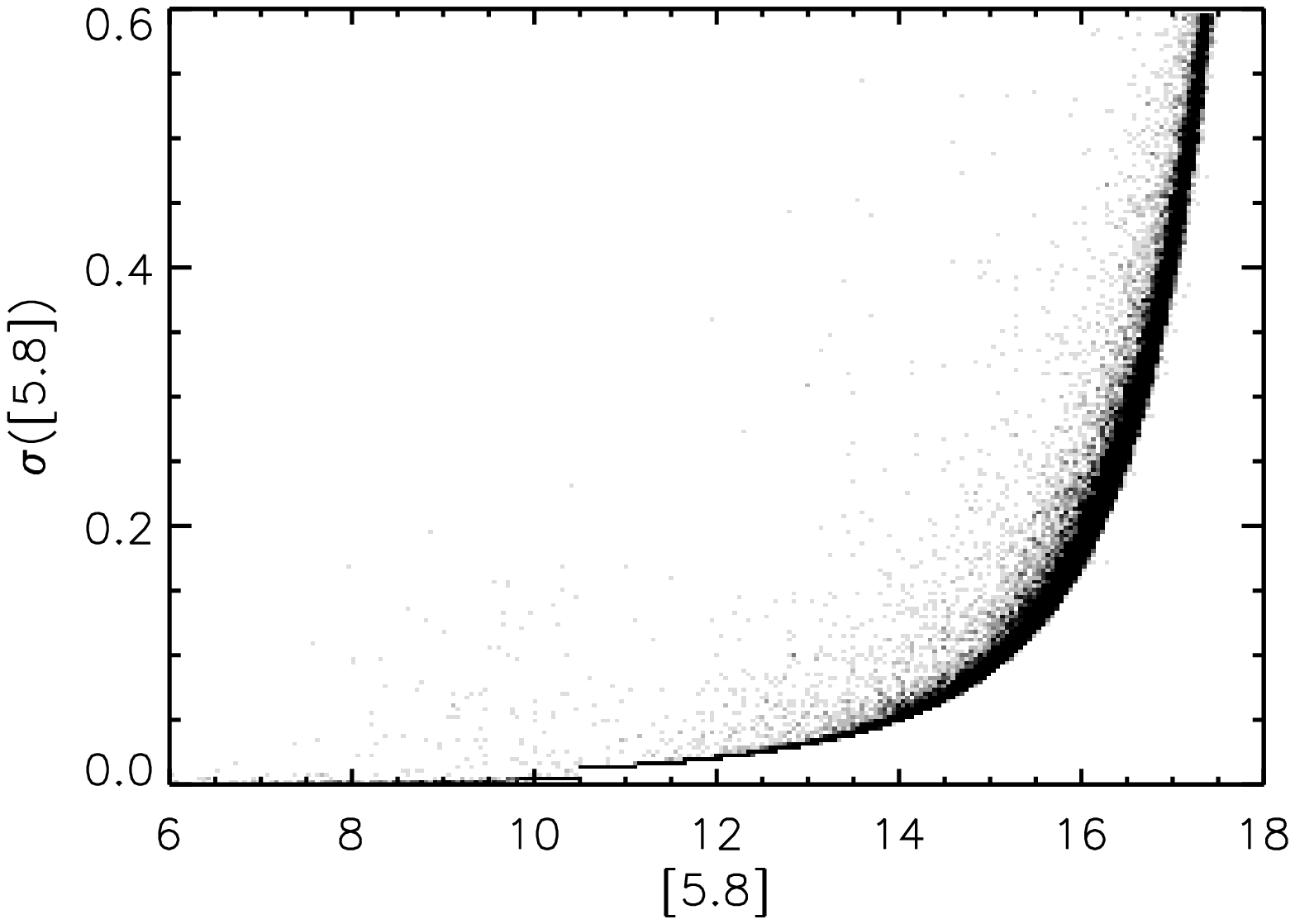}{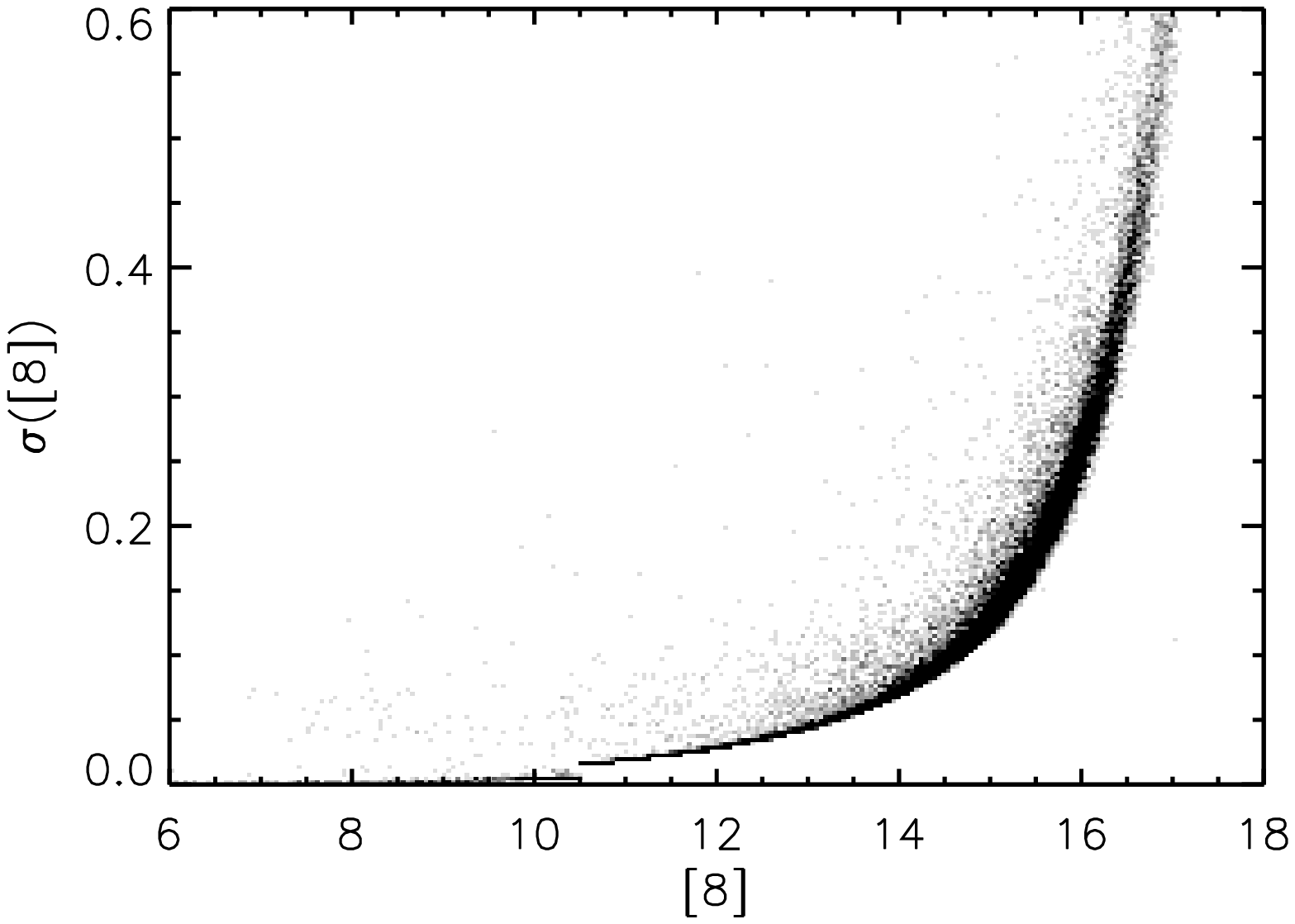}
\caption{Density plot of the magnitude vs. error distribution for IRAC data.} 
\label{magerrorirac}
\end{figure}

Figure \ref{magerrorirac} shows the distribution of IRAC magnitudes vs. 
photometric uncertainties.  Most sources follow a thin distribution 
in photometric uncertainty over the full range of magnitudes.  However, for each filter a 
second, sparser distribution lies on top of the main one; many sources 
have photometric uncertainties $\gtrsim$ 0.1--0.2 mags even though the errors 
from photon noise can be $\lesssim$ 0.05 mags.  Typically, these sources 
are located in regions of higher stellar density or near the PSF wings of 
bright stars, so estimates of the local background flux are more uncertain.
 
\begin{figure}
\plottwo{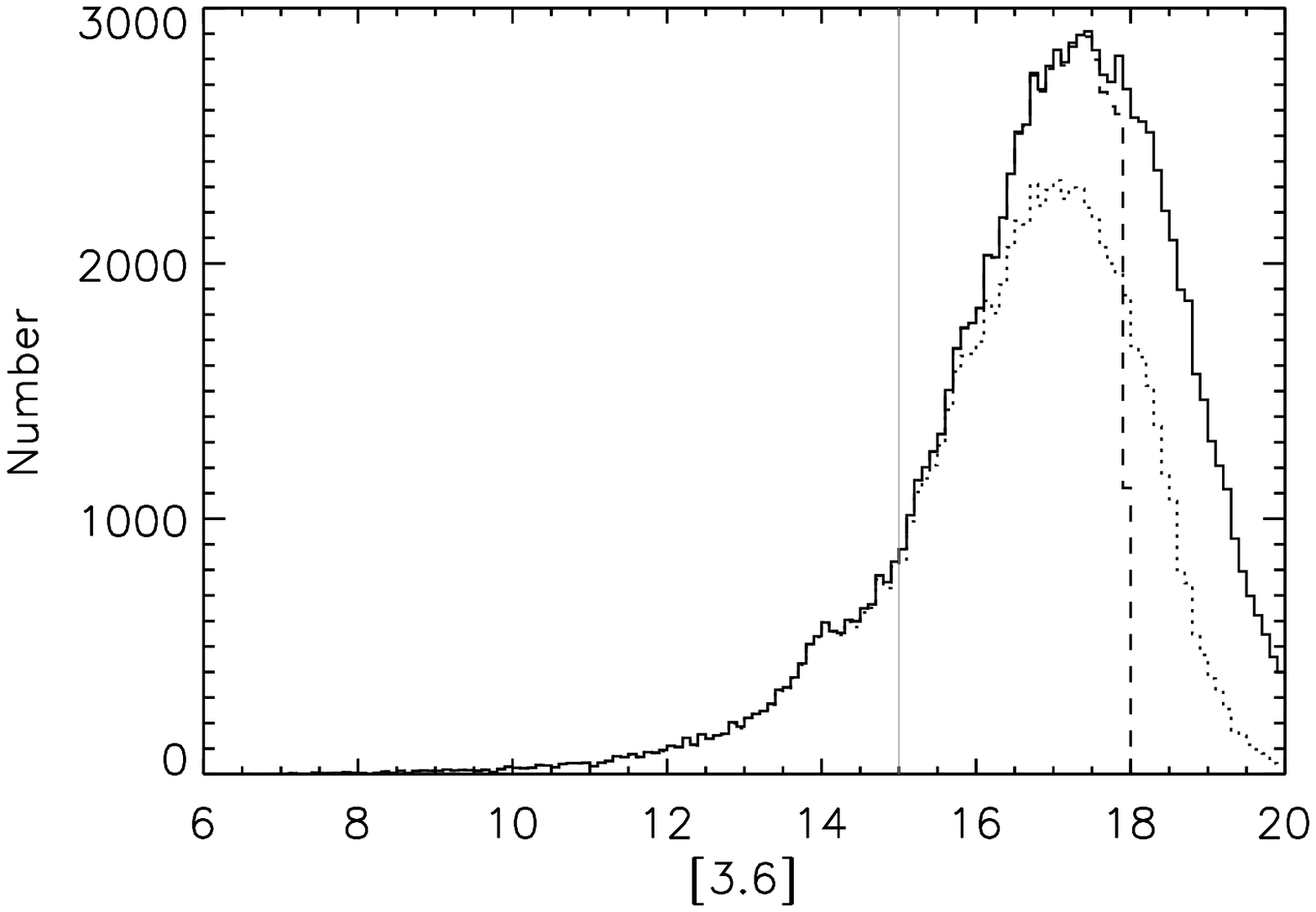}{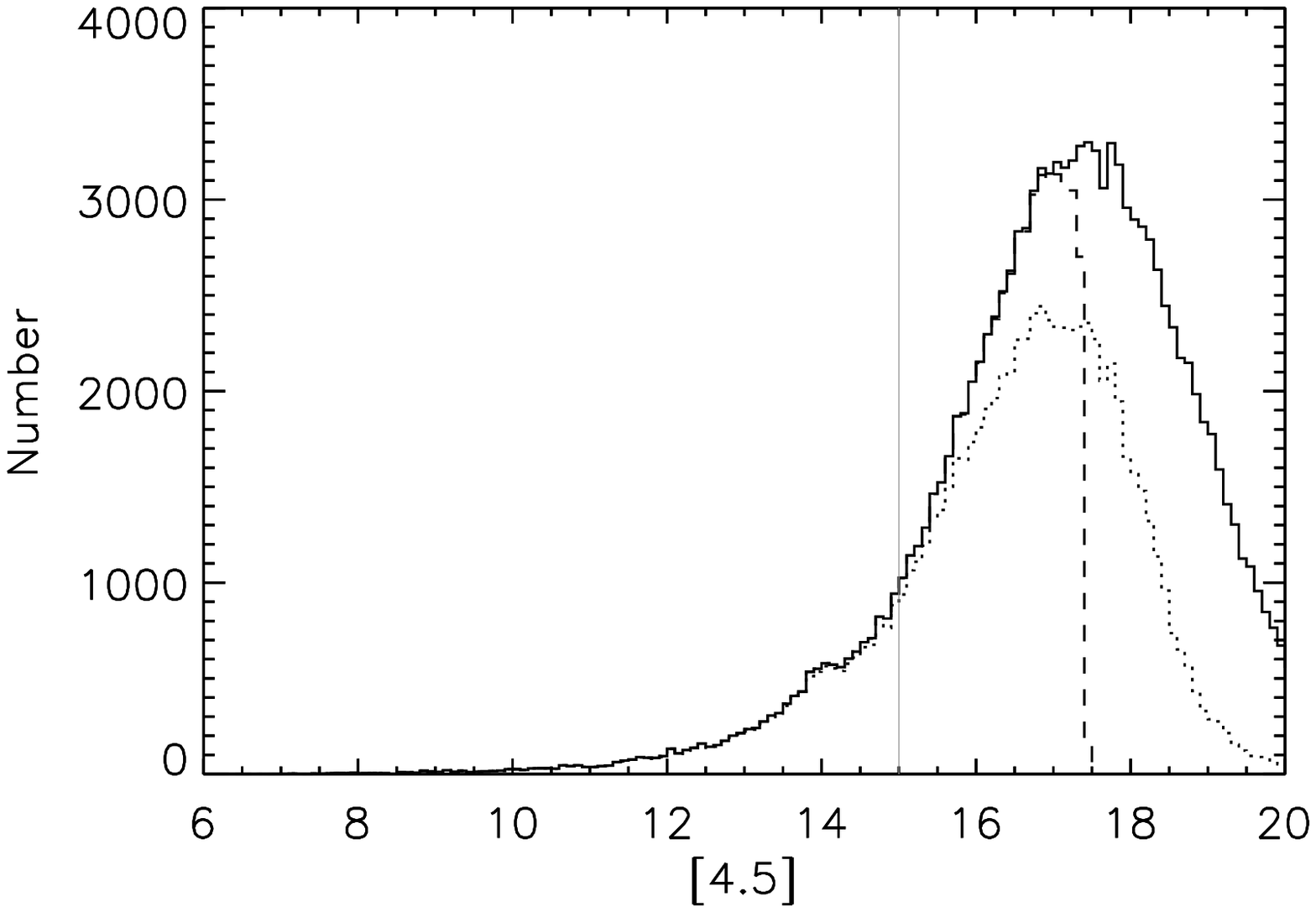}
\plottwo{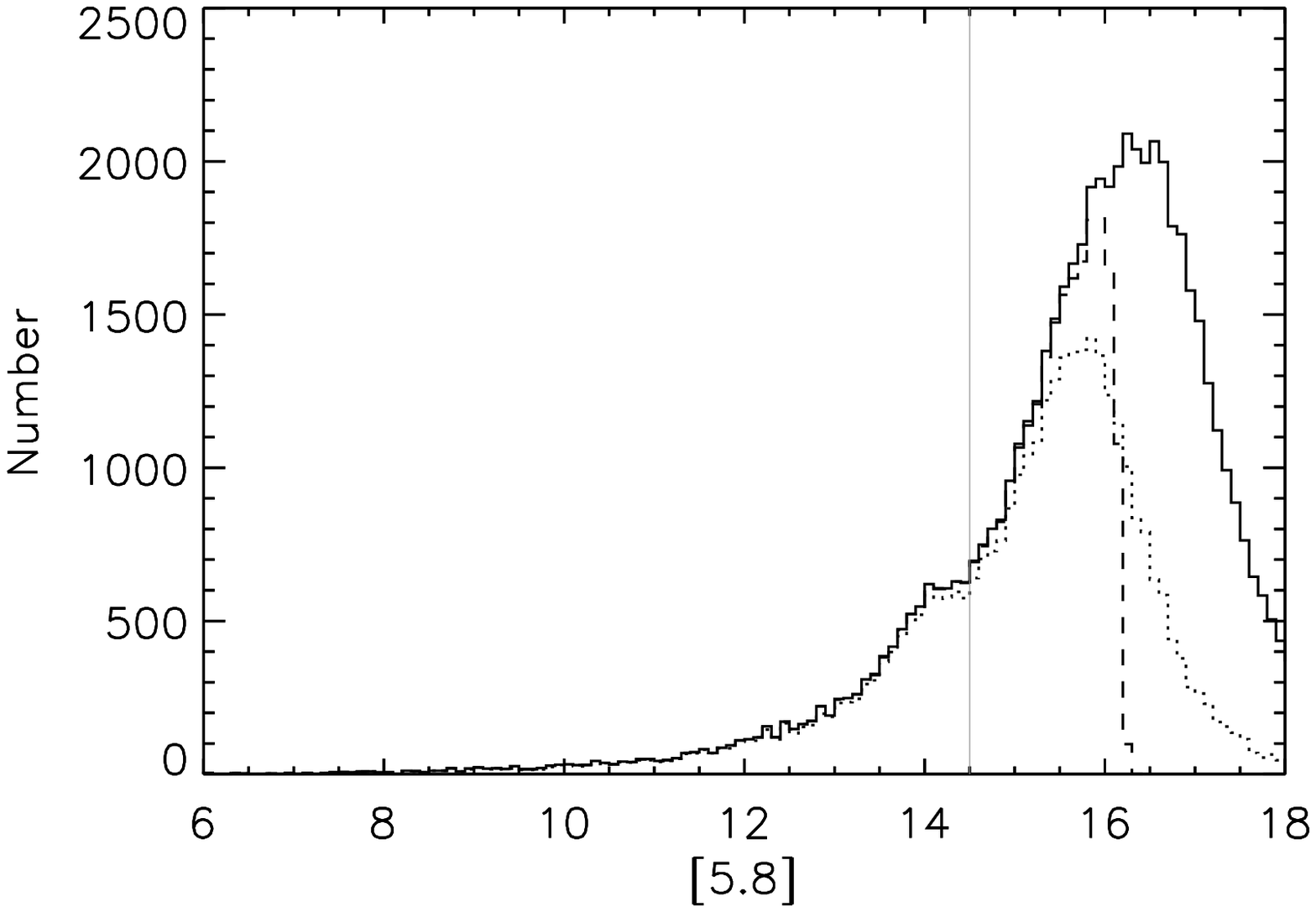}{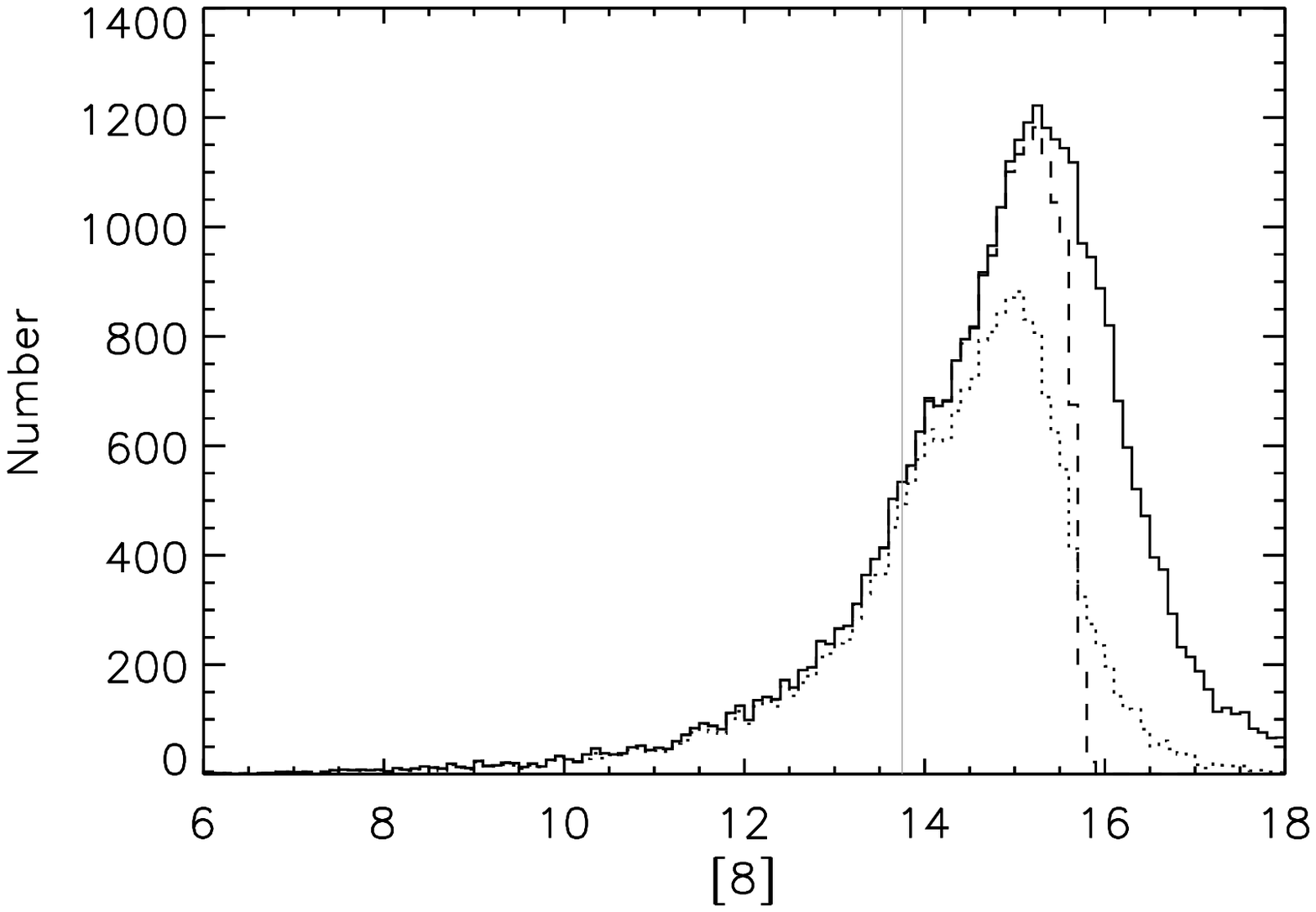}
\caption{Number counts in the IRAC bands.  The solid black lines identify 
all candidate detections, the dashed lines identify 5$\sigma$ detections, 
and the dotted lines identify confirmed point sources (those with detections in 
multiple bands).  The grey vertical lines identify where the source counts peak 
in Cycle 1 data presented by \citet{Currie2007a}.
%In each panel, the 
%vertical dotted line identifies the magnitudes at which the 
%number counts peak based on earlier, Cycle 1 data presented 
%by \citet{Currie2007a}.  }
}
\label{numbercounts} 
\end{figure}

Figure \ref{numbercounts} displays the number counts in each IRAC band as 
a function of magnitude.  In total, we detect $\sim$ 110,000, $\sim$ 120,000, 
$\sim$ 58,000, and $\sim$ 32,500 candidate point sources in the [3.6], [4.5], 
[5.8], and [8] filters, respectively.  
%Of these, 76,021; 67,504; 31,459; 
%and 23,439 are detected at the 5$\sigma$ level (dashed line).  Because they 
Of these, $\sim$ 76,000; $\sim$ 68,000; $\sim$ 31,000; 
and $\sim$ 23,000 are detected at the 5$\sigma$ level (dashed line).  Because they 
have matches in multiple IRAC or 2MASS filters, the vast majority 
%of the candidate detections are confirmed: 80,198 in [3.6], 80,418 in [4.5],
%34,242 in [5.8], and 20,750 in [8].
of the candidate detections are confirmed: $\sim$ 80,000 in [3.6], $\sim$ 80,000 in [4.5],
$\sim$ 34,000 in [5.8], and $\sim$ 21,000 in [8].

Compared to the data presented in \citet[][vertical grey lines]{Currie2007a}, 
our new data are $\sim$ 1.5--3 magnitudes deeper in each filter,
clearly probing far deeper down the cluster mass function and 
detecting fainter background sources.  The number counts peak at [3.6] $\sim$ 18, 
[4.5] $\sim$ 17.8, [5.8] $\sim$ 16.5, and [8] $\sim$ 15.5.  Both the distribution of 
5 $\sigma$ candidate detections and confirmed detections peak at substantially 
fainter limits.  
%While the greatest gain in depth is in the [3.6] and [4.5] channels, 
%we achieve similar detection frequencies in each band.  
%To the limits, we 
%detect 78,453, 69,766, 40,004, and 22,505 sources, about 75\% of the candidate 
%detections in each filter.  
 
%The new data are deeper not only because the integration times/pixel are $\sim$ 8 times 
%longer but also because the data were taken under better conditions and were 
%processed more carefully.  The \citet{Currie2007a} data were taken 
%during high solar activity, whereas these data were taken under normal conditions.  
%Artifact removal in these data is far better than in the Cycle 1 data, which 
%successfully removed cosmic ray hits but often failed to remove column pulldown and 
%muxstriping/muxbleeding effects near bright stars.   Finally, our MOPEX-based 
%overlap correction yielded far better estimates for background fluxes in each frame 
%than in the Cycle 1 data.  Noticeably absent is the 'tiling' effect that is pervasive 
%in the Cycle 1 mosaic, especially near the center of $\chi$ Persei.

\subsection{MIPS 24 $\mu$m Data}
\subsubsection{Image Processing and Photometry}
The Multiband Imaging Photometer for Spitzer \citep[MIPS,][]{Rieke2004} imaged 
h and $\chi$ Persei on March 15-16, 2008, October 25-26, 2008 and March 26 and 29, 2009 
as a part of General Observation Programs 40690 and 50664 (PI: Scott Kenyon).  
The program 40690 observations were focused on the cluster dominated regions with 
a typical integration time per pixel of 2000s, while the program 50664 data imaged 
the low-density halo regions surrounding the h Persei and $\chi$ Persei centers 
with a typical integration time per pixel of 1000s.  Combined, these observations 
provide MIPS 24 $\mu$m data for point-sources over a $\sim$ 0.4 square-degree area, covering 
the core-dominated membership population and much of the known halo population.

Using the Data Analysis Tool (DAT) pipeline, we processed individual BCD frames, removing 
latent image artifacts and bad pixels, and mosaiced the data using standard outlier 
rejection similar to that used in the MOPEX reductions of our IRAC data.  For 
point-source detection, we selected groups of pixels lying $>$ 5 $\sigma$ 
above the local background on the mosaic.  Point-spread 
function (PSF) fitting photometry was performed on detected sources on 
the mosaiced data using model PSFs from the \textit{Spitzer Science Center} webpage.

\subsubsection{MIPS Source Counts and Detection Limits}
\begin{figure}
\plottwo{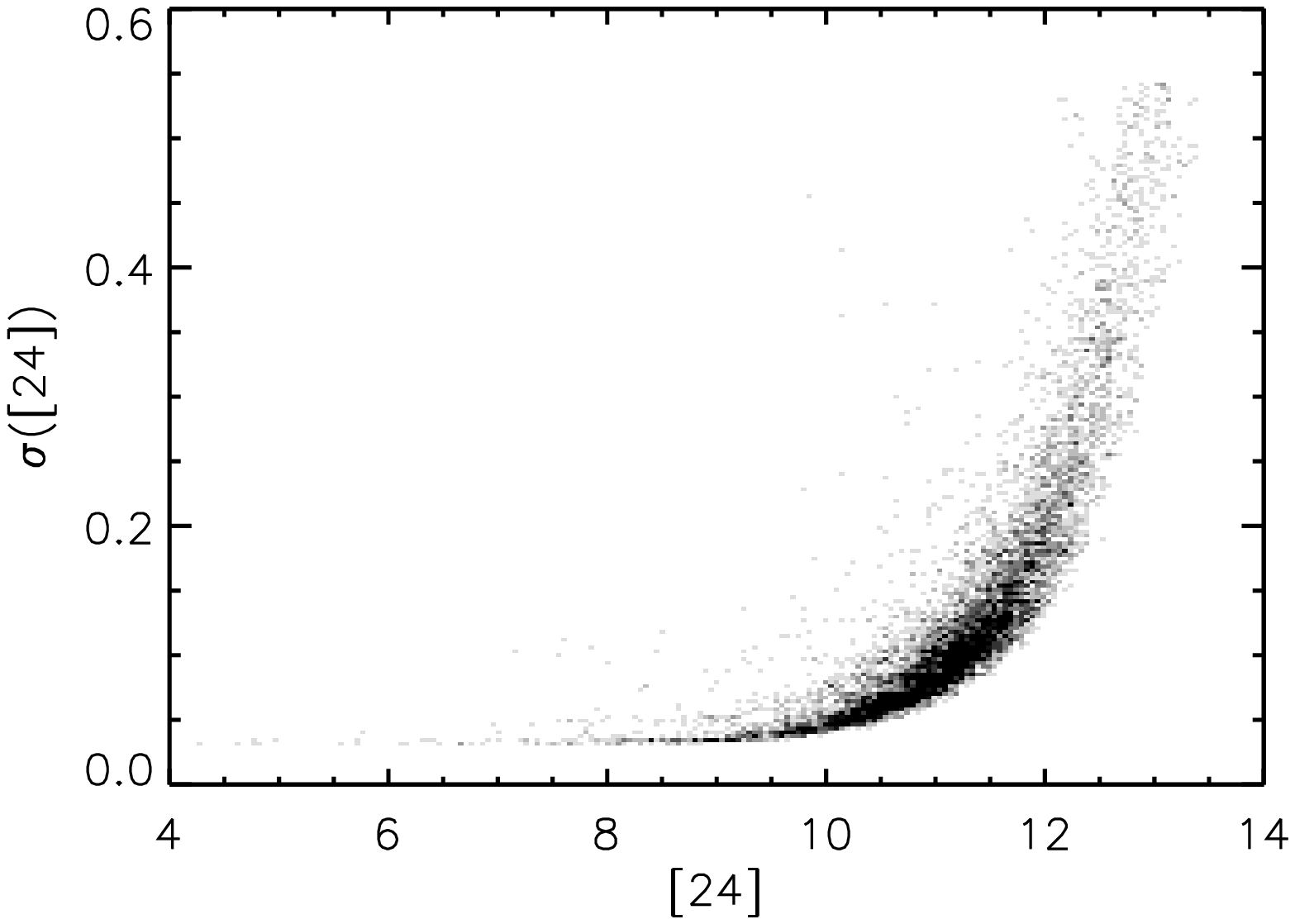}{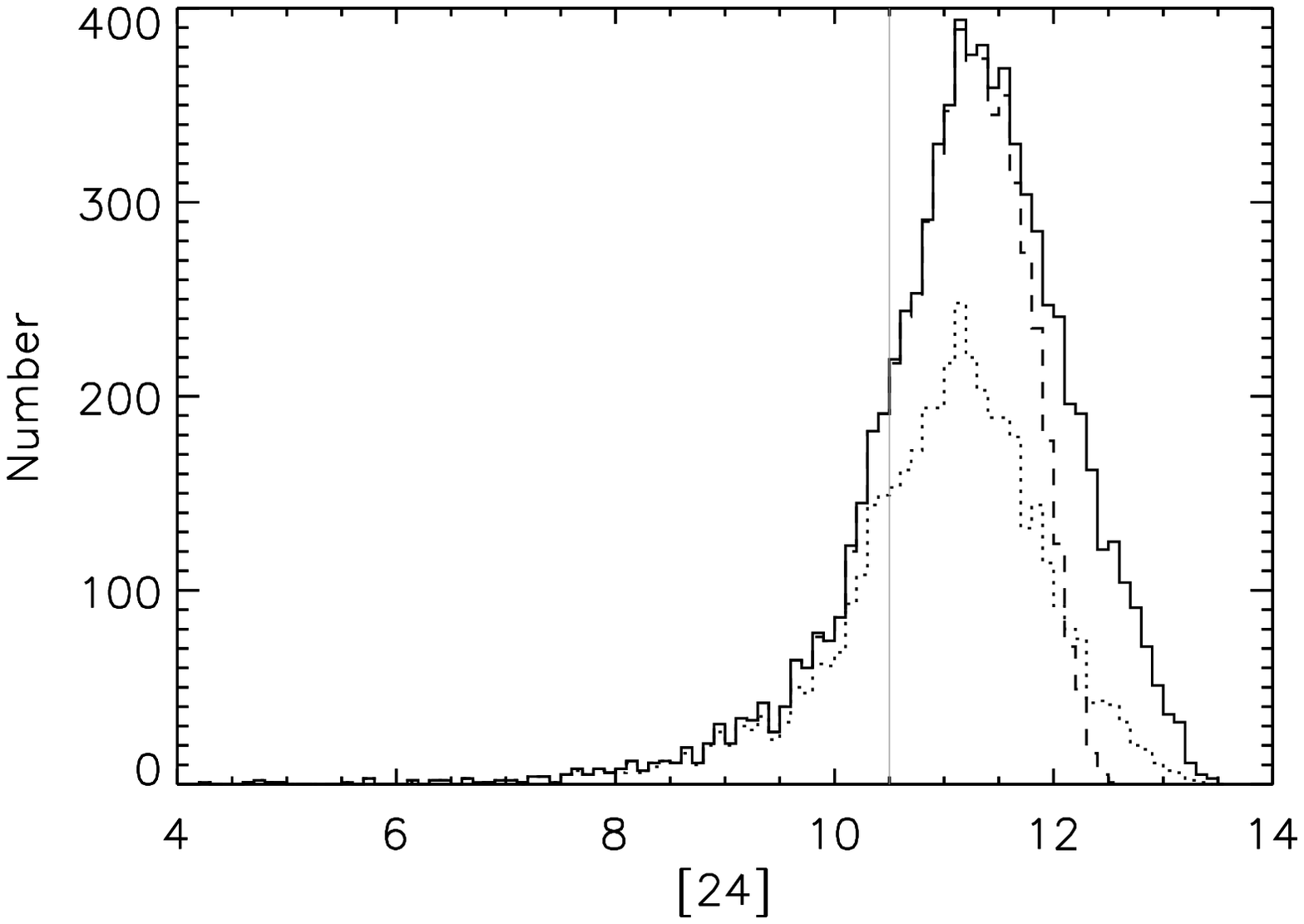}
\caption{Density plot of the magnitude vs. error distribution (left) and 
number counts (right) for MIPS 24 $\mu$m data.  Symbols are the same as in Figures 1 and 2. The grey vertical line identifies where the source counts peak in data presented in \cite{Currie2008a}.}
\label{mag24error}
\end{figure}

The MIPS data contain 7583 point sources. Of these, 6174 are detected at a 5$\sigma$ 
significance or better;
1098 (4270) have counterparts in the 2MASS (IRAC) data.
%We detect 7,583 candidate point sources in MIPS, 6,174 of which are 5$\sigma$ detections, and 
%4,322 have counterparts in the 2MASS and/or IRAC filters.  
%We detect 7,583 candidate point source with MIPS data. Of these, 6,174 are 5$\sigma$ detections and  
%1,098 (4,270) of which have counterparts in the 2MASS (IRAC) data. 
The distribution of 
[24] vs. $\sigma$([24]) is thicker than for the IRAC bands, largely due to the different survey 
depth from the two observing programs (Figure \ref{mag24error}, left panel).  The MIPS source counts 
peak at [24] $\sim$ 11.25 for the shallower program 50664 data, while 
the deeper program 40690 data yield 5-$\sigma$ detections about a magnitude fainter in regions near the cluster centers.
Compared to the MIPS data presented in \citet{Currie2008a} (vertical grey line), our new data are then
0.75--1.75 mags deeper.

\subsection{IRAC and MIPS Colors of All Detected Sources}
\begin{figure}
\centering
%%\plotone{ch14all.ps}
\plottwo{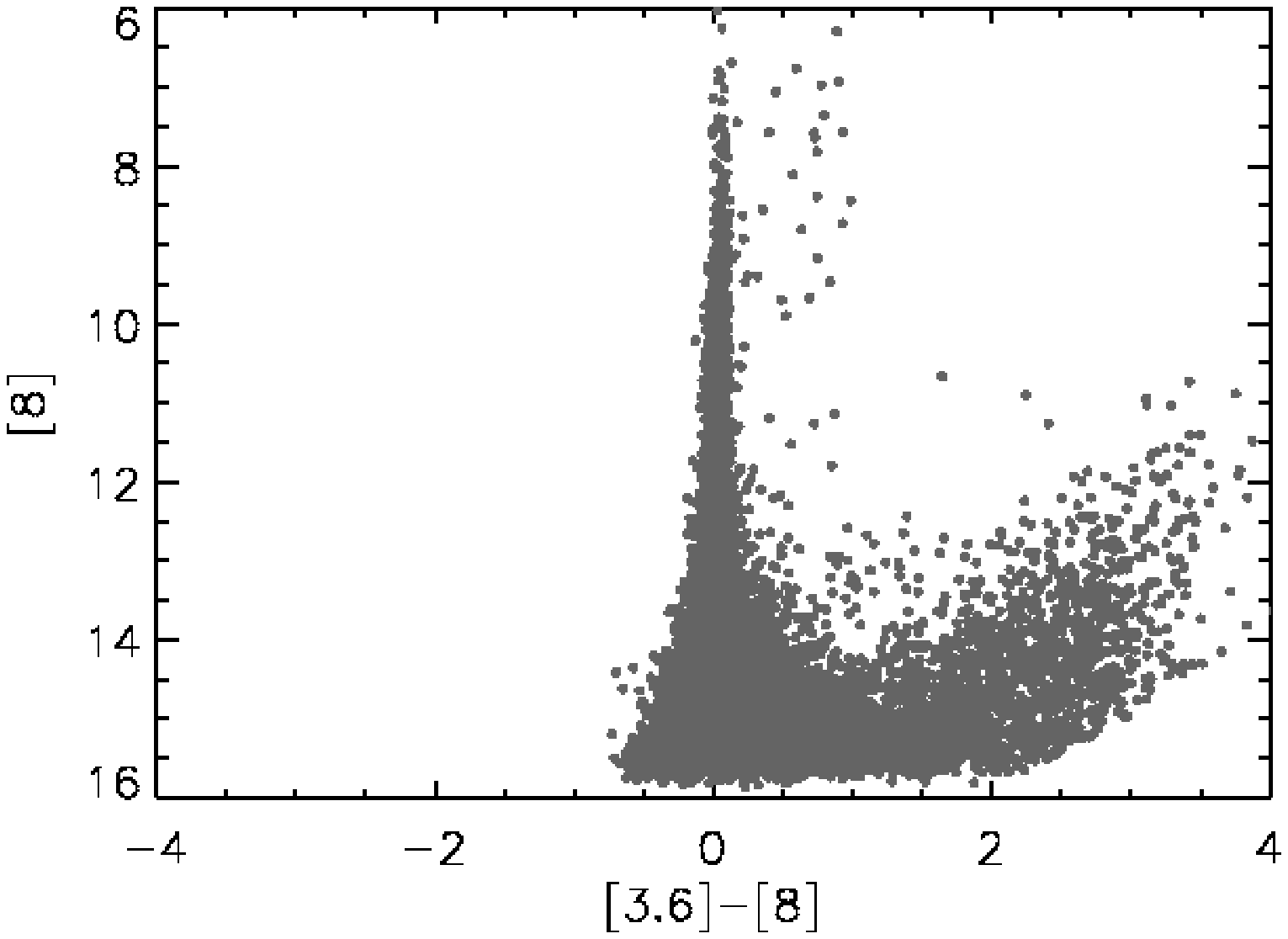}{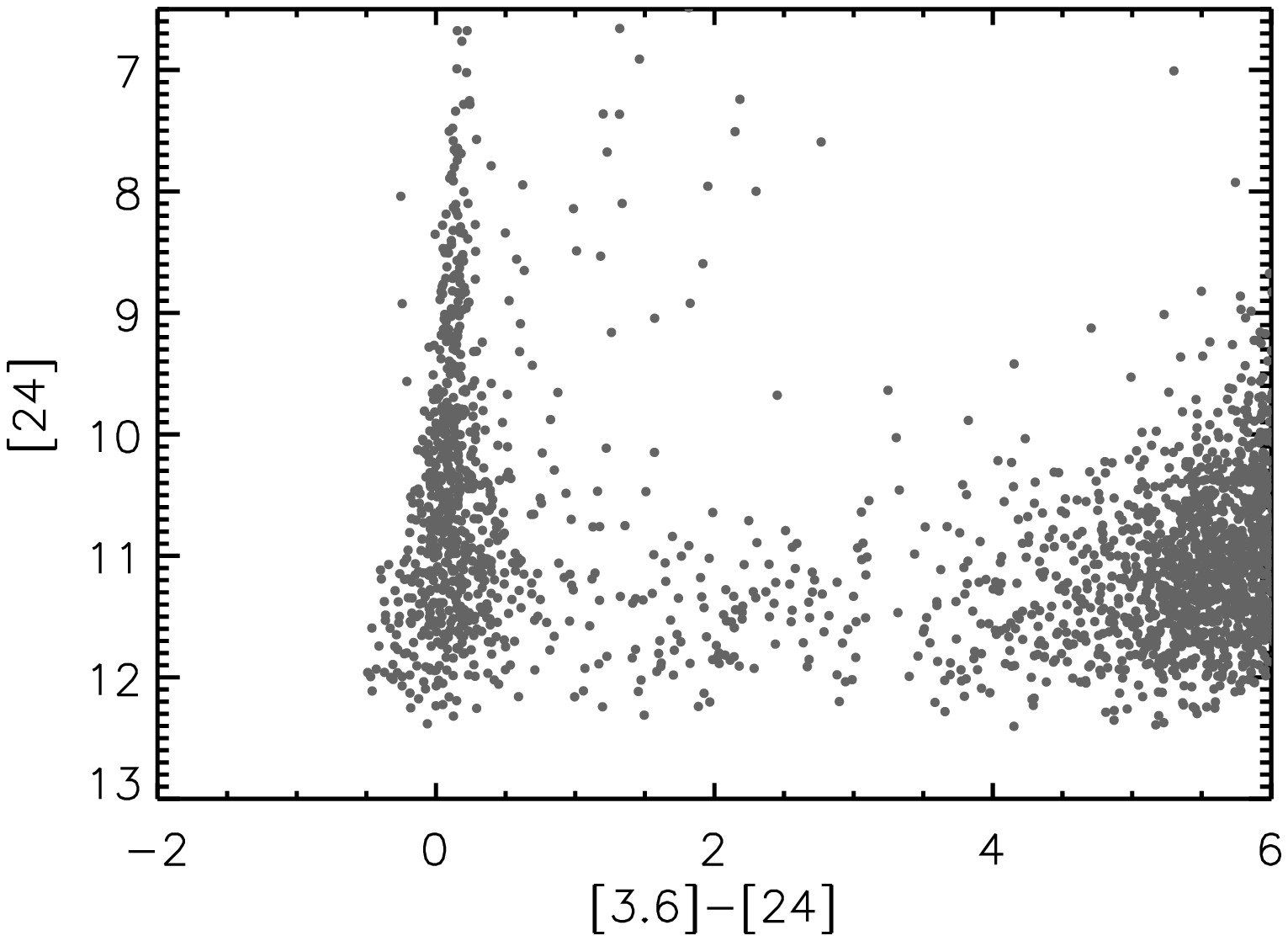}
\plottwo{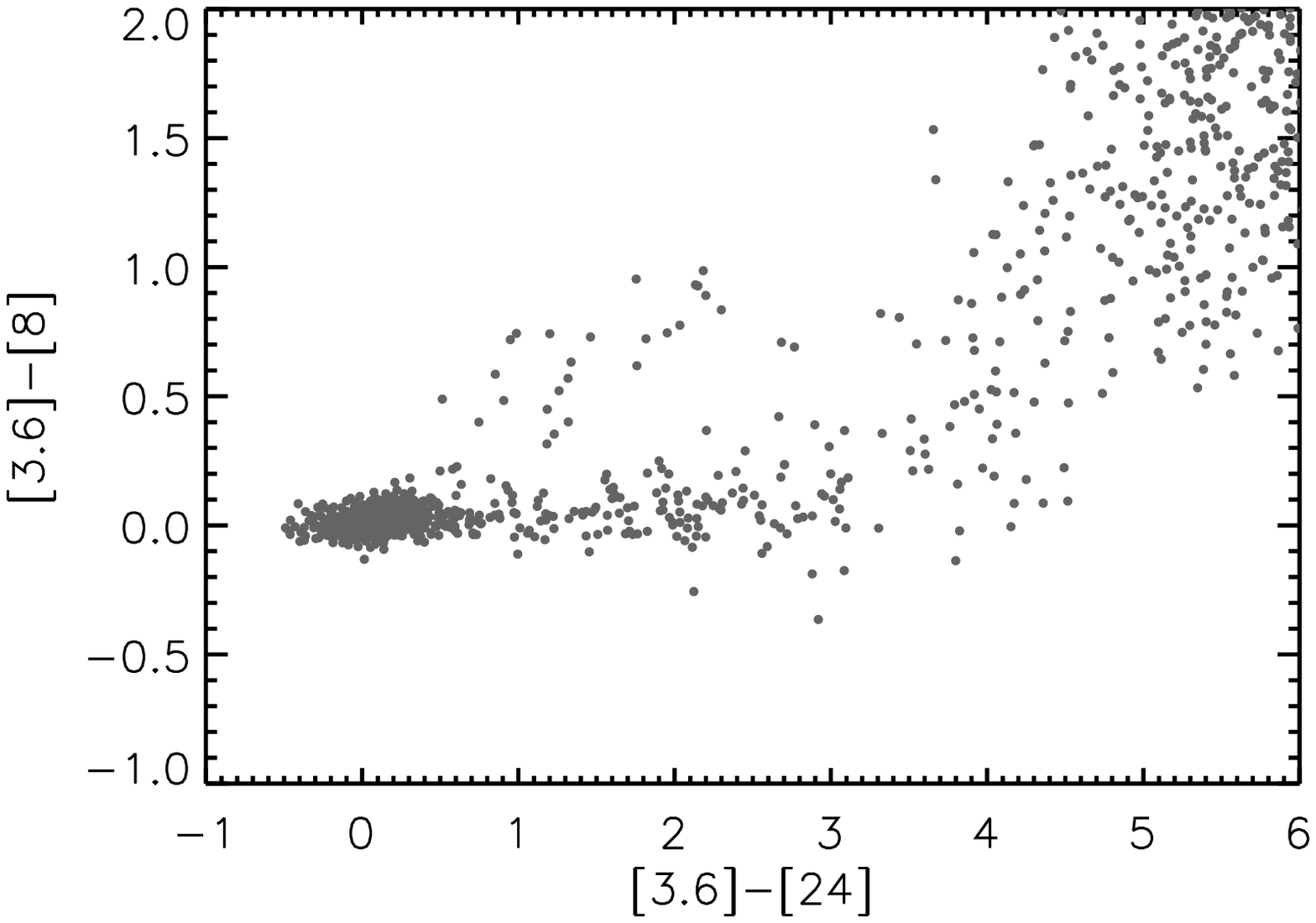}{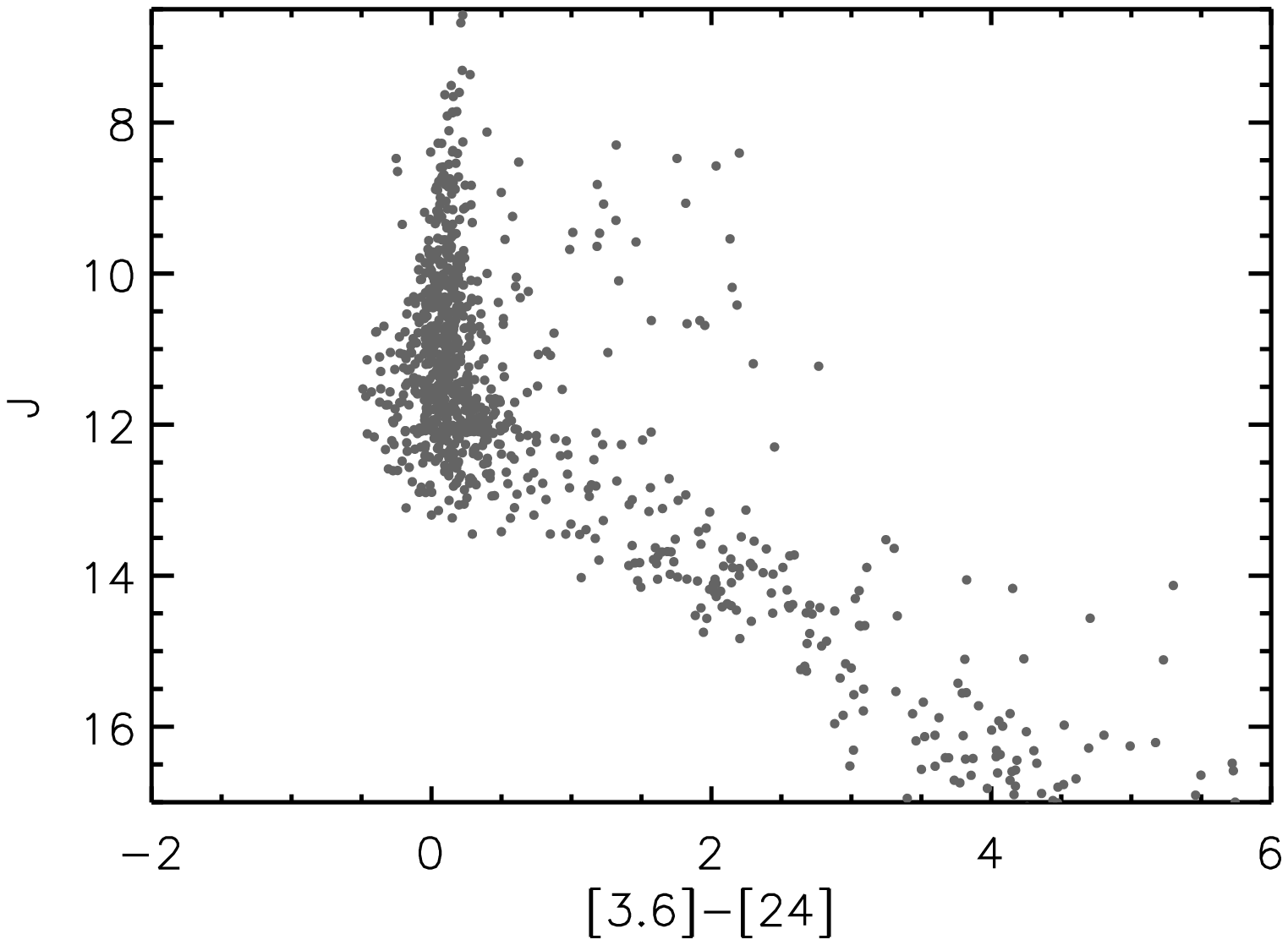}
%%\plottwo{j324all.ps}{ch14124_all.ps}
\caption{IRAC/MIPS color-magnitude and color-color diagrams for objects detected in multiple bandpasses.  
(Top-left) [8]/[3.6]-[8] diagram showing a well-defined main locus of objects with $\sim$ zero color, 
a population of weak, 0.25--1.5 magnitude excesses and a dense population of objects with red colors 
([3.6]-[8] $\sim$ 1.5--4).  (Top-right) [24]/[3.6]-[24] diagram showing more clearly the same 
three major populations: objects with zero color, a small weak-excess population ([3.6]-[24] $\sim$ 0.5--4.5) 
and a dense strong-excess population ([3.6]-[24] $\gtrsim$ 4.5).  (Bottom-left) [3.6]-[8]/[3.6]-[24] diagram 
showing that the objects with the strongest excesses in IRAC likewise have strongest excesses in MIPS, while 
those with [3.6]-[24] $\sim$ 0.5--4.5 include objects with and without IRAC excesses.  (Bottom-right) 
J/[3.6]-[24] diagram showing that we only detect objects with zero [3.6]-[24] color down to a 2MASS 
J magnitude limit of $\sim$ 12.5--13.}
\label{iracmipsall}
\end{figure}

Figure \ref{iracmipsall} displays the IRAC and MIPS colors for all targets 
detected in multiple bandpasses with photometric uncertainties less than 0.2 mags ($\sim$ 5-$\sigma$).  
Our sample includes a well-defined distribution 
of objects centered on zero [3.6]-[8] color (top-left panel), which widens from 
about 0.1 mag to 0.5--0.75 mag from [8] = 12 to [8] = 15.5.  Our sample also includes 
a large population of objects with very red [3.6]-[8] colors ($\gtrsim$ 1.5) 
consistent with those expected for young stellar objects or red, active galaxies 
\citep[i.e.][ see Appendix of this work]{Gutermuth2008,Gutermuth2009}.  The [8] vs. [3.6]-[8] diagram also shows a smaller 
population of objects slightly redder ([3.6]-[8] $\sim$ 0.25--1.5) 
than the main distribution of zero-color objects, consistent with the colors expected 
for highly reddened background stars; red, cool giants; Be stars; and pre-main sequence 
stars with optically-thin emission from remnant protoplanetary disks or warm debris disks 
\citep[][]{Currie2008a,CurrieSiciliaAguilar2011, KenyonBromley2004}.  The sample's distribution of [24] vs. [3.6]-[24] colors  
reveals the same kind of populations: objects with zero color, many with very red ([3.6]-[24] $\gtrsim$ 4.5) MIPS 
colors, and a much smaller population of slightly red objects ([3.6]-[24] $\sim$ 0.5--4.5).

The bottom panels of Figure \ref{iracmipsall} clarify the nature of the many objects with 
strong IRAC/MIPS excesses and the few with weaker excesses.  The objects with strong excesses in 
one band (IRAC or MIPS) must in general be extremely red from 1 $\mu$m to 24 $\mu$m, 
given the dense clustering of objects at [3.6]-[8,24] $\sim$ [1.25,5]--[2,6] (bottom-left 
panel); the bottom-right panels shows that these red objects preferentially have very faint $J$-band magnitudes.
These colors are consistent with those either for young stellar objects or for active galaxies.
Most of the objects with weaker 24 $\mu$m excesses appear to have very weak to 
negligible IRAC excesses, concentrated mainly along a narrow band from [3.6]-[8,24] $\sim$ [0,0.5] 
to [0.5,4.5], and have J $\sim$ 12.5--16.5, comparable to the near-IR brightnesses of 
AFGK members in h and $\chi$ Persei \citep{Currie2010a}.

 \subsection{Ancillary Data Establishing Cluster Members}
To identify and characterize disks surrounding h and $\chi$ Persei stars, 
we combine Spitzer data with optical/near-IR data for likely cluster members, 
updating the list from \citet{Currie2010a} with a more accurate one 
 of 13,956 stars (Table \ref{memberlist}) \footnote{Specifically, further inspection 
of the membership list presented in \citeauthor{Currie2010a} shows that 
our color-magnitude criteria identifying members from $VI$ photometry 
at our sensitivity limit ($V$ $\sim$ 23-24) and/or for the reddest objects ($V$-$I$ $\gtrsim$ 4) 
was insufficiently stringent, including some stars lying more than 0.75 mags above the main 
cluster locus.  We also removed several duplicate entries for the brightest stars, which were 
not previously flagged since their centroid positions from the three optical imaging datasets 
differed enough that they were not identified as the same source.  These steps reduce 
the number of Double Cluster members by $\sim$ 1\%.  Note that removing all of these 
sources has no bearing on the bulk cluster properties (distance, reddening, age) reported in \citet{Currie2010a}.}.
Optical spectroscopy confirms the stellar 
nature of sources, provides a better estimate of the stars' intrinsic 
fluxes by yielding their extinctions, and reveals whether the stars
show evidence for circumstellar gas accretion, which can be present for  
protoplanetary disks but is absent for debris disks.   In total, 12505 ($\sim$ 90\%) 
of the optically-selected cluster members have matches in our Spitzer data.
We briefly review the sources of these ancillary data.

\subsubsection{Optical and Near-Infrared Photometry}
Optical VI$_{c}$ photometry of h and $\chi$ Persei stars was originally taken as a 
part of the MONITOR program \citep{Aigrain2007} and covers $\sim$ 0.6 square 
degrees on the sky.  We supplement this photometry with that from \citet{Slesnick2002}, 
which was a shallower survey covering a wider, $\sim$ 1 square degree area.  
Cluster stars with optical photometry range in mass from bright, $\sim$ 15--20 M$_{\odot}$ 
supergiants to very faint, $\sim$ 0.2 M$_{\odot}$ pre-main sequence M stars 
\citep[see][]{Currie2010a}.

Near-IR JHK$_{s}$ band photometry is drawn from the 2MASS All-Sky Survey \citep{Skrutskie2006}.  
The survey has nominal 10 $\sigma$ limits of J = 15.8 and K$_{s}$ = 14.3.  However, 
the number counts of detections in the h and $\chi$ Persei field suggest 
completion limits of J = 15.5 and K$_{s}$ = 15 \citep{Currie2007a}, where the 
formal photometric uncertainties reach $\sim$ 0.1 mags and 0.2 mags, respectively.

\subsubsection{Optical Spectroscopy: Identifying Accreting Stars}
Finally, we add optical spectroscopy of h and $\chi$ Persei stars.  The 
 survey presented in \citet{Currie2010a} includes $\sim$ 11,000 
stellar spectra with sufficiently high signal-to-noise to determine 
accurate spectral types.
  \citet{Currie2010a} limited their membership analysis to 
spectroscopically-observed stars with V band and 2MASS photometry (7,465), which 
produced a list of 4,702 members.
Many other stars with spectra presented in \citet{Currie2010a} but without 
optical photometry (and thus not analyzed for membership) may in fact be 
members of the Double Cluster and can be confirmed after obtaining 
new optical photomery.   
%Because the focus of \citet{Currie2010a} was characterizing stellar properties 
%and establishing membership, we did not further analyze stars with H$_{\alpha}$ 
%emission and other accretion diagnostics in that paper.  

Here, we consider evidence for stars accreting gas from protoplanetary 
disks amongst the entire set of spectra listed in \citet{Currie2010a}.
We derive $EW(H_{\alpha})$ with SPTCLASS \citep{Hernandez2004}, which uses the standard 
IRAF routine \textit{sbands}.  Manually comparing $EW(H_{\alpha})$ with SPTCLASS-derived 
estimates for this sample and for the previously reported sample from \citet{Currie2007c} 
shows excellent agreement.  
%Also using SPTCLASS, we derive equivalent widths for 
%$H_{\beta}$ and CaII.  

Based on these measurements, we identify accretors from the 
subset of spectra not previously analyzed in \citet{Currie2007c} and revisit 
the \citet{Currie2007c} classifications.  The spectra analyzed in \citet{Currie2007c} are largely 
drawn from 2006 MMT/Hectospec observations focused on $\chi$ Persei, the less massive of the two 
regions of h and $\chi$ Persei \citep[see ][]{Currie2008phd}.  Thus, newly-identified accretors 
are mainly drawn from regions surrounding h Persei and the low-density halo population.  Our
 analyses are described in \S 4.

\subsection{Combined Spitzer and Ground-Based Sample Characteristics}

\subsubsection{IRAC and MIPS Completeness}
To select point sources with dust emission, we rely on the 8 $\mu$m IRAC 
data and the 24 $\mu$m 
MIPS data. Within the far more sensitive [3.6] and [4.5] IRAC images, we 
detect nearly all (90\%--95\%) 
cluster members from $V$ $\sim$  6 ([3.6] $\sim$ [4.5] $\sim$ 3-5) to $V$ = 24 
([3.6] $\sim$ [4.5] $\sim$ 17--17.5).
Members lacking detections fall outside of the IRAC coverage. 
Thus, the ratio of the detection frequencies at [8] ([24]) and [4.5] 
provide a robust estimate of the 
completeness at 8 $\mu$m (24 $\mu$m). Figure \ref{spitcomp_mem} shows the result.

%Figure \ref{spitcomp_mem} depicts the fraction of h and $\chi$ Persei members within the 
%Spitzer survey regions that are detected with IRAC 8 $\mu$m and MIPS 24 $\mu$m, the 
%filters we use to identify evidence for warm and cold dust emission.  The fraction of 
%members detected at the far more sensitive IRAC [3.6] and [4.5] filters is nearly 
%%constant at $\sim$ 90--95\% from the bright magnitude limit (m(V,[3.6],[4.5]) $\sim$ 3--5) to 
%V = 24.  Members lacking detections fall outside of the IRAC coverage.  Therefore, we 
%use the ratio of objects detected at 8 $\mu$m and those detected at 4.5 $\mu$m to 
%define our completeness at 8 $\mu$m.  While the coverage overlap between IRAC and MIPS 
%is not 100\%, we still use the ratio of 24 $\mu$m detections to 4.5 $\mu$m detections 
%to assess completeness.  

Figure \ref{spitcomp_mem} (top panels) assesses our sample completeness.
Our 8 $\mu$m completeness diverges from 100\% at V $>$ 17 and drops below 50\% by V = 22.  
Our 24 $\mu$m completeness drops from a maximum value of $\sim$ 80\% at V = 11 
to 25\% by V = 13.  In \S 4, we derive statistics for disk emission in a given filter 
only for h and $\chi$ Persei stars where we detect at least 25\% of the sample.  Thus, 
at 8 $\mu$m, we focus on stars brighter than V = 22.  At 24 $\mu$m, we derive 
statistics for stars brighter than V = 13.5.  

The bottom panels of Figure \ref{spitcomp_mem} 
%compare the $V$ vs. [8] and $\sigma$([8]) distributions to 
assess the level to which the 5-$\sigma$ cutoff we will impose for analyzing our IRAC detections (Section 3) may bias our 
results.  For stars with $V$ = 11--19, 19--20, 20--21, and 21--22, the typical [8] photometric uncertainties 
are $\sigma$([8]) $\approx$ 0.05--0.1, 0.1, 0.15, and 0.175, respectively.  For these $V$ magnitude bins, the percentage of stars
detected at better than a 5-$\sigma$ level at [8] is 99.5+\%, 95\%, 86.3\%, and 76\%, respectively. 
We obtain similar results for $V$ vs. [24] and $\sigma$([24]).  For the brightest stars, all MIPS-24 detections are 
5-$\sigma$ detections, while the majority of detections near our faint limit ($V$ = 13.5) are likewise 5-$\sigma$.

\subsubsection{Masses of Spitzer-Detected h and $\chi$ Persei stars}

Double Cluster members targeted by Spitzer cover a wide range of optical/IR brightnesses, 
spectral types, and thus stellar masses.  To estimate the typical stellar masses corresponding 
to an h and $\chi$ Persei star of a given optical brightness, we use two sets of 
isochrones -- \citet{Baraffe1998} for stars less than 1.4 $M_{\odot}$ and 
\citet{PallaStahler1999} for more massive stars.  To map between the 
stars' effective temperatures predicted from these isochrones and spectral type and 
median $V$-band magnitude, we first adopt the
effective temperature scale listed in \citet{Currie2010a}\footnote{Note that the stellar mass sampling from the 
\citet{PallaStahler1999} isochrones is quite coarse, and thus values for the upper main 
sequence in h and $\chi$ Persei are less precisely determined than for the 
pre-main sequence.  However, we opt to use these 
isochrones since they yield an age for other young associations -- specifically, Upper Scorpius -- 
that can agree with the age derived from \citet{Baraffe1998} \citep[cf.][]{Preibisch2002}.  Other isochrones with 
better sampling may yield spurious stellar mass-dependent age spreads \citep{Currie2010a}.  Furthermore, 
while the sampling in mass is coarse, the resultant sampling in spectral type and $V$-band magnitude is 
not (see Table \ref{obsint}).  We will further discuss stellar ages in \S 6.}.
We compute the median $V$-band magnitude for stars of a given mass/spectral type directly for BAF stars 
since such stars are well sampled from our spectroscopic survey \citep{Currie2010a} and are on/very near 
the main sequence so the adopted $T_{eff}$ scale should match that of field dwarfs.  For lower-mass stars 
poorly probed by our spectroscopic survey, we simply adopt a distance of 2.34 $kpc$ and a mean Double Cluster 
reddening of E(B-V) = 0.54 (derived mostly from BAF stars).  From these assumptions, we determine the predicted 
$V$-band magnitude for these lower-mass, later-type stars.

Table \ref{obsint} shows the expected spectral type and $V$ band magnitudes for cluster members with 
different stellar masses.  Stars with $V$ $\approx$ 12, 13, 15, 18, 19, 20, and 22 
typically have spectral types of SpT $\approx$ B2.5, B4, A1.5, K0, K3, K5, and M1.5, respectively.
Adopting our hybrid Palla \& Stahler/Baraffe 14 Myr isochrone to map between spectral type/$T_{eff}$ 
and stellar mass, stars with these spectral types typically have stellar masses of M$_{\star}$ $\approx$ 
6, 5, 2, 1.2, 1.1, 0.95, and 0.6 $M_{\odot}$.  Thus, given our IRAC/MIPS completeness limits, we 
will later only derive the 8 $\mu$m excess frequency for stars with M$_{\star}$ $\approx$ 0.6--6 M$_{\odot}$
and the 24 $\mu$m excess frequency for stars with M$_{\star}$ $\approx$ 4--6 M$_{\odot}$ (\S 3.2 and 3.3).

%\section{Infrared Colors of h and $\chi$ Persei Stars}
\section{Circumstellar Dust in Disks Around h \& $\chi$ Persei Stars} \label{sectdust}
Table \ref{spitzerphot} lists photometry for all $\sim$ 12,500 h and $\chi$ Persei members with 
Spitzer data.  Here, we first examine the Spitzer colors for cluster members to 
identify stars with evidence for circumstellar dust.  Next, we 
quantify the frequency of warm dust emission at 8 $\mu$m and set 
limits on the population of members with slightly cooler dust emission probed by MIPS-24 $\mu$m data.
For all analyses, we restrict our sample to stars with 5-$\sigma$ detections.

\subsection{IRAC and MIPS Colors: Evidence for Warm Circumstellar Dust Emission}
\begin{figure}
\centering
\plottwo{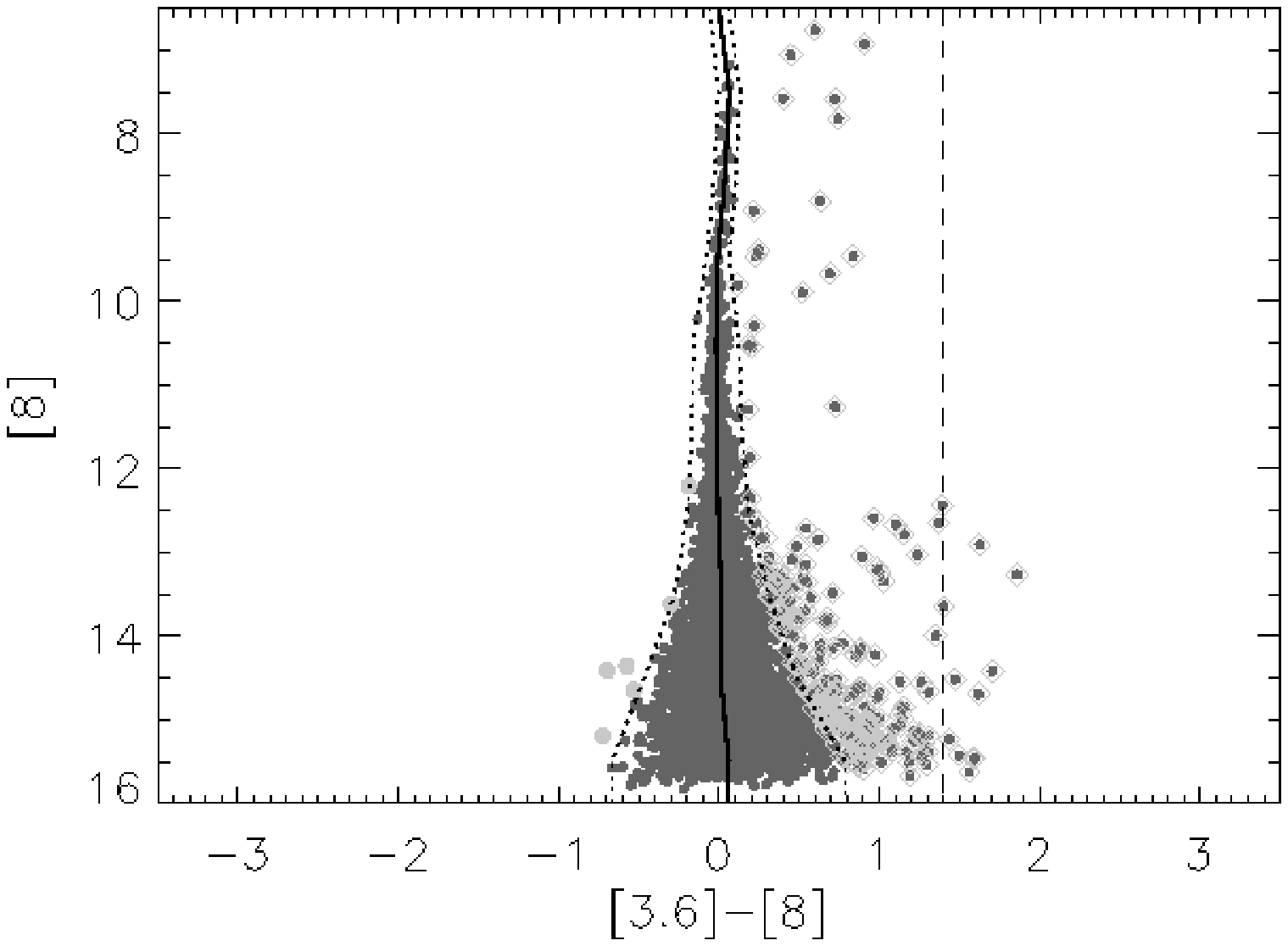}{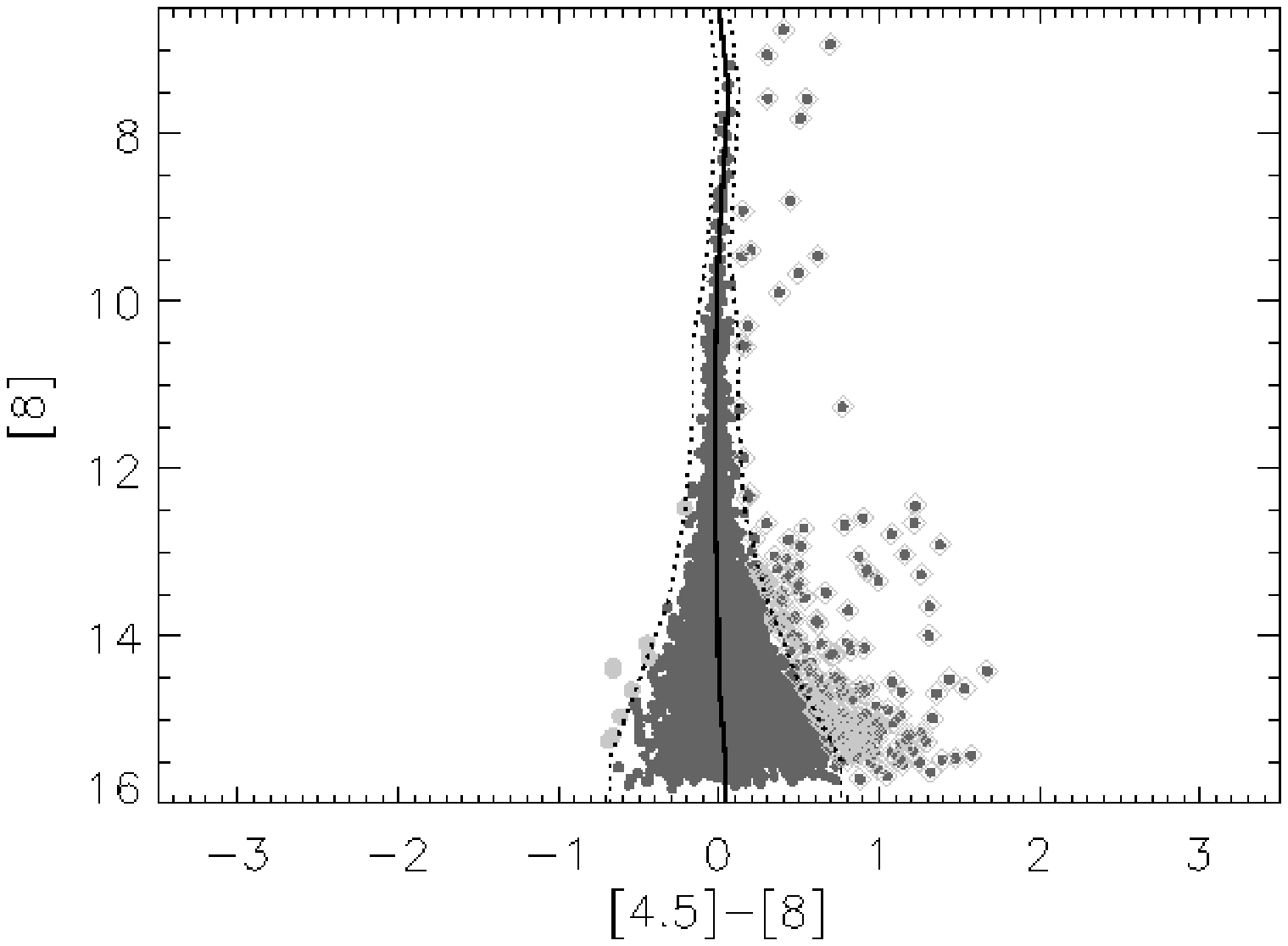}
\plottwo{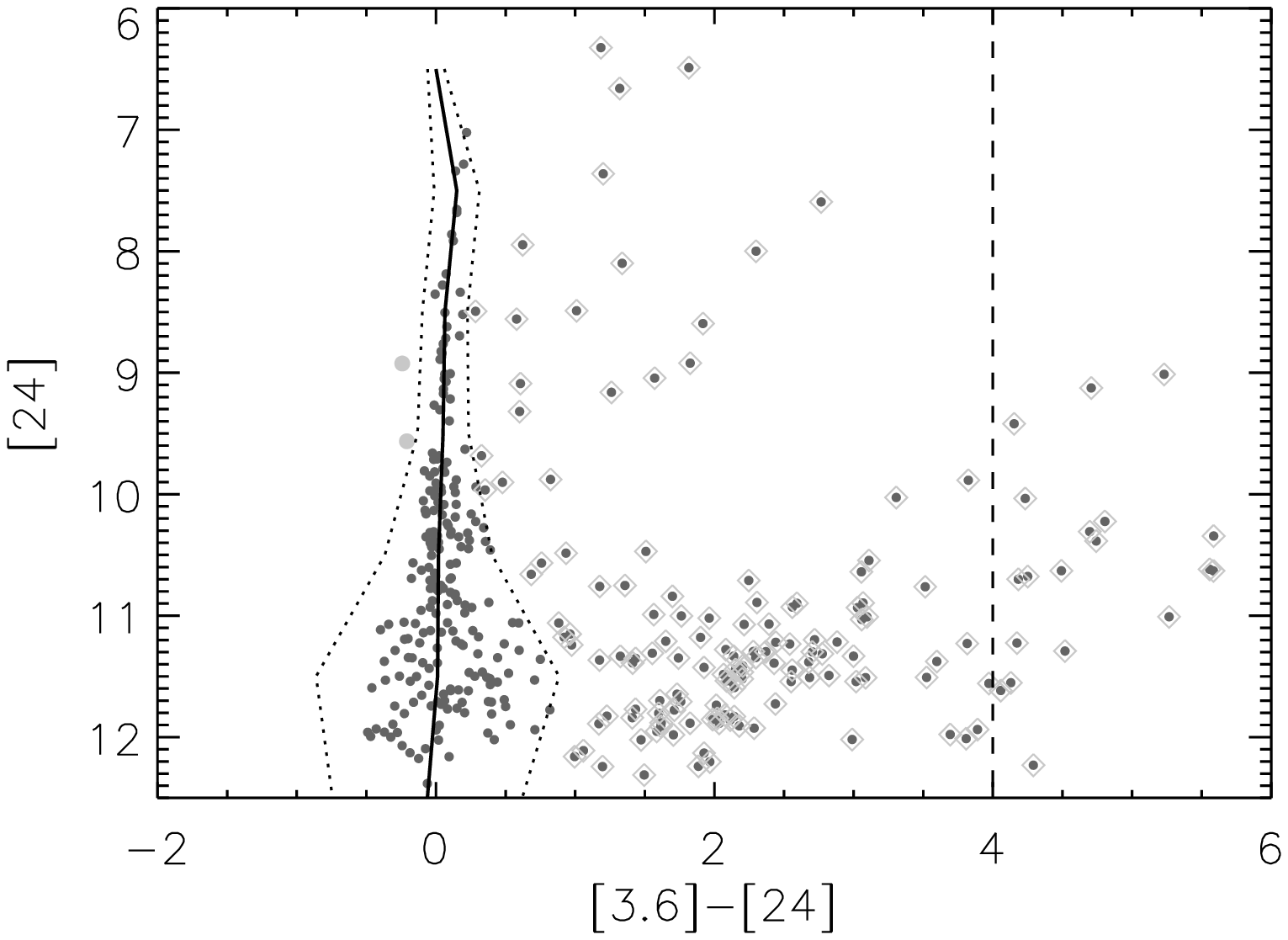}{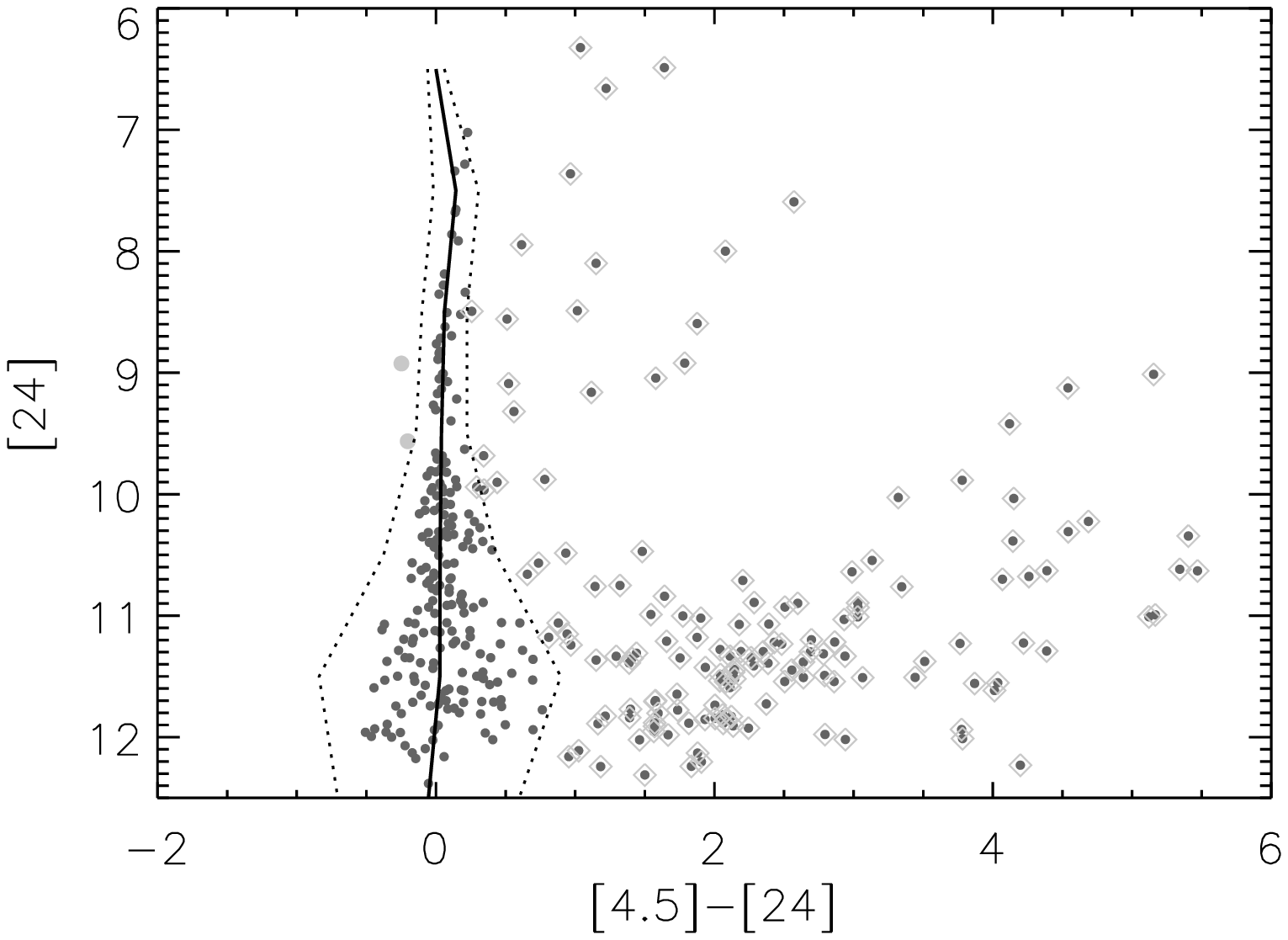}
\caption{Color-magnitude diagram identifying IR excess sources (diamonds) and negative outliers (grey dots) compared to the full 
distribution of h and $\chi$ Persei sources.  The solid lines identify the median color for a given [8] or [24] magnitude and the 
dashed lines separate objects whose colors are consistent with bare photospheres in a given passband and those deviant from 
this distribution by more than $\sim$ 3-$\sigma$.  The vertical dashed lines in the top-left and bottom-left panels depict the typical colors for 
blue protoplanetary disks in Taurus drawn from Figure 5 in \citet{Luhman2010}.  }
\label{iracmipsmem}
\end{figure}

Figure \ref{iracmipsmem} displays the IRAC/MIPS color-magnitude diagrams
for h and $\chi$ Persei members.  The members-only distributions clearly differ 
from those of the population of all objects, as they include very few objects 
with extremely red IRAC and MIPS colors (i.e. [3.6]-[8] $>$ 2, [3.6]-[24] $>$ 5).  
As shown in the Appendix, the objects with very red colors are most consistent with being
PAH-emission galaxies or active galactic nuclei, whereas our members-only sample 
shares little to no overlap with the range of galaxy colors.  There may be 
some other bona fide, hitherto unidentified cluster members with circumstellar 
disks amongst our full sample of Spitzer-detected sources.  However, 
focusing only on those objects previously identified as members via optical photometry/spectroscopy 
allows us to derive statistics on the Double Cluster's disk population with 
significantly less extragalactic contamination.

The top panels of Figure \ref{iracmipsmem} show the IRAC [8] vs. [3.6]-[8] and 
[8] vs. [4.5]-[8] distributions for cluster members.  
The main locus of colors for cluster members is centered roughly about zero with a dispersion that broadens 
from $\pm$ 0.05--0.06 mags to $\sim$ 0.75 mags for the faintest stars detected at the 5-$\sigma$ level.   
The distributions exhibit some evidence for a gap in the excess population 
at [8] = 11, where excess sources brighter than this limit all have [3.6]-[8] $<$ 0.9, while 
fainter excess sources include some with slightly redder colors.
Still, the vast majority of fainter stars with colors redward of the main locus exhibit [3.6]-[8] colors 
$\sim$ 1 magnitude or less. 

To provide a comparison with our sample, we overplot [3.6]-[8] colors typical of blue protoplanetary disks 
in Taurus from Figure 5 in \citet[vertical line][]{Luhman2010} \footnote{The disagreement between excesses around most Taurus stars 
and those around h and $\chi$ Per stars is even stronger than Figure \ref{iracmipsmem} suggests by eye.  
\citeauthor{Luhman2010} sample includes many stars with excess emission 
at [3.6] \citep[cf.][]{Hartmann2005}.  Few, if any, h and $\chi$ Persei stars have excess emission at [3.6] \citep{Currie2007a}.  
Thus, the 3.6--8 $\mu m$ emission for Taurus stars relative to their photospheres is even stronger than h and $\chi$ Per's stars.  
}.  Our sample includes over 10,000 stars and the cluster is only slightly older than clusters/moving 
groups containing some stars with optically-thick, near-IR luminous disks (e.g. TW Hya, Upper Scorpius).
Nevertheless, we identify very few stars whose [3.6]-[8] colors overlap with the range of colors
typical for protoplanetary disks.   
  Instead, IRAC excesses for nearly all h and $\chi$ Per stars are comparable to the level of emission from the two varieties 
of transitional disks \citep{CurrieLada2009,CurrieSiciliaAguilar2011} and 
warm debris disks indicative of terrestrial planet formation \citep{KenyonBromley2004,Currie2007b,
Rhee2007,Lisse2009}.  

The MIPS colors (bottom panels) show similar trends.  Here, the 
distributions of [3.6,4.5]-[24] colors broaden from $\sim$ 0.05 magnitudes at the bright end 
again to a wider distribution for the faintest stars detected at [24].  Stars brighter 
than 9th magnitude have excesses up to $\sim$ 2--2.5 magnitudes.  While many stars fainter 
at [24] likewise exhibit these weak excesses, they also include some with 
[3.6,4.5]-[24] = 3--5.5, comparable to the 24 $\mu$m excesses exhibited by the 
most luminous debris disks \citep{Rieke2005,Low2007,Currie2008a} and some transitional disks 
\citep{CurrieSiciliaAguilar2011}.  Few stars have [3.6]-[24] colors overlapping with most Taurus stars.

\begin{figure}
\centering
%%\plottwo{ch14.ps}{m24324.ps}
\plottwo{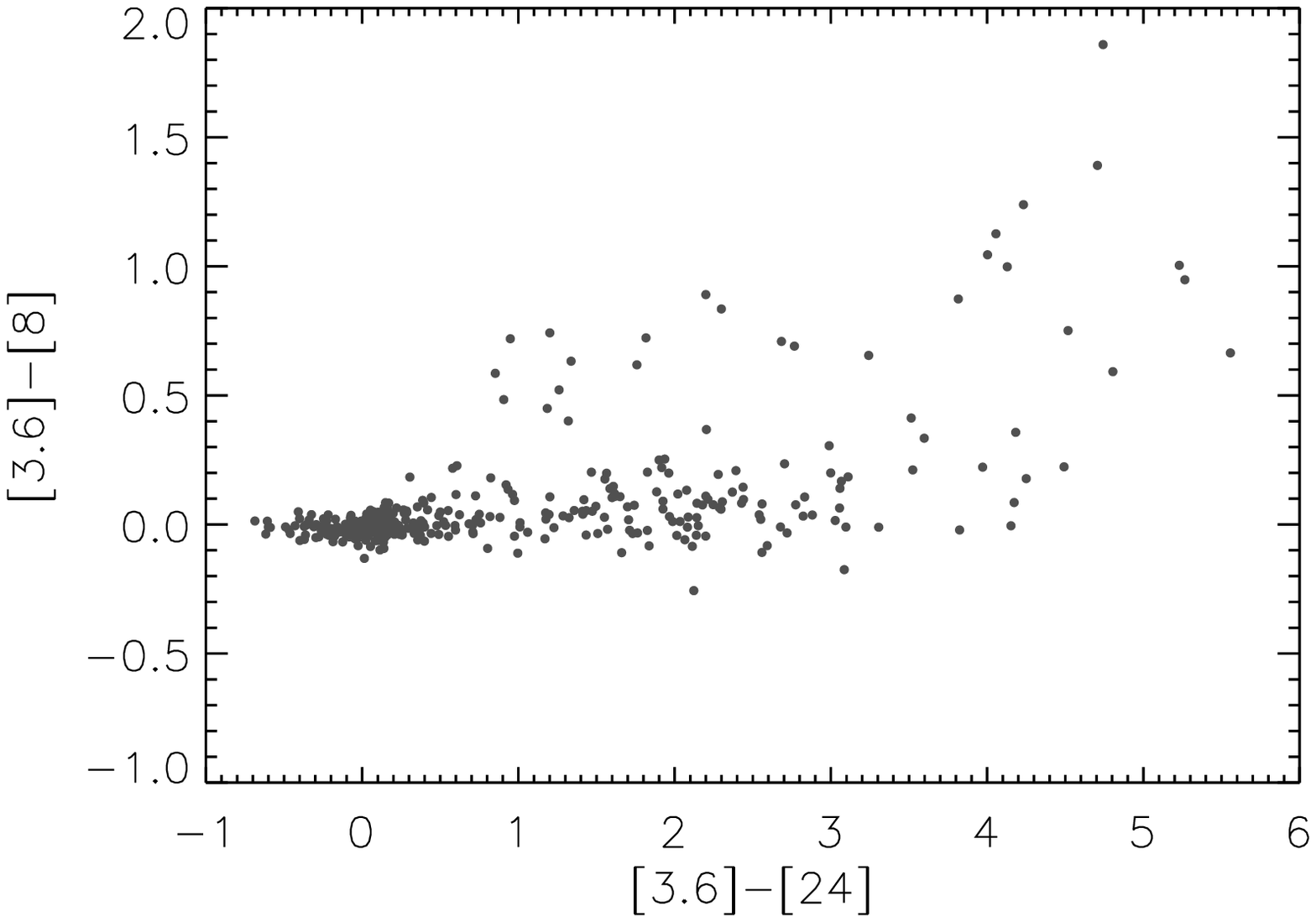}{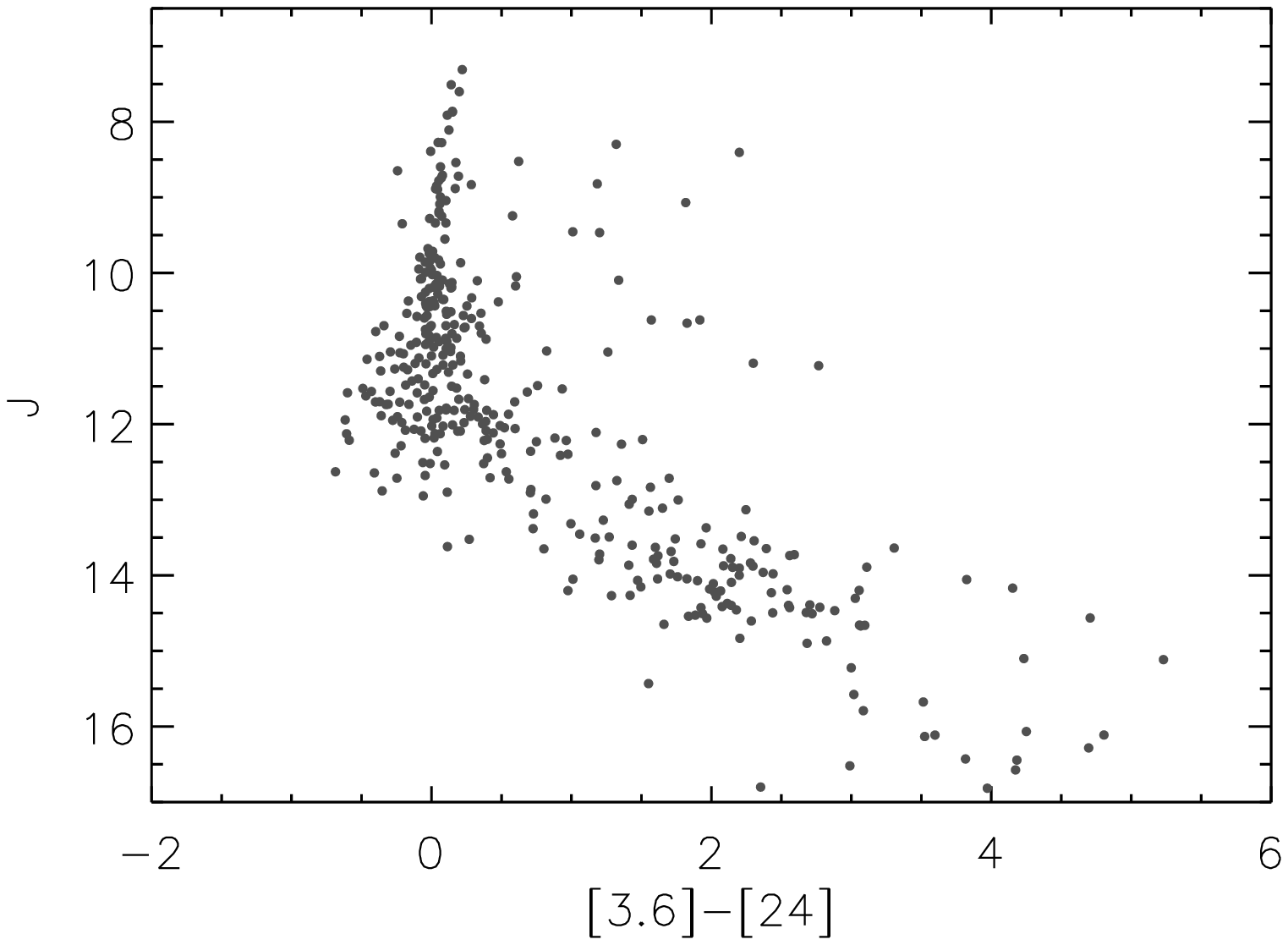}
%%\plottwo{j324.ps}{ch14124.ps}
\caption{Same as Figure \ref{iracmipsall} (bottom panels) except just for h and $\chi$ Persei members.}
%  For the [8]/[3.6]-[8] 
%and [24]/[3.6]-[24] diagrams we overplot the extent of the main locus of photospheric colors (dotted lines) and 
%identify stars flagged either as having excesses consistent with them having circumstellar 
%disks (or, perhaps in a few cases, contamination from red, active galaxies) or unphysically blue colors indicative of bad photometry (grey circles).}
\label{iracmipsmemexc}
\end{figure}

\begin{figure}
\plottwo{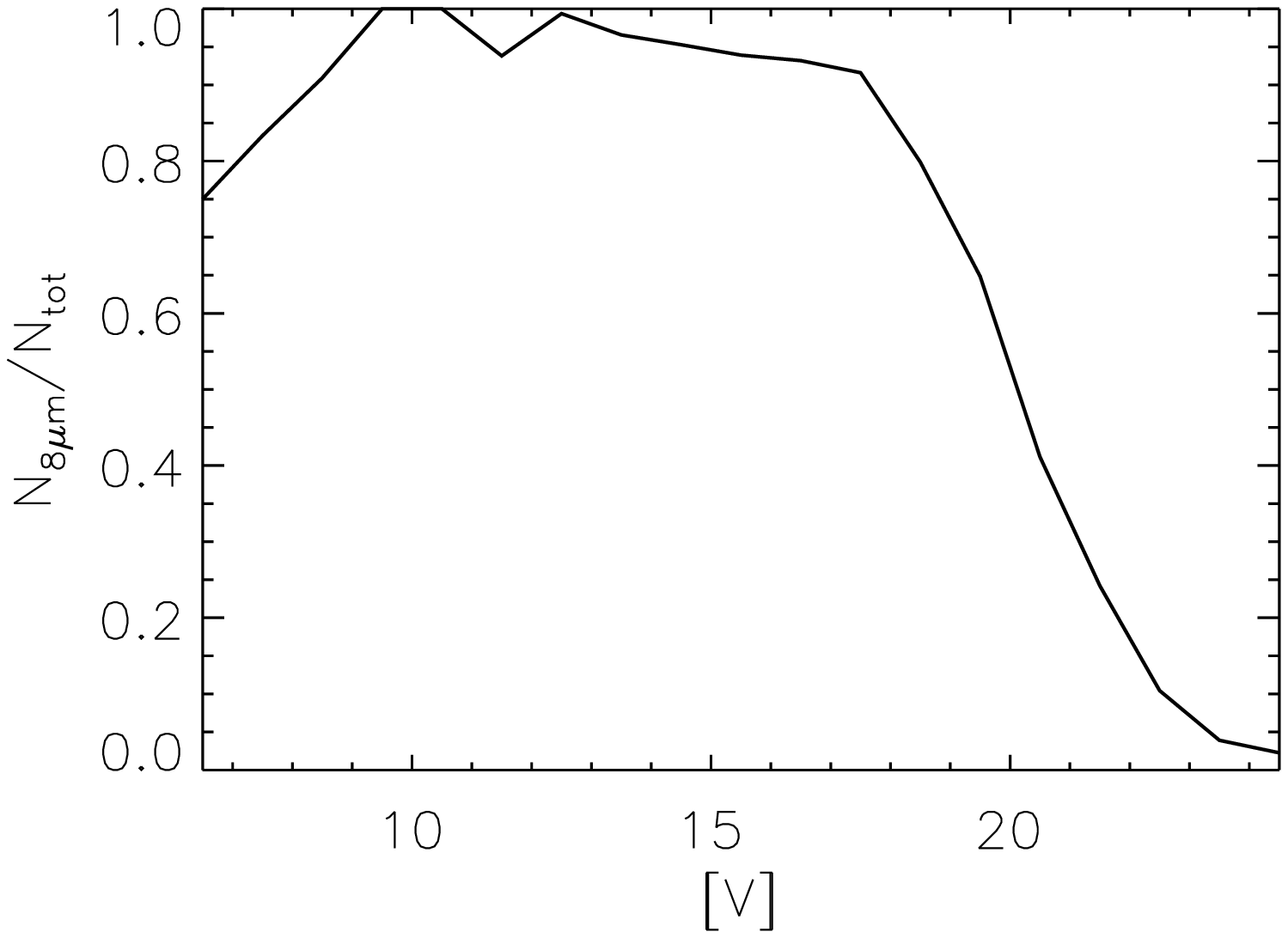}{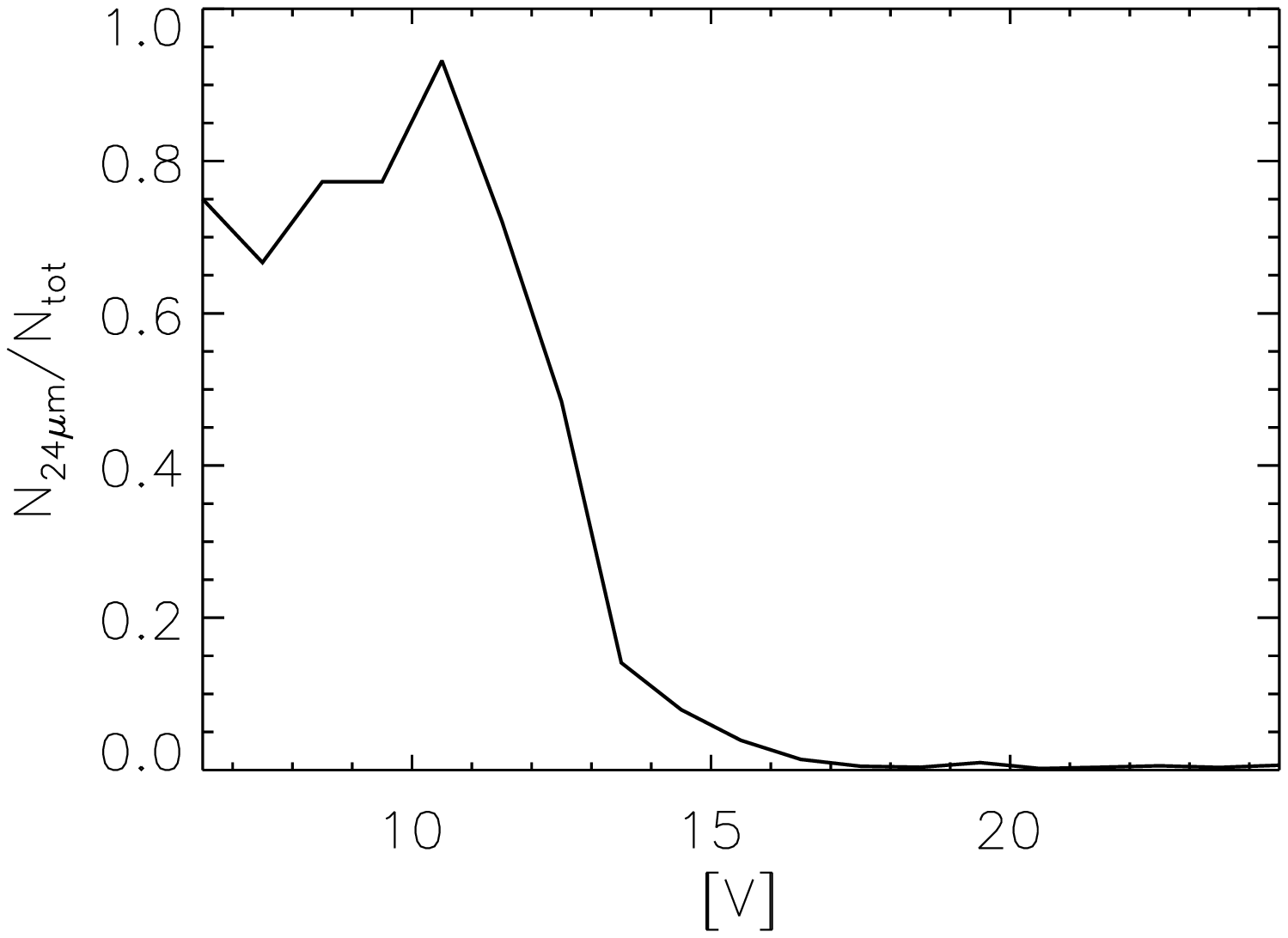}
\plottwo{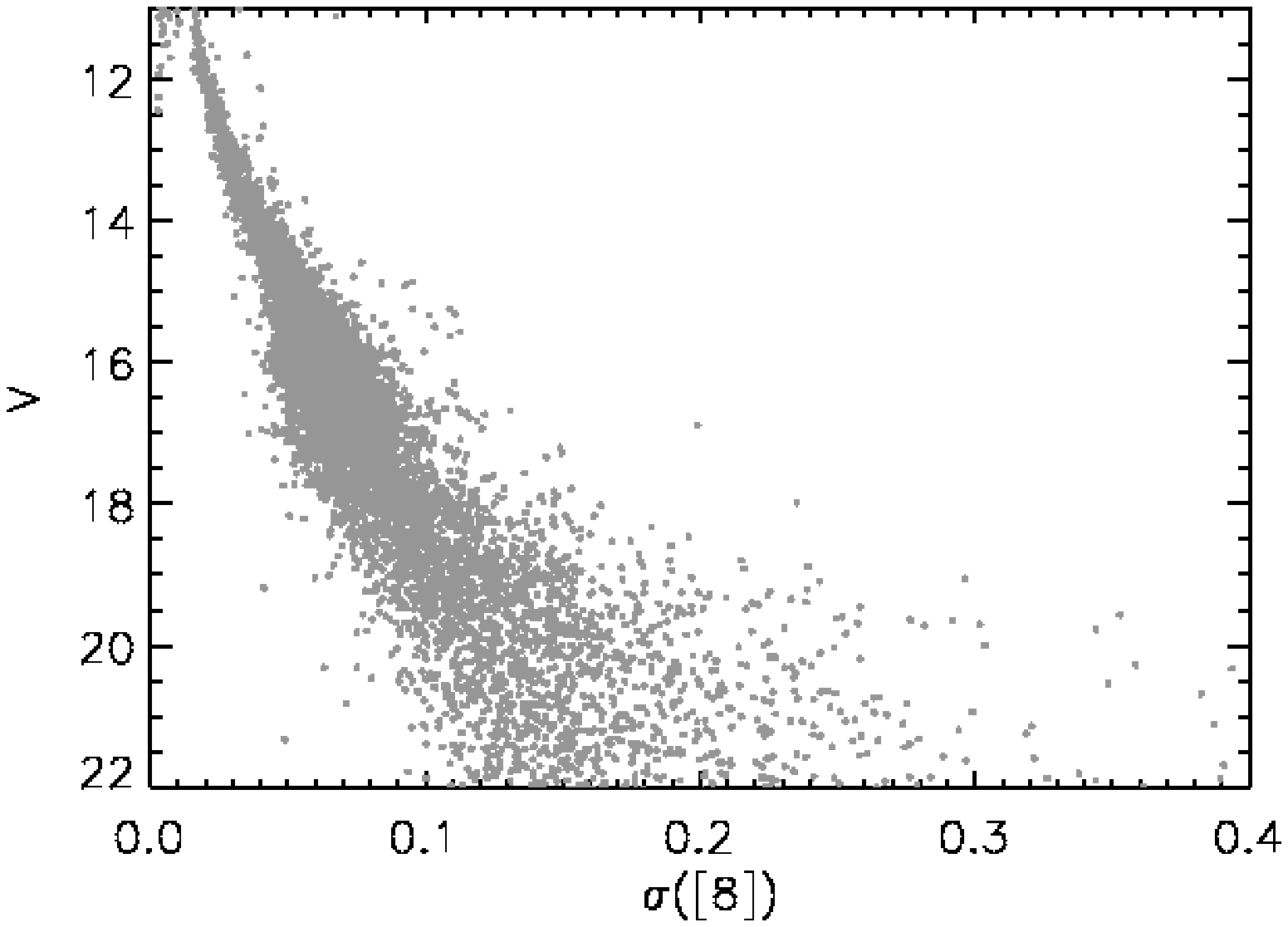}{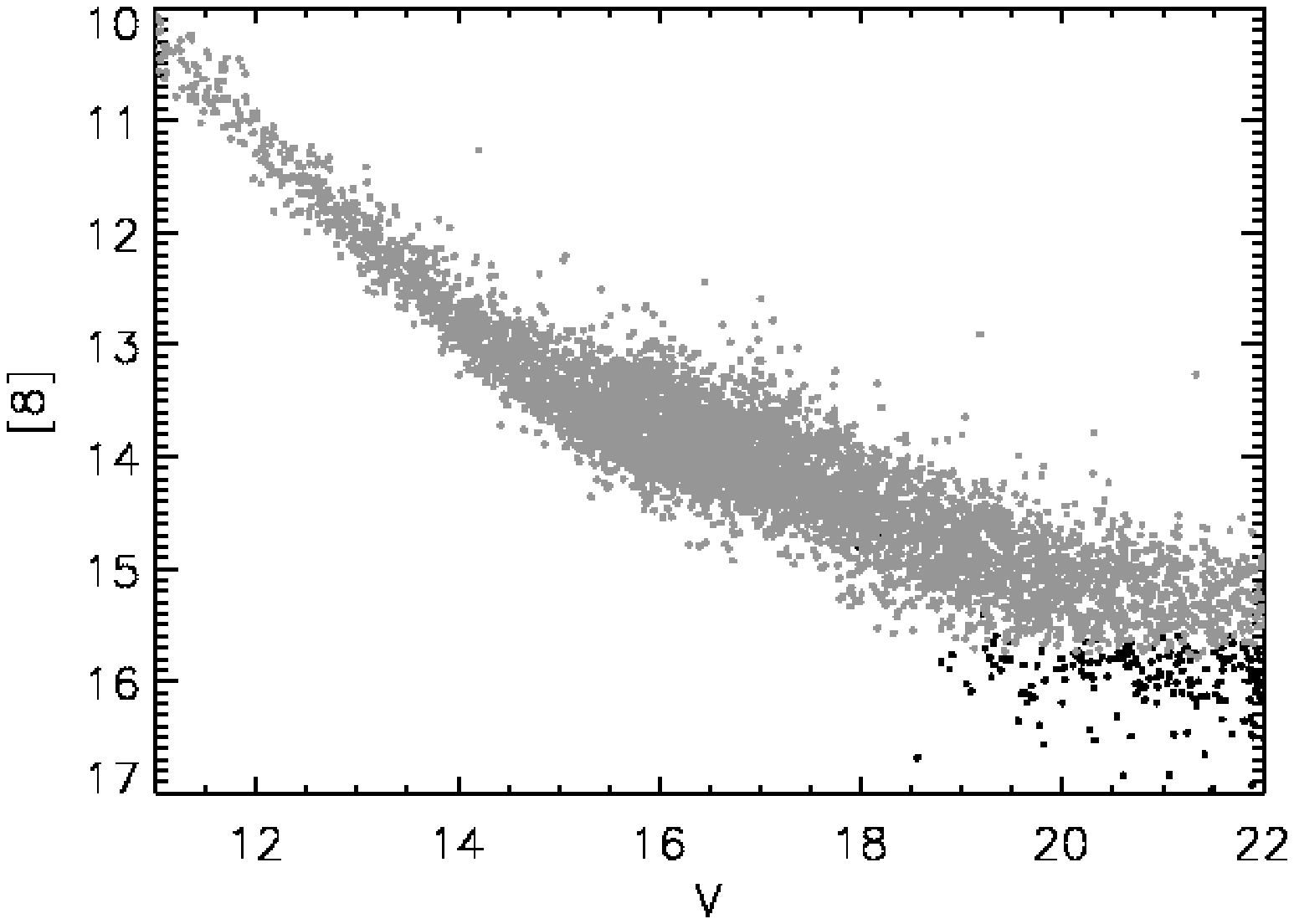}
\caption{(Top) Completeness at [8] and [24] for cluster members.  We restrict our statistical analysis 
to stars with $V$ $\le$ 22 at 8 $\mu$m and $V$ $<$ 13 at 24 $\mu$m. 
(Bottom) $V$ vs $\sigma$([8]) and $V$ vs. [8] for our sample.  In the righthand panel, the light grey dots identify stars with 5-$\sigma$ [8] detections.  Nearly all stars with lower signal-to-noise ratios are fainter than [8] = 15.5.}
\label{spitcomp_mem}
\end{figure}

Figure \ref{iracmipsmemexc} further elucidates the IR excess population for h and $\chi$ Persei.  
As shown in the left-hand panel, most of the stars with 24 $\mu$m excess emission 
have nearly zero color at [3.6]-[8], though most of the ones with the largest 24 $\mu$m excesses 
are likewise red at [3.6]-[8].  All of the MIPS-detected stars with [3.6]-[8] excess have clear 24 $\mu$m excess.  
Thus, there are many more stars with excess at [24] than at [8]; stars 
with bona fide [8] excesses are virtually certain to have excesses at [24] as well, although 
they may fall brow our [24] detection limit.
%with bona fide [8] excesses not detected at 24 $\mu m$ likely fall below the 24 $\mu m$ detection 
%limit rather than lacking excesses at this wavelength.

The right-hand panel of Figure \ref{iracmipsmemexc} further demonstrates that for much of our sample 
we are only identifying the stars with the most luminous 24 $\mu$m excess emission.  
From the results in Section 2.5, the 24 $\mu m$ data only reach the photospheres of B-type stars.  
Any later-type stars detected in this band must have excess emission.
%Here, all the stars fainter than $\sim$ 12th magnitude in $J$ have a 24 $\mu$m excess, whereas 
%brighter stars detected at 24 $\mu$m primarily have photospheric emission.   Thus, we can 
%only derive statistics for the frequency of 24 $\mu$m excess emission for the brightest 
%stars in the Double Cluster as was suggested in \S 2.5.

%\subsection{Frequency of IRAC-Excess Emission from Warm Dust}
\subsection{IRAC-Excess Emission from Warm Dust}
\subsubsection{Identifying IRAC-Excess Sources}
To identify stars with statistically-significant excess emission at 8 $\mu$m, we follow methods 
similar to those described in \citet{Hernandez2006} and later work, defining the threshold for an infrared 
excess source based on the source color compared to the intrinsic dispersion of colors at a given 
magnitude.  The Double Cluster members analyzed here have [8] magnitudes as bright as $\sim$ 
5th magnitude and as faint as $\sim$ [8] = 15.5, where the former's photometric errors are negligible 
and the latter's are as large as 0.2 magnitudes.  While the IRAC colors for young stellar photospheres show 
little variation for stars earlier than mid-M \citep[cf.][]{Luhman2010,Pecaut2013}, binarity 
 could in principle artifically redden the system by up to $\approx$ 0.05 mags (e.g. see Table 13 in Luhman et al. 2010).  
Additionally, while reddening for most Double Cluster 
members falls within a range of E(B-V) = 0.4--0.7 with a median value of 0.52--0.56, \citet{Currie2010a} 
identified some stars best modeled as being far more heavily reddened and some with low reddening.  
Given various infrared extinction laws \citep{Indebetouw2005,Flaherty2007}, this range in 
reddening could induce a spread in color of up to $\approx$ 0.05--0.06 magnitudes in 
[3.6, 4.5]-[8].  Finally, the IRAC detector also suffers 
small gain variations across the field depending on whether bright point sources landed on 
a given pixel in the observations taken immediately prior \citep{Knutson2007}, and small 
errors in centroiding position due to bright nearby stars may also affect flux and background 
determinations.  

Therefore, instead of defining our threshold for an IR excess source based on intrinsic 
photometric errors, we empirically derive the spread in photospheric colors and identify excess 
sources as statistical outliers.  For each color-magnitude diagram (e.g. [8] vs. [3.6]-[8]), 
we first divide the distribution  into 
0.5 magnitude bins and derive the median color and standard deviation for each bin.  We then use a 
quadratic interpolation to determine the median color and standard deviation at each 
magnitude.  Nominally, we identify 3-$\sigma$ outliers as those with evidence for IR excess.  
To be conservative, we include an additional 0.06 magnitude buffer to account for color variations 
due to binarity and differences in reddening, an amount that also roughly corresponds to the spread in IRAC color 
for the brightest cluster members with negligible photometric errors. 

Thus, IR excess sources are defined as those with color $\ge$ median(color) + 3$\times$$\sigma_{color}$ + 0.06. 
%Similarly, those with bad photometry have colors $\le$ median(color) - 3$\times$$\sigma_{color}$ - 0.06. 
The empirically-derived dispersion in color is always greater than the quadrature-added 
uncertainties in the photometric measurements.
Compared to the criterion previously used in \citet{Currie2007a}, our threshold is less stringent 
for brighter stars but far stricter for faint stars since it varies as 3$\times$$\sigma_{empir}$, not 
1$\times$$\sigma_{phot}$.

We first compute the frequency of the IRAC-excess population as a function of 
%observed properties, namely the 
%deep 
the stars' $V$ and $I$ band photometry.  
%While nearly all stars brighter than $V$ = 19 and $I$ = 17 are detected at 8 $\mu$m,
%the detection rate drops to below 40\% for stars fainter than $V$ = 22 and $I$ = 19 (Figure \ref{spitcomp_mem}).  The similar limit for 
%MIPS-24 $\mu$m is signficantly brighter ($V$ = 13).  Therefore, 
%As described before, 
We restrict our analysis to stars brighter than $V$ = 22, where
 our completeness drops well below 50\% (see \S 2.5).  Furthermore, the 
percentage of stars whose 
[8] detections whose high photometric uncertainties ($\sigma$([8]) $>$ 0.2) preclude us from considering them here 
grows to 14--24\% for $V$ = 20--22.
Therefore, our excess frequencies for the faintest stars ($V$ = 20--22) are significantly more uncertain 
than for brighter stars.

Second, we compute the frequency of the IRAC-excess population as a function of 
the spectral types derived in \citet{Currie2010a}.  Stars with $V$ $\sim$ 20--22 typically 
correspond to K5--M1 stars (cf. Table \ref{obsint}).  Therefore, only our excess frequencies for stars 
earlier than mid-K are robust against completeness/photometric biases.

\subsubsection{Frequency of IRAC-Excess Sources}
\begin{figure}
\centering
\plotone{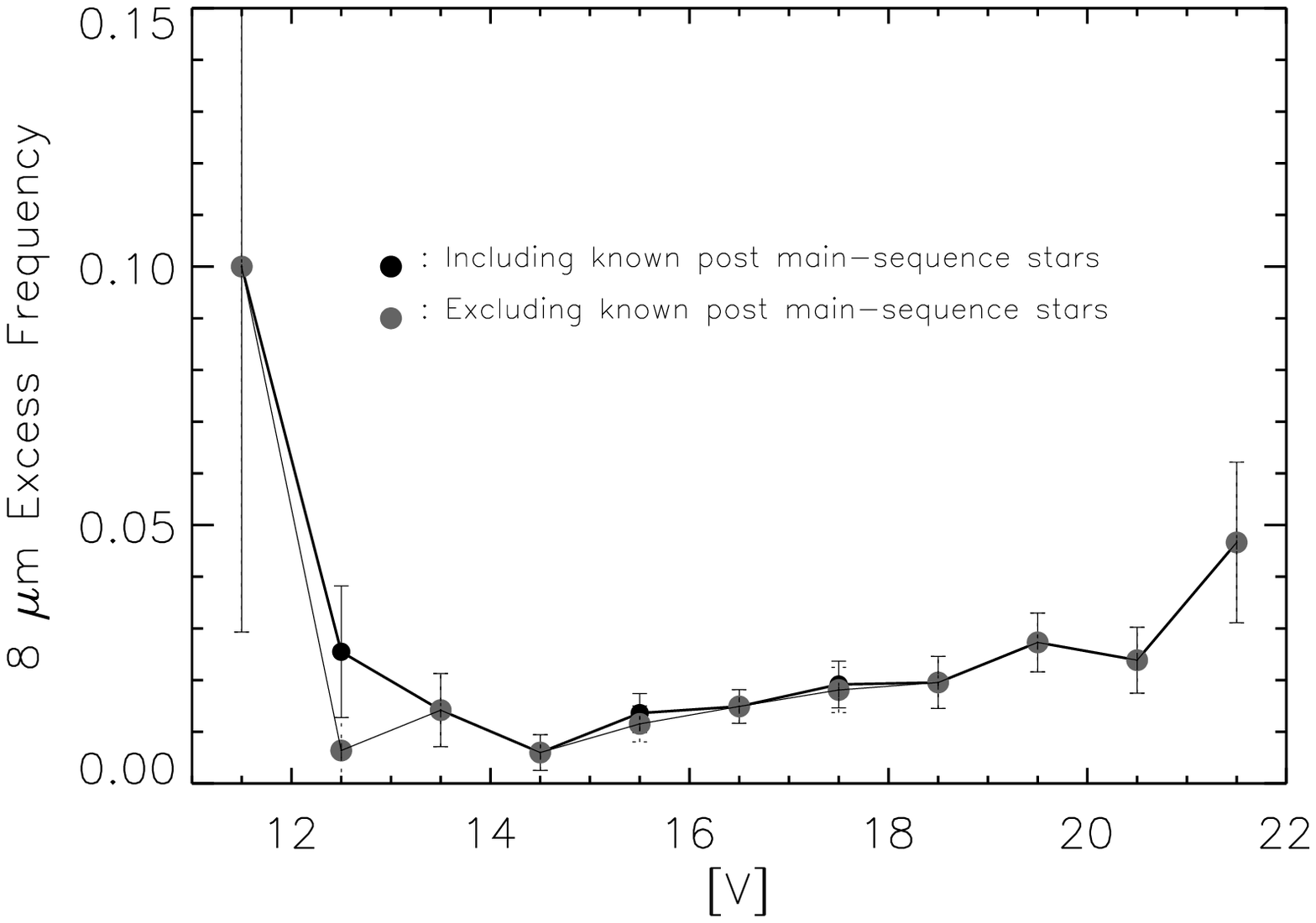}
\plottwo{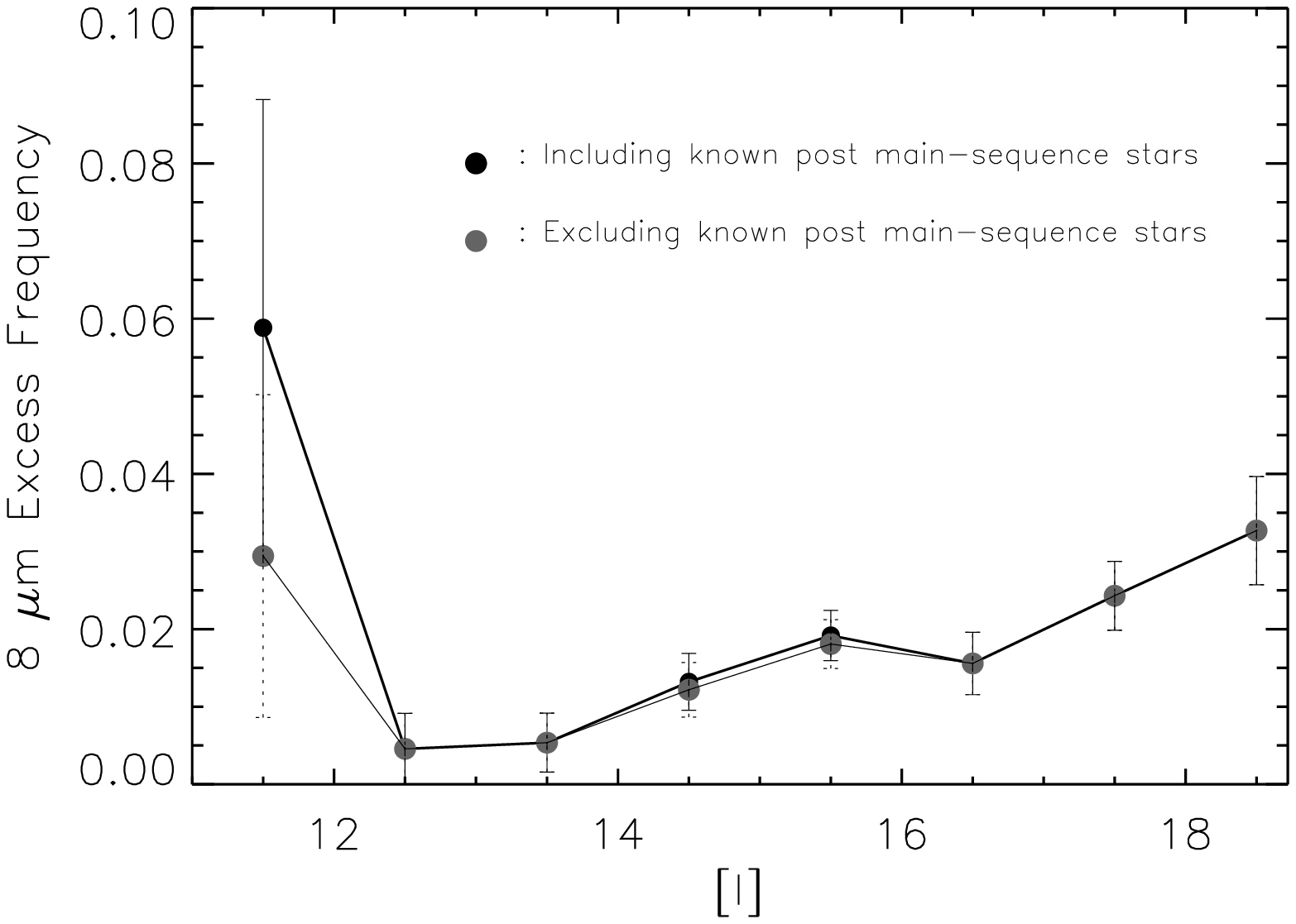}{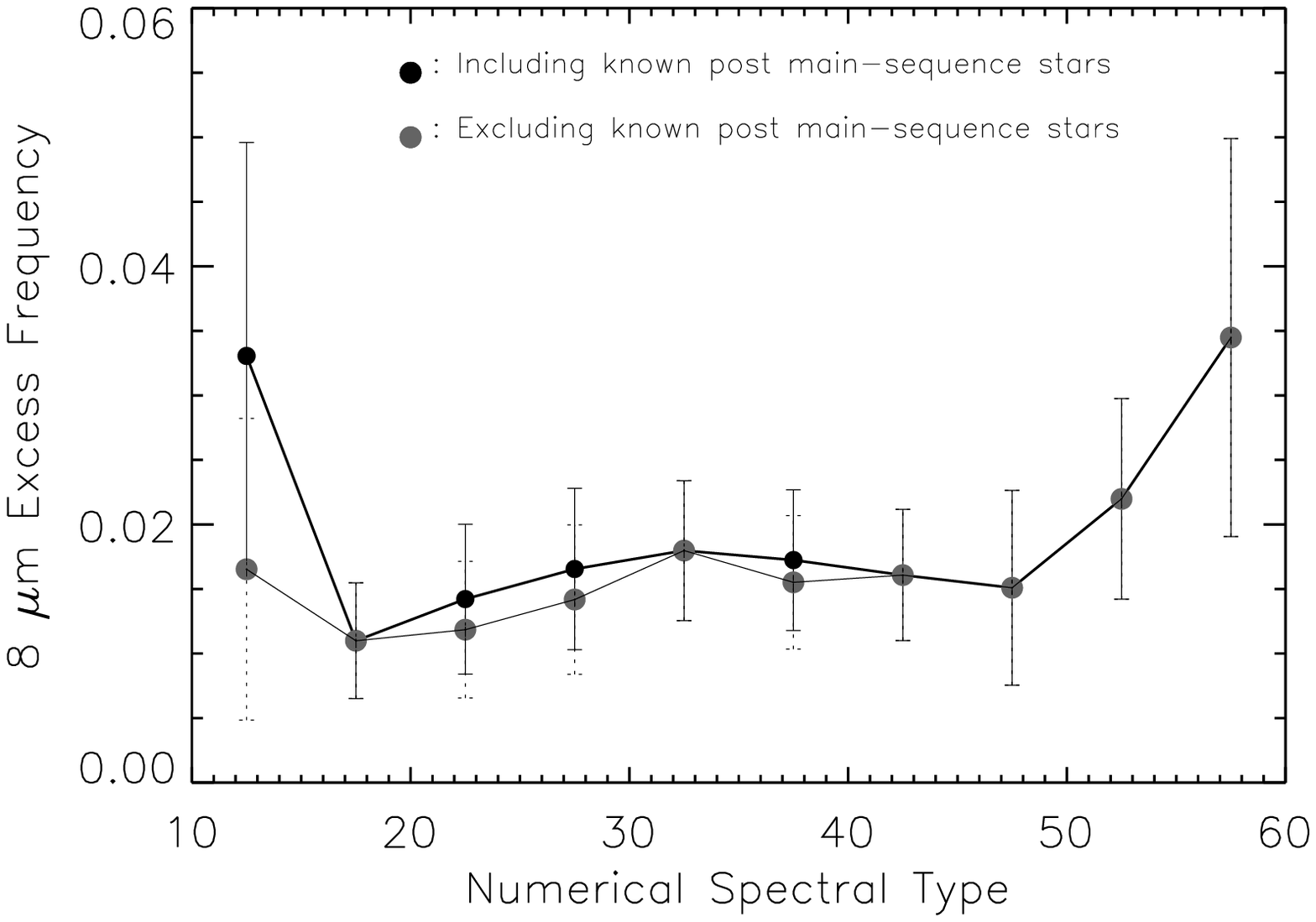}
\caption{Frequency of 8 $\mu$m excess emission from disks vs. $V$-band magnitude (top), $I$-band magnitude (bottom-left), 
and (numerical) spectral type (bottom-right).  As described previously, the numerical spectral type has the following 
formalism: 10=B0, 11=B1, ... 68=M8.  In all plots, report frequencies including and excluding post-main sequence stars 
like giants and Be stars whose excess emission is likely not indicative of planet formation.}
\label{excfreq}
\end{figure}

Figure \ref{excfreq} shows the frequency of the IRAC-excess population as a function of $V$, $I$, and spectral type.  
Overall, 263 h and $\chi$ Persei members or $\sim$ 2.2\% $\pm$ 0.3\% show evidence for 8 $\mu$m excess 
likely due to warm dust.  From $V=14$ to our 
completeness limits, the frequency of warm dust appears to steadily rise from $\sim$ 0.5--1\% (e.g. 
3/502 and 11/955 at $V$ = 14--15 and 15--16, respectively) to $\sim$ 3\% (e.g. 23/843 at $V$ = 19-20) and 
$\sim$ 4--5\% (9/193 at $V$ = 21--22).  The IRAC excess frequency vs. $I$ shows similar trends. 
The frequency as a function of spectral type appears roughly flat from A0 to G5 at $\sim$ 1.5\% but rises to 
3.5\% for K stars, albeit with considerable uncertainty.

If we consider all stars regardless of evolutionary status (black dots in all figure panels), the frequency 
of warm dust for the brightest stars 
($V$ $<$ 11, $I$ $<$ 10, B0-B5) is as high as in any magnitude/spectral bin covering lower-mass stars.
However, many of these stars are known Be stars or other post-main sequence stars whose apparent IRAC excess 
is likely due to free-free emission or to warm dust from a circumstellar excretion disk and not a circumstellar accretion disk or 
debris disk connected to planet formation. To estimate the frequency of solely circumstellar disks related to 
planet formation, we then trim our sample of known Be stars, giants, and supergiants using lists of such stars 
from \citet{Currie2010a}, \citet{Marsh2012}, and from the \textit{SIMBAD} database.  In the 
latter case, various sources contribute to the Be star and (super)giant identifications \citep{Bragg2005,Slesnick2002}.  

The revised IRAC-excess frequencies show that nearly all of the previously-identified excess sources are in fact 
Be stars or post-main sequence stars.  With these stars removed, the excess frequencies for the brightest, 
earliest stars are $\sim$ 1--1.5\%, comparable to or slightly smaller than their frequencies for 
slightly fainter, later and (presumably) lower-mass stars ($\sim$ 1.5--2\%).  Thus, in agreement with 
previous results \citep{Currie2007a,Currie2008phd}, we identify a slight brightness/spectral-type dependent 
frequency of warm dust indicating that dusty circumstellar disks are less frequent around the most massive stars 
($M$ $\sim$ 1.5--5 $M_{\odot}$).  
\subsection{MIPS-Excess Emission from Cooler Dust}
%\subsection{Frequency of MIPS-Excess Emission from Cooler Dust}
To quantify the frequency of stars with MIPS excesses, we follow 
the same procedure used in \S 3.2.1 to flag IRAC excess sources, 
defining the excess sources as those whose $K_{s}$-[24], [3.6]-[24], 
or [4.5]-[24] colors are redder than the median color
 by more than 3-$\sigma$+0.06.  Overall, we identify 152 h and $\chi$ Persei stars 
with evidence for excess emission at 24 $\mu$m.  As explained in Sect. 2.5, we 
restrict our analysis to stars with $V$ $<$ 13.5 due to the poorer 
MIPS completeness compared to IRAC.

  Following \S 3.2.1, we also derive the frequency in 
terms of spectral type and mass, using the rough conversion between 
$V$ magnitude and these two properties as listed in Table \ref{obsint}.
Because of poorer sampling in mass and $T_{eff}$ from 
the \citet{PallaStahler1999} isochrones and the large number of 
post-MS stars dominating spectral types B0 and B1, we restrict 
the range of spectral types considered to B2--B6.
The corresponding mass range is $M_{\star}$ = 4--6 $M_{\odot}$.

\begin{figure}
\centering
%%\plotone{hafield.ps}
\includegraphics[scale=0.33]{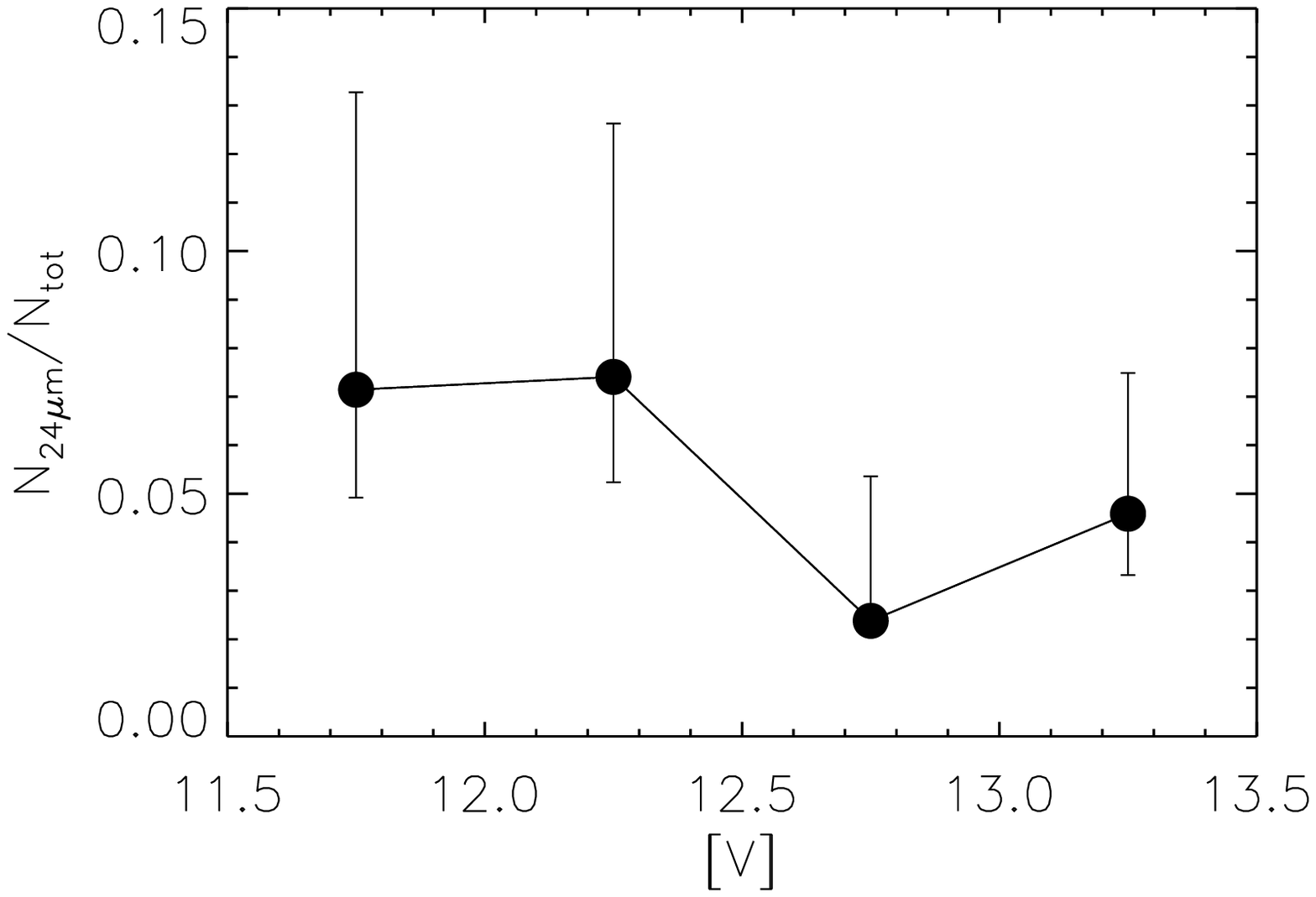}
\includegraphics[scale=0.33]{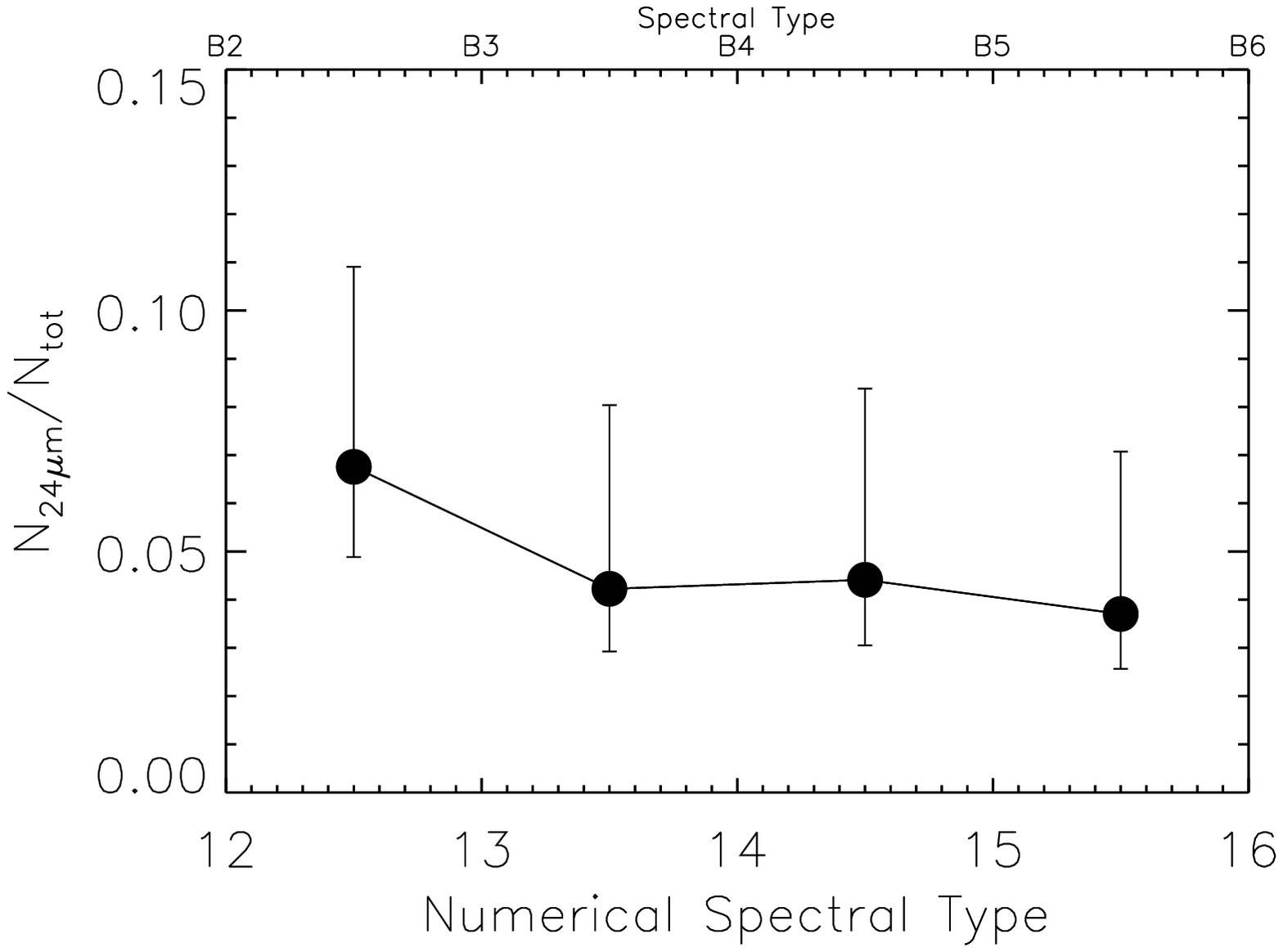}
\includegraphics[scale=0.33]{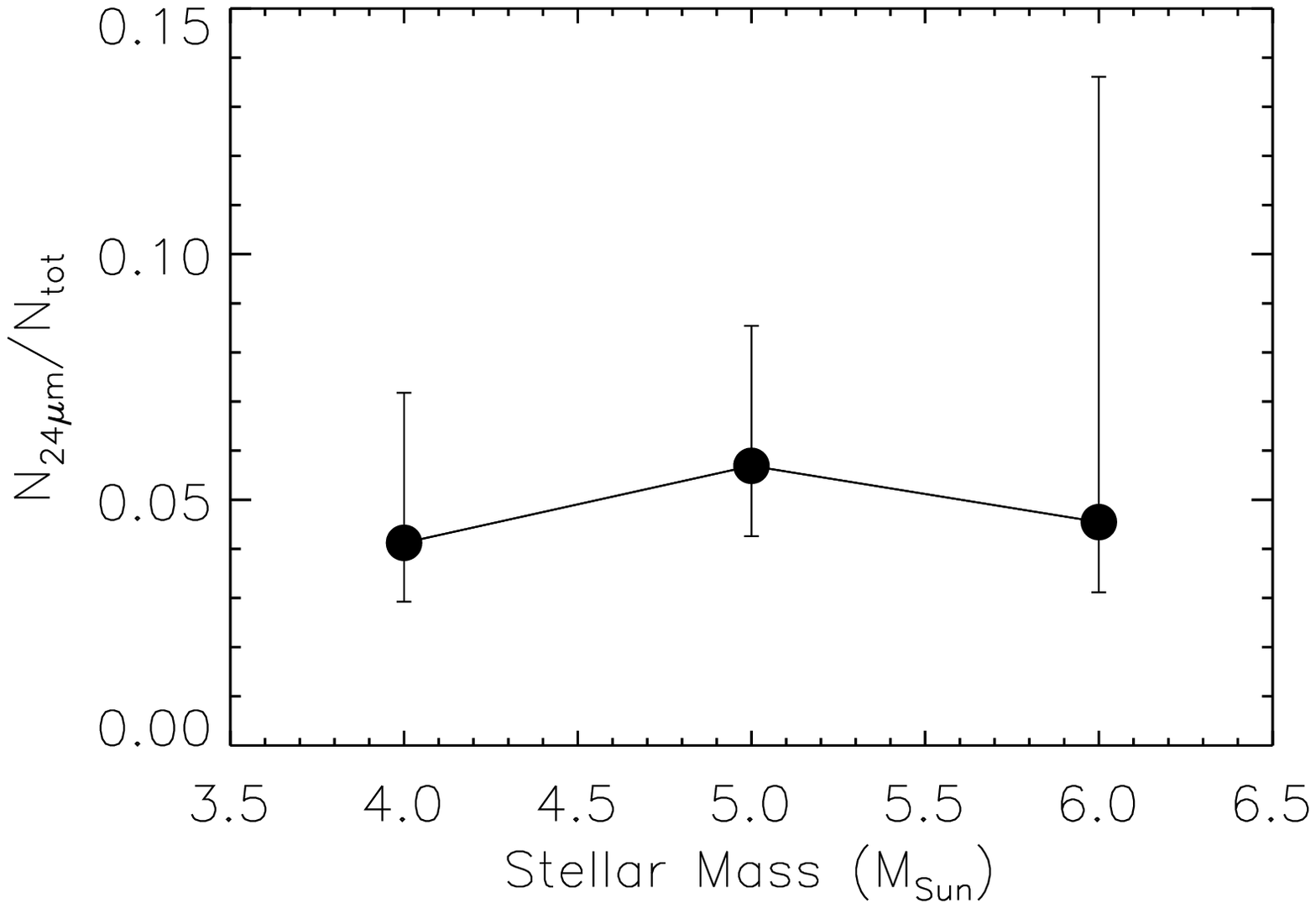}
\caption{(Left) Frequency of h and $\chi$ Persei stars with MIPS-24 $\mu$m 
excess as a function of $V$-band magnitude (left), spectral type (middle), and 
inferred stellar mass (right).  About 5\% of the 4--6 $M_{\odot}$ h 
and $\chi$ Persei stars show evidence for 24 $\mu$m excess.}
\label{mipsfreq}
\end{figure}

Figure \ref{mipsfreq} displays the frequency of MIPS-24 $\mu$m excess 
emission as a function of $V$ magnitude, spectral type, and mass 
for these massive, early-type stars.   About 5\% of 
these stars show evidence for MIPS-excess emission from cooler 
dust compared to the 1--3\% that show IRAC excess emission 
over the same range in $V$ magnitude ($\sim$ 12--13.5), 
spectral type ($\sim$ mid B), and mass (4--6 $M_{\odot}$) (see Figure \ref{excfreq}).
Although there is a slight decline in excess frequency 
with increasing $V$ magnitude and later spectral types, the 
error bars are large enough to easily be consistent with 
no trend, indicative of a constant frequency of MIPS-excess 
emission.  Any (small) correction for confusion with faint galaxies will 
not change this conclusion.

\section{Circumstellar Gas in Disks Around \lowercase{h} and $\chi$ Persei Stars} \label{sectgas}
\subsection{Identifying Accreting h and $\chi$ Persei Stars}
Our criteria for identifying possible accreting systems follows \citet{White2003}'s 
spectral-type dependent threshold based on the observed $H_{\alpha}$ 
equivalent width. For the stars with 
spectral types of K0 and later, the $EW(H_{\alpha})$ provides a robust measure of 
accretion \citep{White2003}. Nominally, for K0--K7 stars, we identify those with 
$EW(H_{\alpha})$ $\ge$ 3 \AA\ as accretors.  For K7--M2.5 (M2.5--M5.5) stars, 
$EW(H_{\alpha})$ $\ge$ 5 \AA\ (20 \AA) are flagged as accretors. 
These criteria are similar to those adopted in \citet{CurrieKenyon2009} but 
are more precise than those adopted for h and $\chi$ Persei stars in \citet{Currie2007c}, 
where stars with $EW(H_{\alpha})$ $>$ 5--10 \AA\ were identified as candidate accretors. 

It is more difficult to identify possible accretors among the AFG-type 
stars which comprise 
the bulk of our spectroscopic sample. 
Within this group of intermediate mass stars, much larger 
$H_{\alpha}$ emission line fluxes are required 
to produce $EW(H_{\alpha})$ $>$ 3 \AA\ \citep[e.g.][]{White2003}.
While these stars have additional clear accretion signatures 
\citep[i.e. U band excess, optical veiling, $H_{\alpha}$ emission;][]{Calvet2004},  
they lack a clear $EW(H_{\alpha})$ criterion for separating weakly accreting stars 
from chromospherically active ones using low-resolution spectroscopy. 
Because $H_{\alpha}$ absorption is often 
much broader than 
the absorption lines in later-type stars, some stars show clear emission 
line reversals even though $EW(H_{\alpha}) < 3$ \AA 
\citep{Dahm2008,CurrieKenyon2009}. Devising a reliable metric for 
identifying accreting AFG stars is beyond the scope of this paper, so for 
AFG stars we nominally adopt the same $EW(H_{\alpha})$ threshold applicable for K0--K7 
stars and simply flag the few others showing core line reversal and $H_{\alpha}$ clearly 
in emission (no matter what the equivalent width).

Table \ref{speccatalog} lists the 42 new candidate accreting stars and 30 
previously-identified candidate accretors from \citet{Currie2007c}.  
Of these 72 objects, 47 (51) are among our members with Spitzer (optical) 
photometry. Three were initially rejected in \citet{Currie2010a} as likely members.

Despite the preferential sampling in favor of BAF stars, most of the 
new candidate accretors are G0 or later (38/42).  Of the BAF
stars listed in Table \ref{speccatalog}, two have $H_{\alpha}$ in absorption and thus formally are not 
identified as accretors.  However, they have a strong core emission line reversal. 
These include a B9 star, which may be the earliest known accreting star older than 10 Myr.  While the 
$H_{\alpha}$ equivalent widths generally are far smaller than for 1--5 Myr-old T Tauri stars and 
thus more indicative of low accretion rates \citep{White2003,Dahm2008}, three early K stars have much stronger emission 
(EW($H_{\alpha}$) = 25-83 \AA) reminiscent of their younger counterparts.  As our spectroscopic sample 
is very incomplete for the h and $\chi$ Persei mid K to M star population \citep{Currie2010a}, it is unclear whether any of the kinds of stars 
comprising most members of the youngest star-forming regions show frequent $H_{\alpha}$ emission indicative 
of accretion in the Double Cluster.

\subsection{Frequency of Accreting h and $\chi$ Persei Stars}
\begin{figure}
\centering
%%\plotone{hafield.ps}
\plottwo{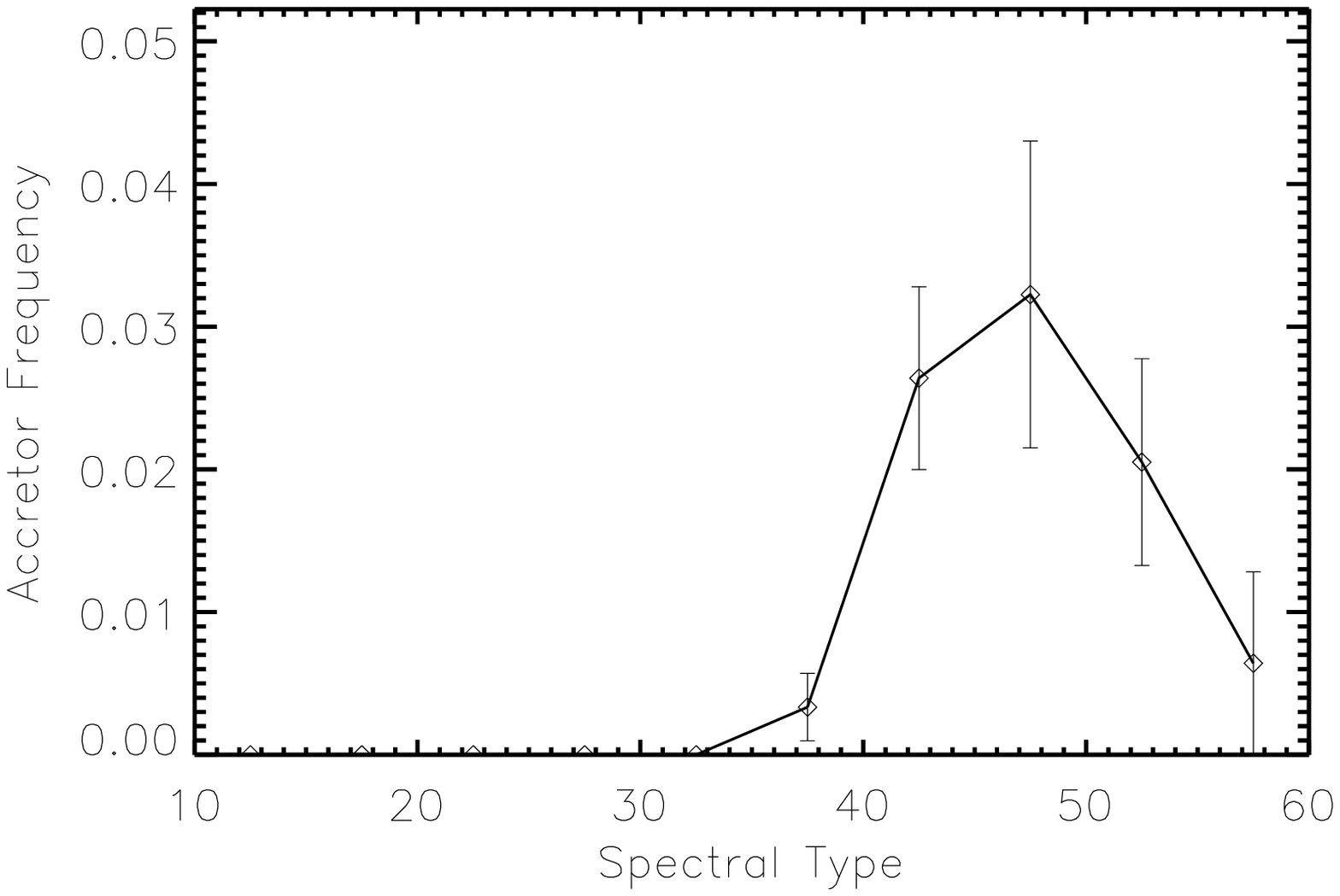}{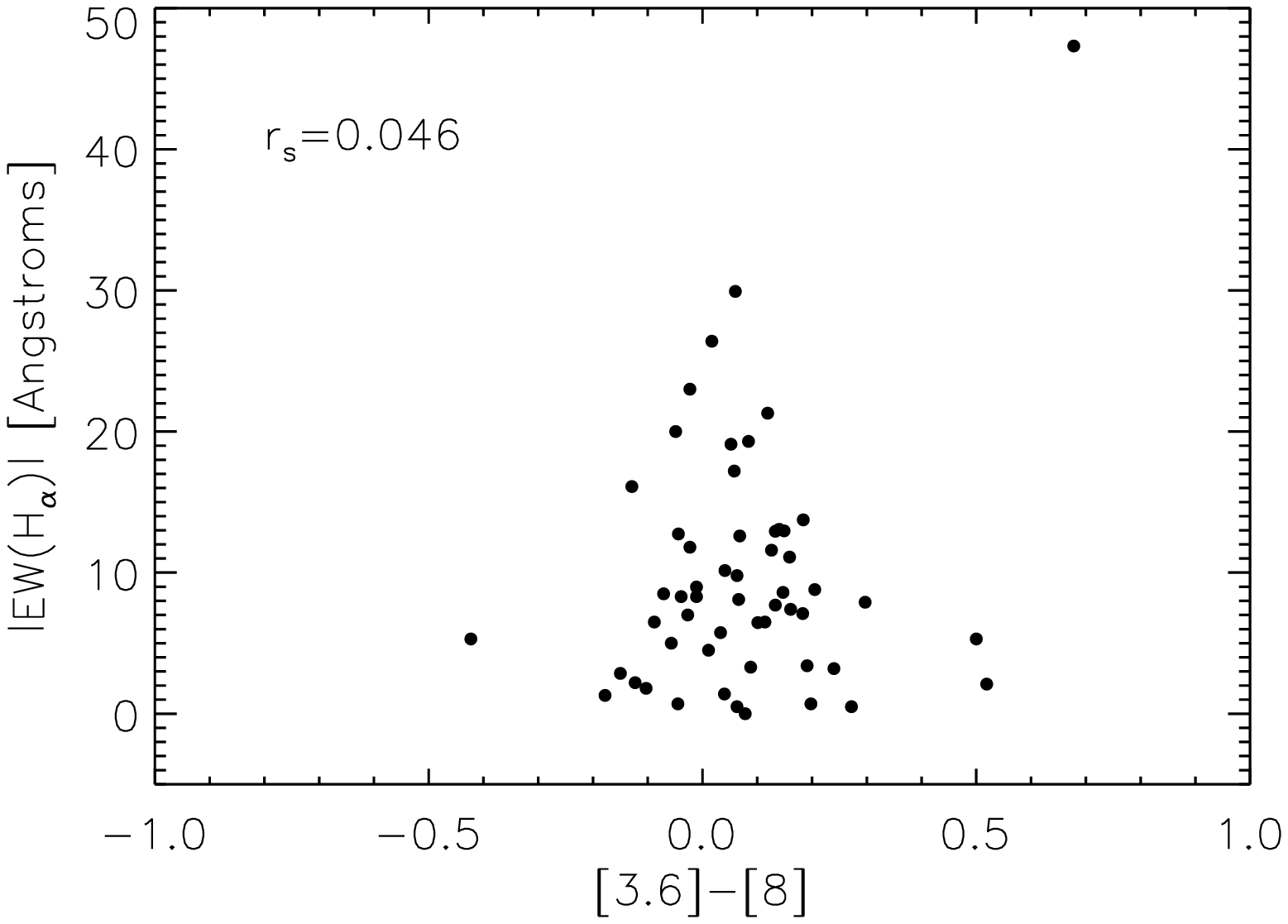} 
\caption{(Left) Frequency of accreting h and $\chi$ Persei stars as identified from their 
$H_{\alpha}$ equivalent widths.  (Right) The $H_{\alpha}$ equivalent width vs. the [3.6]-[8] 
color, showing that there is no clear correlation between accretion diagnostics and those of 
warm circumstellar dust.}
\label{accfreq}
\end{figure}

Figure \ref{accfreq} shows that the accretion frequency for likely h and $\chi$ Persei members 
with optical spectra is highest for GK stars ($M$ $\sim$ 0.8--1.3 $M_{\odot}$) and lowest for earlier stars
($M$ $>$ 1.4 $M_{\odot}$).
Our sample of accretors includes two F-type stars, while all the remaining 
accretors are G0 or later.  As a result, the accretion frequency is consistent with zero 
from the earliest stars through F5, while the frequency rises above 2\% for G0--K5 stars, 
peaking at $\sim$ 3.3\% for late G stars.   
This trend is consistent with a stellar mass dependent frequency of gas accretion \citep{CurrieKenyon2009},
although our critera for identifying accretors may slightly underestimate this population around the earliest (BA) stars.
The frequency is lower for K5--M0 stars, although 
here again our sampling is poor.

\subsection{Accretion and Circumstellar Dust Emission}
Recent results imply that the frequency of dust and gas in protoplanetary disks in most young clusters 
decreases on a timescale of $\sim$ 3 Myr \citep{Hernandez2007, 
CurrieKenyon2009,Fedele2010,CurrieSiciliaAguilar2011}.  
However, the presence of dust in older, $\sim$ 10-20 Myr old clusters does not necessarily 
imply detectable protoplanetary disk gas and vice versa.  Many disks around 10-20 Myr old stars 
with dust emission are instead optically-thin, gas-poor debris disks \citep{Chen2005,Roberge2008,Lisse2008,Lisse2009}.  
Conversely, at least some disks showing evidence for protoplanetary disk gas lack strong broadband infrared excess 
emission from dust that would be detectable in our sample \citep{Currie2007c,Bitner2010}.  
While preliminary results showed a weak correlation between protoplanetary disk gas and IR excess in h and $\chi$ Persei 
stars, our expanded sample of accretors and more precise IR photometry allows us to reinvestigate this connection.

Figure \ref{accfreq} compares the $H_{\alpha}$ equivalent width to the [3.6]-[8] color amongst our sample of candidate 
accretors.  Most strikingly, the distribution of IR colors appears nearly symmetrical about zero for a range of 
EW($H_{\alpha}$) with only 1-2 outliers.  This result is in slight contrast with previous studies of 
h and $\chi$ Persei's accreting population \citep{Currie2007c}, which found a very weak correlation between 
$H_{\alpha}$ equivalent width and 8 $\mu$m excess. Specifically, \cite{Currie2007c} derive a Spearman rank 
correlation coefficient of $r_s=0.60$ across their full dynamic range in H$_{\alpha}$ equivalent width 
%(\|EW(H$_{\alpha}$)\|$ \in [0,50]$ Angstroms). 
($\vert$ EW(H$_{\alpha}$) $\vert$ = 0--50 \AA).
Conversely, over the same dynamic range, we derive a Spearman 
rank correlation coefficient of $r_s=0.046$ implying that the relationship between protoplanetary disk 
gas and IR excess cannot be described by a monotonic function.
The disagreement probably is due to 1) poorer photometric precision in the \citet{Currie2007a} IRAC data and 
2) the decision in \citet{Currie2007c} to use $K_{s}$-[8] instead of [3.6]-[8] as a tracer of warm dust, where the former exhibits a 
much stronger spectral type dependence \citep[cf.][]{Luhman2010}.

%\section{Evolutionary States of Disk-Bearing Stars with Excesses in Multiple Bandpasses} \label{sectseds}
\section{Spectral Energy Distribution Modeling of Disk-Bearing Stars}\label{sectseds}

Analysis in \S 3 as well as previous work from \citet{Currie2007a,Currie2008a} demonstrates 
that h and $\chi$ Persei includes a small population of stars with 1) warm, terrestrial zone 
dust identifiable from broadband 8 $\mu$m excess emission and 2) very strong ($\gtrsim$ 3 magnitude) 
excess emission at 24 $\mu$m that in many cases is likely due to cooler dust.  
As our observations are not sensitive to (pre-)main sequence stars lacking broadband 8 $\mu$m 
excess but producing modest 24 $\mu$m excesses ($\sim$ 0.25--1.5 magnitudes) more typical 
around stars of comparable age \citep{Chen2011}, they cannot identify cooler debris disks 
and many homologously depleted transitional disks.   Typical cold debris disks yield negligible broadband 
excesses at 24 $\mu m$ and will also not be detected around h and $\chi$ Persei stars.  While our study does not probe the full extent 
of the disk population, we still can analyze data for stars with IRAC and MIPS excesses 
or those that are accreting
and compare the emission to that predicted for disks covering a range of evolutionary states.  
The disks around such stars represent those that are least evolved and provide an upper limit to the 
relative frequency of normal protoplanetary disks at 14 $Myr$.

\subsection{Disk Terminology and Classification Method}
%Based on the model comparisons and using similar analysis in \citet{CurrieSiciliaAguilar2011} as a 
%guide, 
Following \citet{CurrieSiciliaAguilar2011}, we consider three different classes of disks:
\begin{itemize}
\item \textbf{Primordial Disks} -- We identify optically-thick primordial disks, characteristic of disks in the 
youngest star-forming regions \citep{Gutermuth2009}, generally as those with excess emission above our model disk limit from 4.5 $\mu$m 
to 24 $\mu$m.

\item \textbf{Transitional Disks with Inner Holes/Gaps} -- We identify transitional disks with inner holes 
as those with photospheric emission or weak excess emission through 5.8 $\mu$m but emission in excess of 
our fiducial model at 24 $\mu$m, indicative of a positive $\lambda$$F_{\lambda}$ slope from 8 to 24 $\mu$m and 
optically-thick outer disk emission at 24 $\mu$m.  Those with gaps like LkCa 15's may appear to have emission comparable to 
an optically-thick disk at some near-IR wavelengths but have a steep flux slope ($\alpha$ $\approx$ -3) from the 
near to mid IR (e.g. through 8 $\mu$m) due to an absence of dust between a hot optically-thick population near 
the dust sublimation radius and a cooler optically-thick population at $AU$-scale distances.  %Thus, we identify 
%disks with gaps as those with a near-stellar flux slope through 5.8 $\mu$m but again a sharp jump in flux density 
%at 24 $\mu$m consistent with optically-thick outer disk emission.

\item \textbf{Warm Debris Disks/Homologously Depleted Transitional Disks} -- Objects with weak excess, consistently 
lying below our fiducial disk model have SEDs comparable to homologously depleted transitional disks in 
Taurus, the Coronet Cluster, IC 348 and other young star-forming regions \citep{CurrieSiciliaAguilar2011} 
\textit{and} warm debris disks \citep{Currie2007b,Rhee2007,Cpk2008}.  For all stars 
modeled here with spectra, none are shown to be accreting and thus none have clear evidence for circumstellar gas.  However, 
our optical spectra were obtained at low resolutions, where weakly accreting disks may not be identifiable and/or may be difficult to flag.
Although h and $\chi$ Persei's age may favor objects in this category as having warm debris disks \citep{Currie2007b,Currie2008a}, 
the case is not clear-cut.  Thus, to be conservative we note that these objects may be \textit{either} warm debris disk-bearing 
stars or those surrounded by homologously depleted transitional disks.

Note that our identification of candidate homologously depleted transitional disks is simpler 
than in \citet{CurrieSiciliaAguilar2011} who focus on nearby clusters with richer IR data sets.  Specifically, 
they used Taurus optical to submm data to identify the range of disk masses covered by optically thick primordial disks.  
In addition to the criteria listed above, they then required homologously depleted disks to have disk masses 
below the Taurus optically thick disk range.   Some range in IR colors (e.g. $K_{s}$ - [8] vs. $K_{s}$ - [24]) 
includes both primordial disks and transitional disks, but for clusters much more distant than Taurus (e.g. NGC 2362, 
h and $\chi$ Persei), we lack far-IR/submm data that is required to estimate a disk mass.  Therefore, based on the Taurus 
modeling results, they correct for the frequency of transitional disks derived from IR SED modeling alone (as in this 
paper) by the `contamination rate' of primordial disks with the same infrared excess emission.  We do not introduce such a 
correction factor here since a) it is small to begin with and does not alter our results presented later and b) 
very few stars we model have IR colors that, based on Taurus data, correspond to multiple disk states.
\end{itemize}

\subsection{Models Used to Classify Disks}
\subsubsection{Fiducial Disk Model Comparisons}
We first compare the SEDs of disk-bearing h and $\chi$ Persei stars to geometrically flat, optically thick reprocessing disk 
models to conservatively identify those disks with a reduced optical depth of emitting dust ($\tau_{IR}$ $\lesssim$ 1): i.e. transitional disks.  
Following \citet{CurrieSiciliaAguilar2011}, we adopt models produced from 
the Whitney-Monto Carlo radiative transfer code \citep{Whitney2003a,Whitney2003b,Robitaille2006}.  
The models assume no flaring ($H$/$r$ = constant), no accretion, 
and no protostellar envelope emission.  
%Although the total disk mass is $\sim$ 0.05 $M_{\odot}$, 
The disks are optically thick to their own radiation over spatial scales relevant to our study 
($r$ $<$ 5-10 AU) even for disks over an order of magnitude lower mass than that which we nominally adopt (0.05 $M_{\odot}$).   

As shown in \citet{CurrieSiciliaAguilar2011} and in other earlier publications \citep[e.g.][]{Dalessio2006,Kenyon1987}, making the disk 
flared instead of flat or considering accretion can only result in a more luminous disk\footnote{Inclination likewise has a negligible effect
 except for nearly edge-on cases \citep{CurrieSiciliaAguilar2011}.} and a higher transitional disk frequency.  
 Therefore, our fiducial model represents a conservative limit separating the bluest, lowest-luminosity primordial disks from transitional disks.
 
\subsubsection{Model SED Fitting}
As a separate investigation, we model source SEDs using the grid of radiative transfer disk models from \citet{Robitaille2006}.   As our data are limited to detections at 
$\lambda$ $\le$ 24 $\mu m$, these model comparisons cannot investigate the mass of emitting dust in the disks \citep{CurrieSiciliaAguilar2011}, and our disk classification scheme does not turn on whether or not the disk is flared.  Instead, we simply use these model comparisons to investigate whether each source's best-fit models require an inner hole substantially larger than the dust sublimation radius.  Thus, they can show whether some disks we find lack inner holes (primordial disks, homologously depleted transitional disks) based on fiducial comparisons nevertheless are best-fit by a disk models with inner holes.

To identify the best-fitting disk models, we restrict ourselves to those models satisfying the $\Delta$$\chi^{2}$ criterion of $\chi^{2}$-$\chi^{2}_{best}$ $<$ 3, where 
$\chi^{2}$ refers to the minimum $\chi^{2}$ per data point \citep[see also][]{CurrieSiciliaAguilar2011,CurrieLada2009,Ercolano2009}.  
We adopt a reddening uncertainty of 10\% and a distance uncertainty of 10 $pc$.
To be conservative, we set the division between a disk requiring an inner hole and one without at 10 times the dust sublimation radius.

\subsection{Results for Disks with Excesses in Multiple Bandpasses}
From our sample of $\approx$ 270 stars with infrared excess, 18 show clear excess at 
both 8 $\mu$m and 24 $\mu$m\footnote{Note again that this mismatch in numbers is largely 
driven by our poor sensitivity at MIPS-24 $\mu$m}, are later than B5 (fainter than $J = 13$) and   
thus not likely post-main sequence stars which can therefore be compared in more detail to 
disk models.   One of these -- $\alpha$ = 34.7824, $\delta$ = 57.2346 -- was previously identified in \citet{Currie2007b,Currie2008a}, labeled as 
``Source 5" since it was the fifth candidate h and $\chi$ Persei member with excesses in multiple IRAC bands, and modeled by comparing its SED to 
predictions from terrestrial planet formation calculations \citep{KenyonBromley2004} 
and a simple model assuming two, single-temperature blackbodies.   The other seventeen 
targets listed are new detections.

Of the 18 sources, 7 have optical spectroscopy from \citet{Currie2010a}. The remaining   
11 lack optical spectroscopy %in \citet{Currie2010a} and thus 
and thus do not have a confirmed spectral type or extinction estimates.  To determine the appropriate 
$T_{eff}$ for these 11/18 stars, we first fit the $V$ and $I$ photometry to predicted colors for h and $\chi$ 
Per stars for a range of extinctions using intrinsic dwarf colors and reddening laws listed in
 \citet{Currie2010a}.  The seven other targets had previously-derived spectral types and extinctions from 
\citet{Currie2010a}.  We used these spectral types and extinctions as a starting point 
and explored whether alternative estimates provided a better fit to the optical/near-IR photometry.  
Our derived values agree with these previous determinations to within one subclass 
in spectral type and $\sim$ 10\% in reddening, supporting the parameters derived in \citet{Currie2010a}.  

\subsubsection{Results for Fiducial Disk Model Comparisons}
\begin{figure}
\centering
\plotone{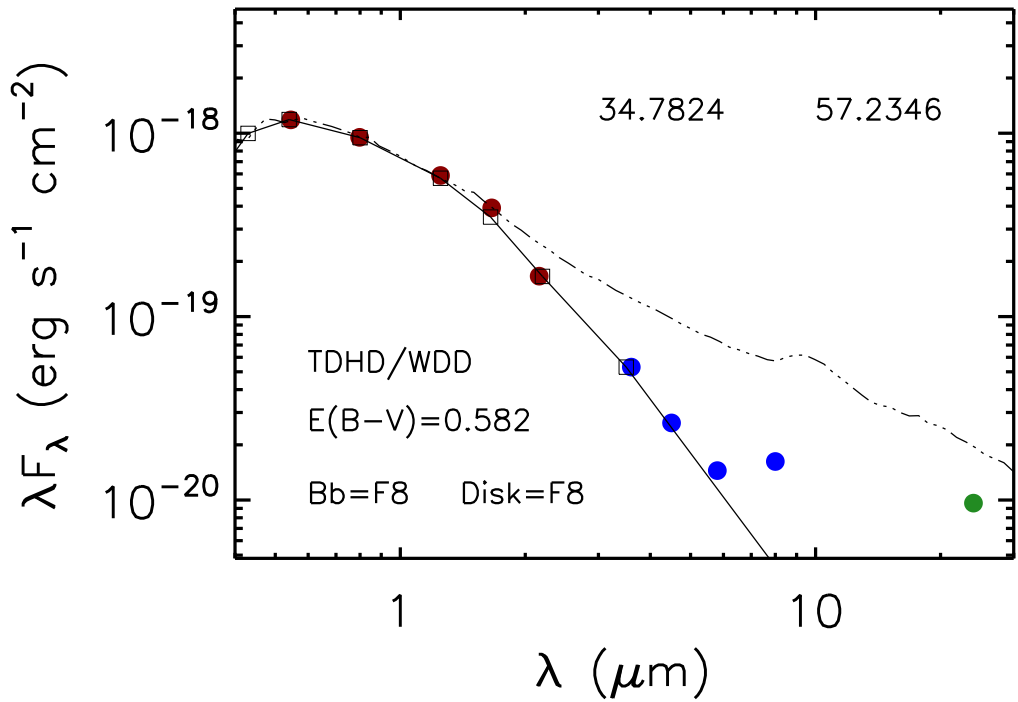}
\caption{Sample spectral energy distribution a for h and $\chi$ Persei star with clear excesses in IRAC \textit{and} MIPS bandpasses compared to 
photometric predictions for a bare stellar photosphere (solid line) and an optically-thick, flat reprocessing disk (dashed line).  
This object was previously identified as a terrestrial zone (warm) debris disk candidate from \citet{Currie2007b}.}%, while the 
%right-hand object is a new detection.}
\label{sampleseds}
\end{figure}

Table \ref{diskevostate} records the known or derived spectral types and reddening values along with IRAC/MIPS colors, 
and disk types of the 18 h and $\chi$ Persei stars with IRAC and MIPS excesses that we model.  Figure \ref{sampleseds} displays a typical model 
comparison with the model spectral type for the photosphere (Bb) and the disk (Disk) directly plotted along with 
its J2000 coordinates.  We display an atlas of all modeled sources in the Appendix.  

The majority (10/18) of these h and $\chi$ Per sources have SEDs most consistent 
with warm debris disks and/or homologously depleted transitional disks.  Of the remaining sources, 
6/18 have SEDs most characteristic of transitional disks with inner holes and only two objects have SEDs 
most consistent with primordial disks.  Depending on whether we classify weak-excess sources as warm debris disks 
or homologously depleted disks, the relative fraction of protoplanetary disks in a transitional state is either 6/8 or 16/18.  
In either case, transitional disks outnumber the primordial disk population. 

\subsubsection{Results for SED Model Fitting}
\begin{figure}
\centering
\plottwo{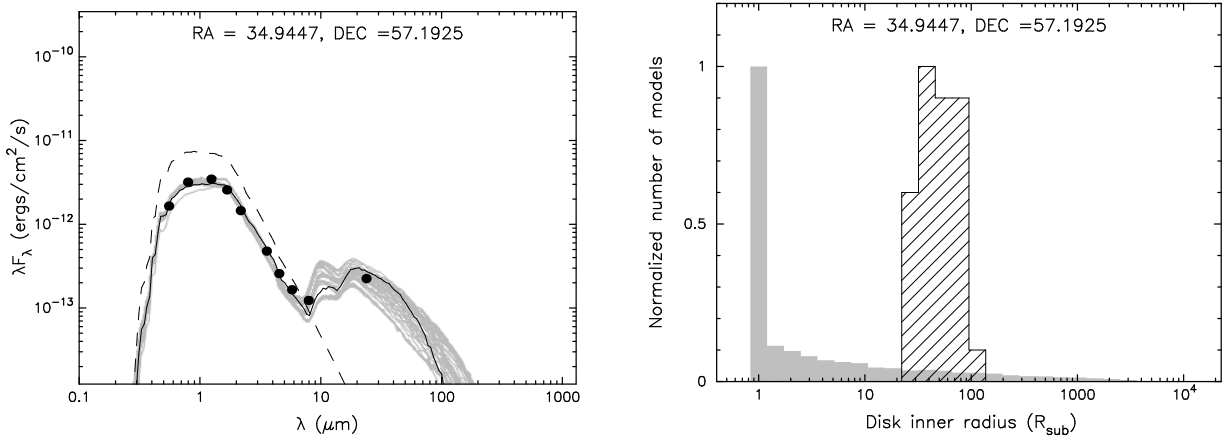}{hxPer_8_rsub.eps}
\caption{Sample SED modeling results using the \citeauthor{Robitaille2006} grid.  The left panel displays the set of best-fitting SED models.  The right panel displays the distribution of disk inner radii (in units of the dust sublimation radius) for these SED models.  For most of our modeled targets, the inferred disk classifications agree with our analysis based on fiducial comparisons.  For this object and 3 others, SED modeling favors reclassifying the disks as having inner holes.
}
\label{sedmodelgrid}
\end{figure}

Table \ref{diskevostate} (second column from the right) records the best-fit disk inner radius for our sources using the \citeauthor{Robitaille2006} grid.  
Figure \ref{sedmodelgrid} displays example analysis results: the set of best-fit SEDs (left panel) and distribution of disk inner radii (right panel)
 for the eighth excess source in our list ($\alpha$ = 34.9447, $\delta$ = 57.1925).
In most cases, our results based on fiducial model comparisons and model fitting agree.   Primordial disks and homologously depleted 
transitional disks identified from fiducial model transitional disks, are not best-fit by a disk model with an inner hole.  Disks with weak IRAC excess but much stronger 
MIPS-24 $\mu m$ excesses are best-fit by disk models with large (R$_{disk}$ $>$ 10 R$_{sub}$) inner holes.  

SED modeling suggests that we may be slightly underestimating the population of transitional disks with inner holes.  
Three sources, identified as having homologously depleted transitional disks/warm debris disks, are best fit by a 
disk model with an $R_{disk}$ $>$ 10 R$_{sub}$ hole.  Two of these sources are early-type (B8.5, A5) stars with very weak, 
debris disk-like excess emission (0.2--0.6 mags in the IRAC bands; 1.6-1.75 mags in MIPS): the inner holes identified 
by SED modeling may instead reflect the location of debris belts.  The third of these (F5.9) has a 4.7 magnitude excess 
in MIPS-24 $\mu m$ characteristic of only the most luminous debris disks \citep[e.g. HR 4796A][]{Chen2005} and many transitional 
disks with inner holes \citep[cf.][]{CurrieSiciliaAguilar2011}.  One source identified as having a primordial disk 
from fiducial model comparisons is best-fit by a disk model with a 30 R$_{sub}$ inner hole.  

In summary, our h and $\chi$ Persei disk analysis shows evidence for substantial evolution, even amongst 
the protoplanetary disk population.  Fiducial model comparisons and SED fitting 
reveal a large fraction of transitional disks, either those with 
inner holes or ones that are homologously depleted, compared to primordial disks.  
%This result is consistent with those for other clusters with ages $\approx$ 5 Myr or greater \citep{CurrieSiciliaAguilar2011,Fang2012}.
%As shown in \S 4, if we instead use gas accretion, not dust, as our diagnostic of disks, the protoplanetary disk population 
%still contains more transitional disks than primordial disks.  

\subsection{Results for Accreting Disks}
\begin{figure}
%%\plottwo{sedacc_accretors_with8excess0.ps}{sedacc_accretors_with8excess1.ps}
%%\plottwo{sedacc_accretors_with8excess2.ps}{sed_accretor1.ps}
\plottwo{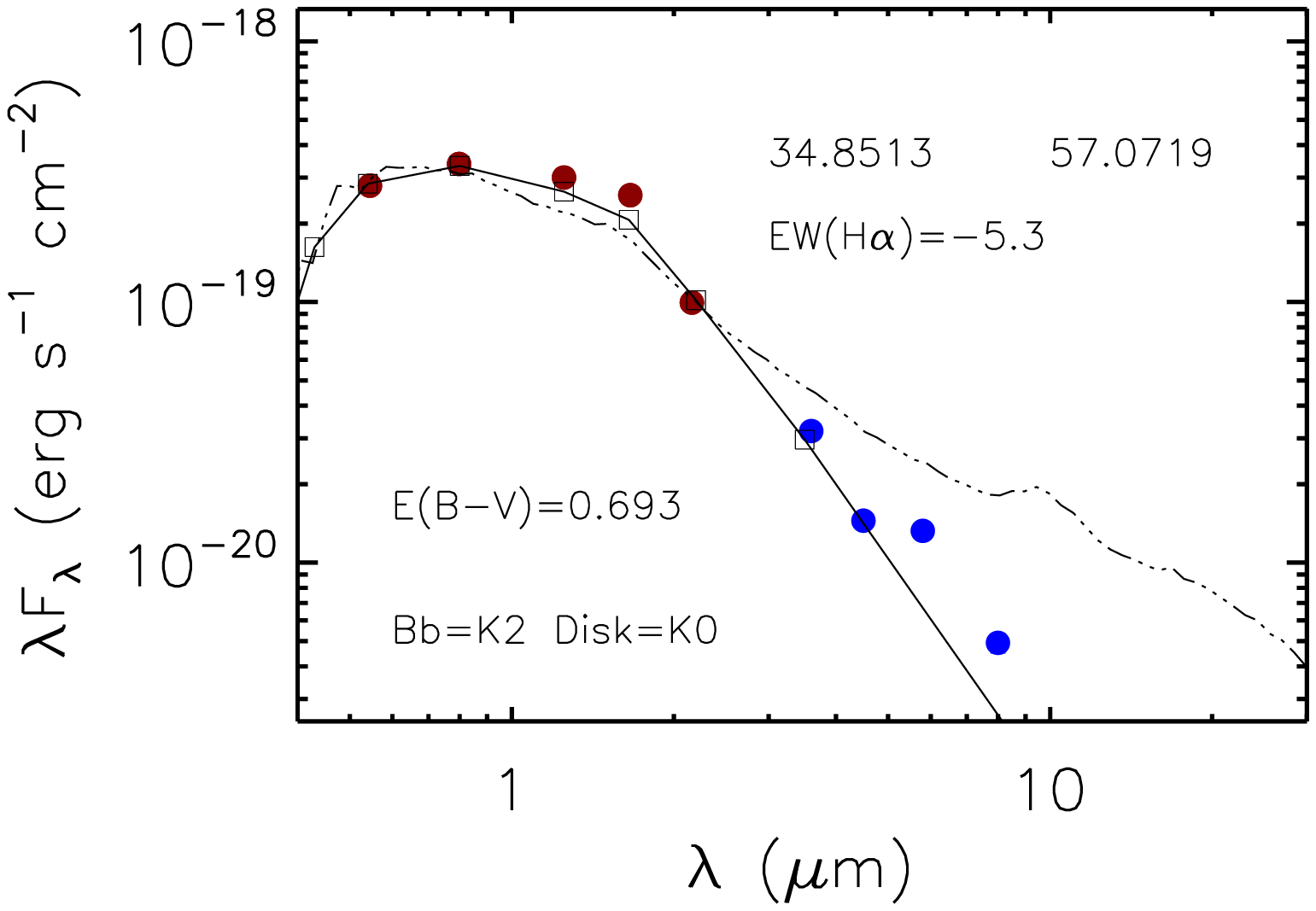}{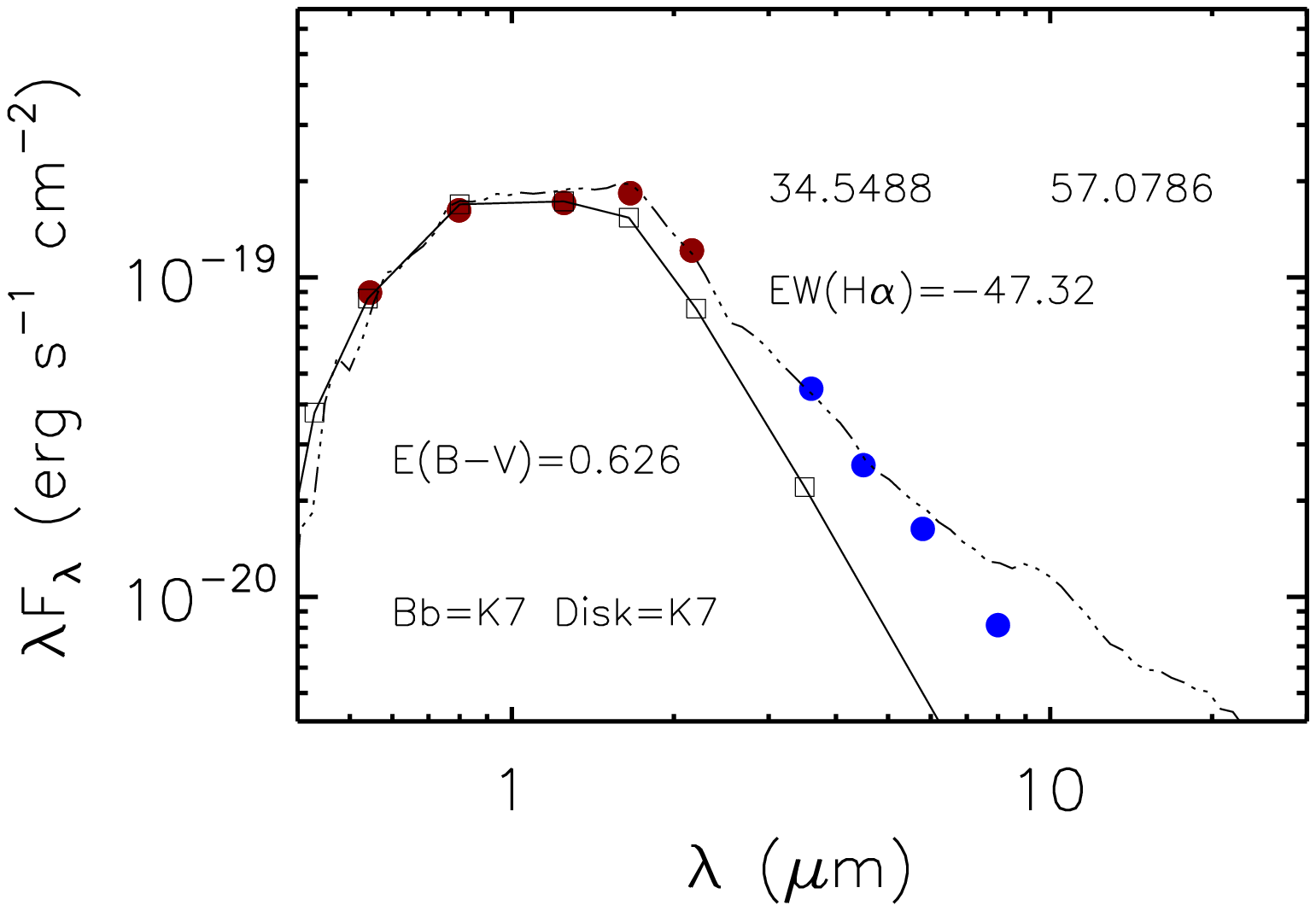}
\caption{SEDs of two likely h and $\chi$ Persei members with $H_{\alpha}$ emission indicative of 
circumstellar gas accretion.  The differences in spectral types adopted for the model stellar photosphere 
(BB) and disk (``Disk") reflect our sampling of the disk model grid: they do not affect our interpretation 
of either object.}
\label{accsed}
\end{figure}

%To supplement 
In addition to our modeling of members with IRAC and MIPS excesses, we also briefly compare the SEDs of accreting 
members identified in \S 4 to the same fiducial disk model.  
While accreting circumstellar disks are often associated with optically-thick ``primordial" disks that are  
typical of the youngest star-forming regions, Figure \ref{accsed} shows that the situation is more complicated.  
%Compared to the optically-thick, geometrically flat disk limit, 
The star with weaker $H_{\alpha}$ emission ($|$EW(H$_{\alpha}$)$|$=5.3 \AA\; left panel)
clearly has weaker warm dust emission relative to the optically-thick, geometrically flat disk limit. 
The other star, the outlier in Figure \ref{accfreq} (right panel),   
has significantly stronger $H_{\alpha}$ emission ($|$EW(H$_{\alpha}$)$|$ = 47.32 \AA) but still has IRAC excess emission from dust lying at or below 
the flat disk limit (Figure \ref{accsed}, right panel).  %Neither source is detected in the MIPS-24 $\mu$m filter. 
The SEDs of the remaining accreting stars that also maintain an IRAC excess, resemble the left-hand panel star, as the strength of the 
excess is always weaker than the flat, optically-thick disk limit.  

Therefore even the long-lived accreting protoplanetary disks in h and $\chi$ Persei show evidence for significantly reduced 
broadband IR dust emission more indicative of transitional disks than the optically thick 
\textit{primordial} disks that dominate the youngest star-forming regions like Taurus.
%The large fraction of transitional disks in h and $\chi$ Persei points to an extended transitional disk lifetime 
%\citep{CurrieLada2009,CurrieSiciliaAguilar2011}.  
In the next section, we compare the frequency and morphologies of 
disks in h and $\chi$ Persei to those in other clusters in a global analysis of disk evolution from 1 Myr to 20 Myr.

\section{The \lowercase{h} and $\chi$ Persei Disk Population in Context: Empirical Constraints on Disk Evolution and Planet Formation} \label{sectdiskevol}
\subsection{Methodology and Goals}
Comparing the disk population in h and $\chi$ Persei to that for clusters/associations 
of comparable or younger ages allows us to constrain how the frequency of warm
circumstellar dust and gas changes with time 
\citep[see]{Hernandez2007,Fedele2010,CurrieSiciliaAguilar2011}.  Circumstellar dust and gas comprise the
building blocks of planetary cores and gaseous envelopes \citep{Pollack1996,KenyonBromley2009}; 
dust may also be second-generation, produced via collisions between planetesimals in young debris disks 
\citep{Wyatt2008,KenyonBromley2008}. Modeling the time evolution of dust and gas then 
sets some limits on the timescale for converting these raw materials into gas giant planets and/or 
the early time evolution of debris disks \citep{Rieke2005,Su2006,Currie2008a,CurrieLada2009,Chen2011,Chen2012}.

In this section, we combine h and $\chi$ Persei data with that from other clusters to 
explore the timescale for protoplanetary disk evolution and the 
properties of debris disks, especially those around high-mass stars (4--6 $M_{\odot}$).
Although we use specific, absolute (and hence uncertain) ages to describe the clusters, our 
arguments are chiefly based on the \textit{relative} ages and the evolutionary sequence 
they establish, which \citet{Soderblom2014} argue are robust.
Table \ref{discfreqtable} lists the clusters/associations we consider and their frequencies 
of circumstellar dust and gas.  The samples probe well into the 
M dwarf, subsolar-mass regime.  Due to the extreme sample size of h and $\chi$ Persei stars 
we achieve robust estimates of the warm dust and gas fractions at $\sim$ 10--15 $Myr$.

First, we follow \citet{CurrieSiciliaAguilar2011}, using a 
Monte Carlo simulation to predict the frequency of circumstellar gas and dust from 
protoplanetary disks for various characteristic dust/gas evolution e-folding timescales. 
We then identify the timescale that best matches the data using a simple reduced 
$\chi^{2}$ criterion.  We use the best-fitting dust evolution timescale as input 
for a second Monte Carlo simulation tracing the relative frequency of transitional disks 
with time for various bulk transition disk timescales ranging between 0.1 $Myr$ and 1 Myr 
using the subset of well-characterized clusters in \citet[][]{CurrieSiciliaAguilar2011} -- 
Taurus, the Coronet Cluster, IC 348, NGC 2362, $\eta$ Cha -- and h and $\chi$ Persei.
Finally we explore debris disk evolution for stars with masses 
$\sim$ 4--6 $M_{\odot}$, a mass range not previously studied, 
by comparing the frequency of MIPS-excess sources in h and $\chi$ Persei to those for 
members of the Sco-Cen Association (Upper Scorpius, Lower Centaurus Crux, 
and Upper Centaurus Lupus) \citep{Chen2011,Chen2012}.  All data probing dust evolution 
come from \textit{Spitzer} observations, whereas data probing 
gas evolution originate from various ground-based spectroscopic surveys.

While our data probing circumstellar dust draw from Spitzer/IRAC observations only, some differences 
between each cluster sample impede our ability to set more robust limits on the typical 
timescale for protoplanetary disks to disappear.  For instance, the completeness limits for 
each cluster data set are different.  
%While most samples are heavily weighted 
%towards K and M spectral types consistent with solar and subsolar-mass pre main sequence stars, but 
%the completeness limit and characteristic stellar masses probed may slightly differ from sample to sample.  
This is especially true for h and $\chi$ Persei, which is more distant than the other clusters 
analyzed here.  Although our methods for identifying excess sources follow those standard in 
 the literature \citep{Hernandez2007}, there may still be small differences between methods 
used to find disk-bearing stars amongst the clusters we analyze.  

With these caveats in mind, we focus simply on understanding the typical evolution timescales for warm dust 
and gas and how these timescales change depending on the assumed ages for clusters.  
For example, recent work by \citet{Pecaut2012} argues that Upper Scorpius may be about twice as old as its 
nominal 5 Myr age \citep{Preibisch2002}, although other work seems to indicate 
 results intermediate between these extremes \citep{Bowler2011}.
\citet{Bell2013} reinvestigate the ages of many young clusters -- e.g. IC 348, $\sigma$ Orionis, and NGC 2362
 -- often used to define these 
evolutionary timescales and argue that cluster ages have generally been underestimated. 
However, \citet{Bell2013} find an age for $\chi$ Persei (14.5 $Myr$) in excellent agreement with 
previous estimates \citep[14 $Myr$,][]{Currie2010a}.

To bracket the error range for young cluster ages \citep{Soderblom2014}, we fit for the warm dust and gas evolutionary timescales assuming nominal cluster ages and those 
revised in \citet{Bell2013} and \citet{Pecaut2012}, while keeping the age of h and $\chi$ Persei fixed 
at 14 $Myr$.  We then use the best-fit timescales as input to investigate the 
transitional disk lifetime as in \citet{CurrieSiciliaAguilar2011}, including our new 
results for h and $\chi$ Persei.

\subsection{Evolution of Warm Dust from Inner Disks}
\begin{figure}
\centering
\epsscale{1}
\includegraphics[trim=14mm 0mm 9mm 0mm,clip,scale=0.38]{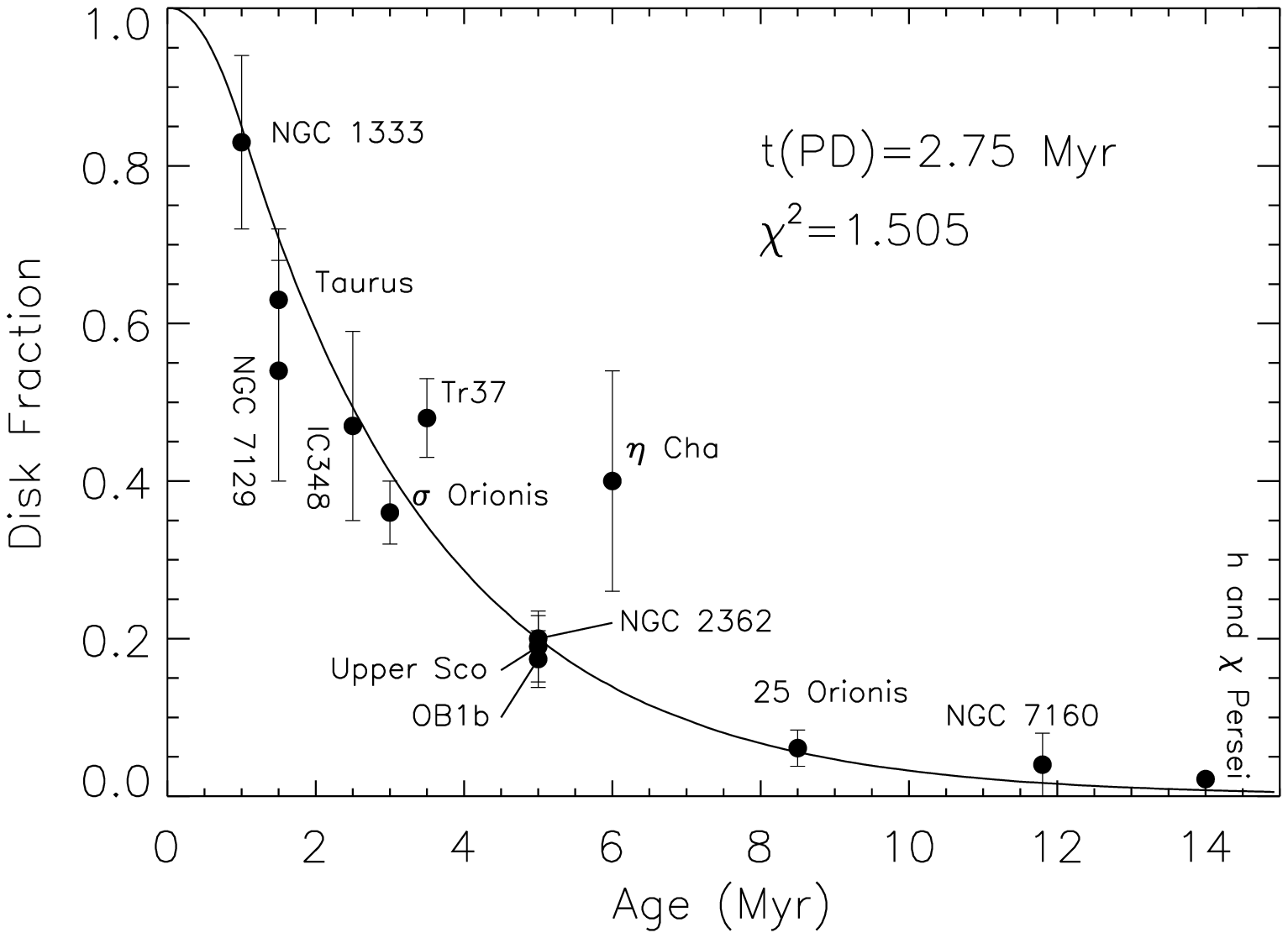}
\includegraphics[trim=14mm 0mm 9mm 0mm,clip,scale=0.38]{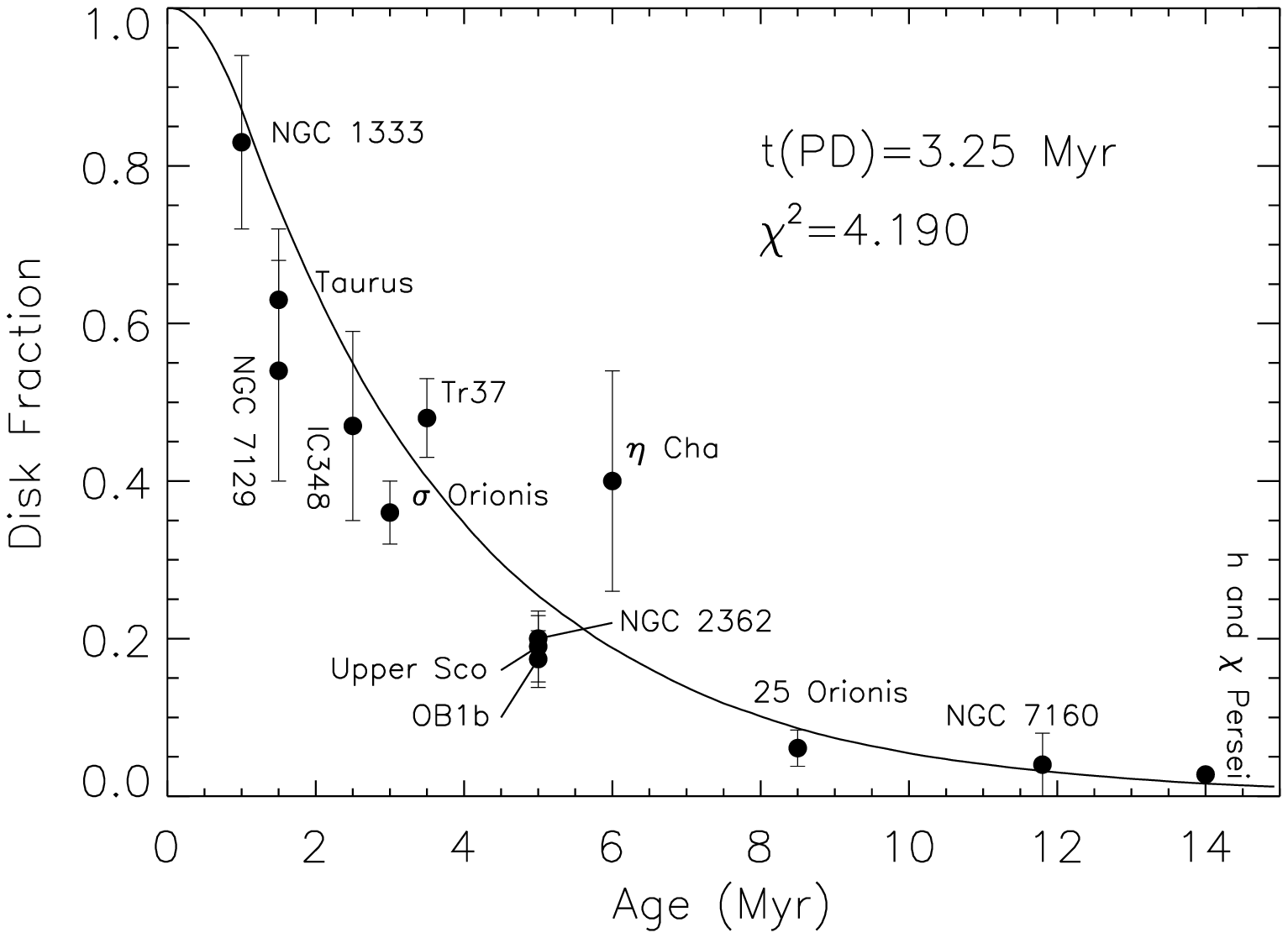}
\includegraphics[trim=14mm 0mm 9mm 0mm,clip,scale=0.38]{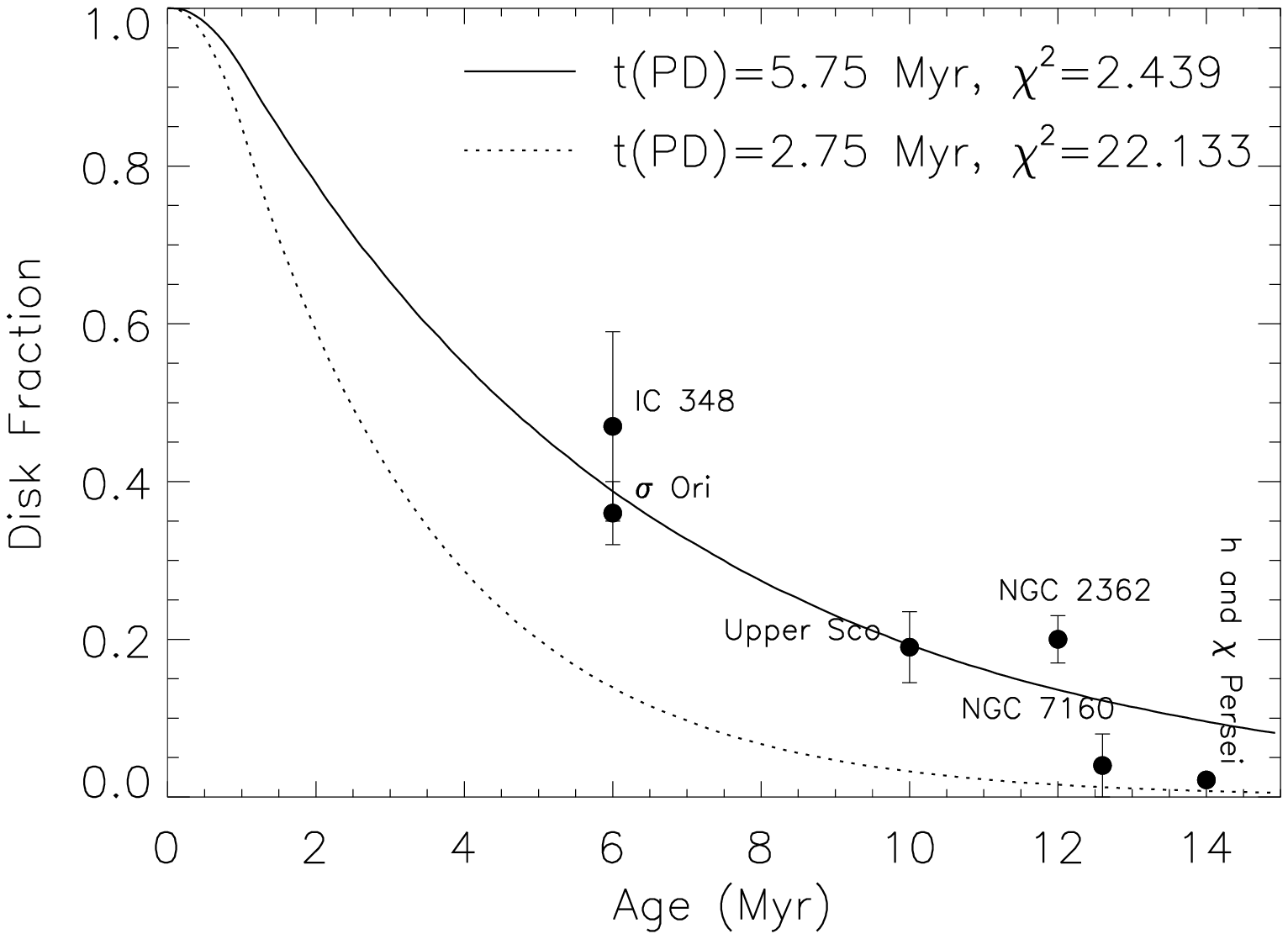}
\caption{Frequency of warm circumstellar dust vs. time for h and $\chi$ Persei 
and many 1--20 $Myr$-old clusters/associations for nominal cluster ages 
(left/middle) and for clusters with revised ages from \citet{Bell2013} (right).  In the 
left and right panels, the solid line depicts 
the best-fit timescale to the dust lifetime from our parametric modeling 
assuming nominal ages and revised ages from \citet{Bell2013}, respectively.  
The middle panel compares the frequencies to a timescale of 3.25 $Myr$ assuming 
nominal ages. 
In the right panel, the dotted line shows the predicted curve for a 2.75 $Myr$ 
lifetime to compare with the best-fit value of 5.75 $Myr$.}
\label{diskfreqevo}
\end{figure}

 Figure \ref{diskfreqevo} plots the frequency of warm dust emission (as probed by 
\textit{Spitzer/IRAC}) for our sample, where we focus on clusters with ages between 
1 and 15 $Myr$.   The best-fit timescale is 2.75 $Myr$. 
We obtain similar results regardless of whether we focus on all clusters or just those
nominally younger than 10 $Myr$.
While this timescale best reproduces our combined sample
there are clearly outliers, most notably Trumpler 37 and the very sparse association $\eta$ Cha 
\citep{SiciliaAguilar2006,Megeath2005,SiciliaAguilar2009}.  The frequencies for 
Taurus and NGC 7129 also appear somewhat low, though in Taurus's case 
binarity may play a role in affecting disk frequencies \citep[cf.][]{KrausIreland2012}.
In both cases, h and $\chi$ Persei lies slightly above the trendline we expect for 
warm dust originating only from protoplanetary disks.

The middle panel displays a model with 
a slightly longer timescale (t(PD) = 3.25 $Myr$), which together 
with 2.75 $Myr$ 
bracket the value considered by \citet{CurrieSiciliaAguilar2011}.  
The near-quadrupling of
the reduced $\chi^{2}$ as the model timescale increases to 3.25 $Myr$ is largely due to 
the inability of the model to reproduce the dust frequencies found in regions such 
as $\sigma$ Orionis \citep{Hernandez2007}, Upper Scorpius \citep{Carpenter2006}, 
and NGC 2362 \citep{DahmHillenbrand2007,CurrieLada2009}.

Adopting revised ages for clusters listed in \citet{Bell2013} and \citet{Pecaut2012} clearly 
yields a longer timescale (right panel) for warm dust in disks.  Our best-fit timescale 
($\chi^{2}_{\nu}$ $\sim$ 2.4) more than doubles to 
5.75 $Myr$, a result driven by some of the same regions ($\sigma$ Orionis, Upper Scorpius, 
and NGC 2362) whose frequencies determine the best-fit timescale for nominal ages. 
%implying a much longer duration of warm circumstellar dust in disks.  
When assuming the revised cluster ages of \cite{Bell2013}, the 2.75 $Myr$ timescale 
found to best fit the combined sample when nominal ages are adopted clearly underpredicts 
the disk frequencies for IC 348, $\sigma$ Ori, Upper Sco, and NGC 2362 ($\chi^{2}_{\nu}$ $\sim$ 22). 
For h and $\chi$ Persei however,  
an intermediate timescale may be more consistent.

\subsection{Evolution of Protoplanetary Disk Gas}
\begin{figure}
\centering
\epsscale{1.1}
\plottwo{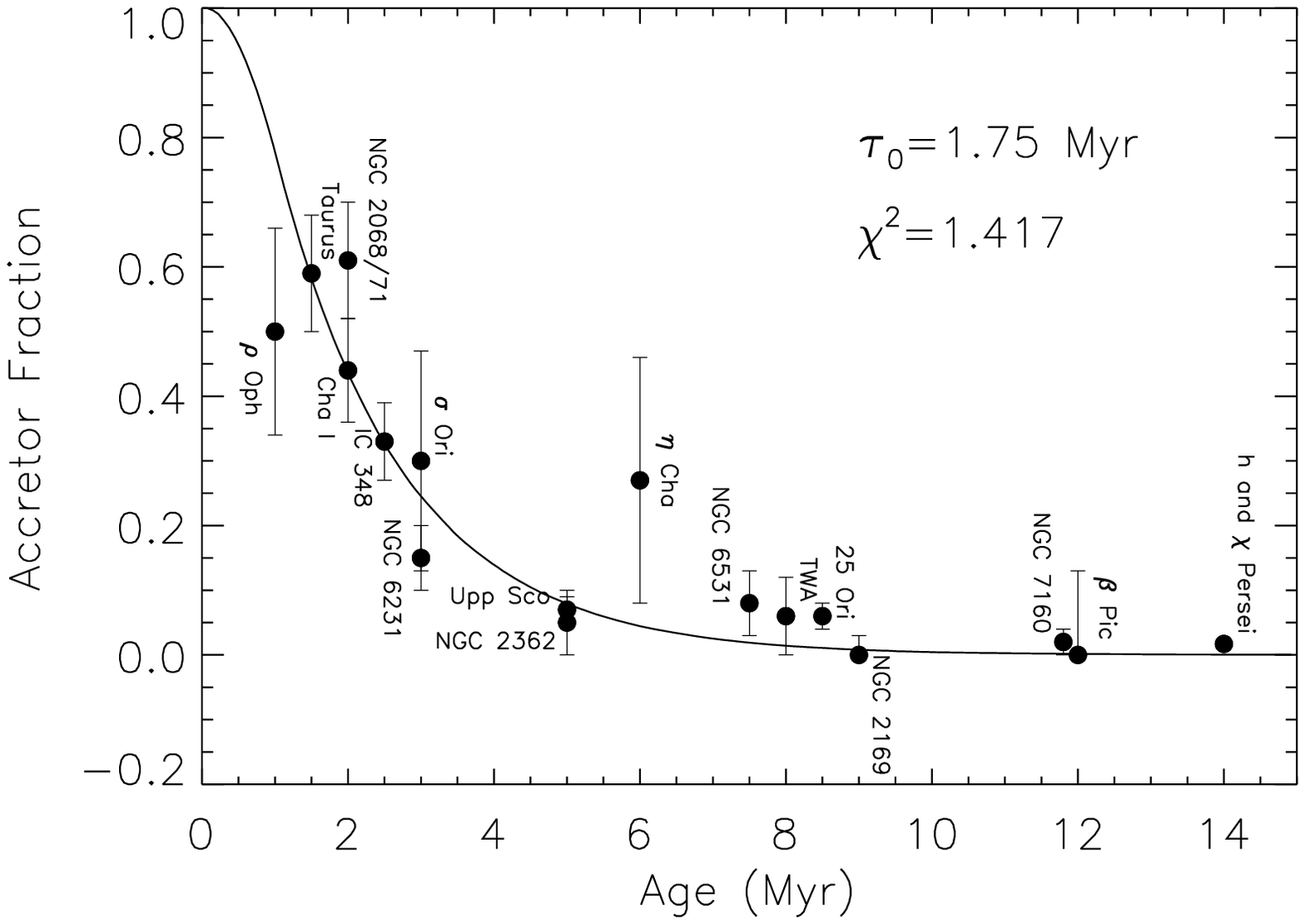}{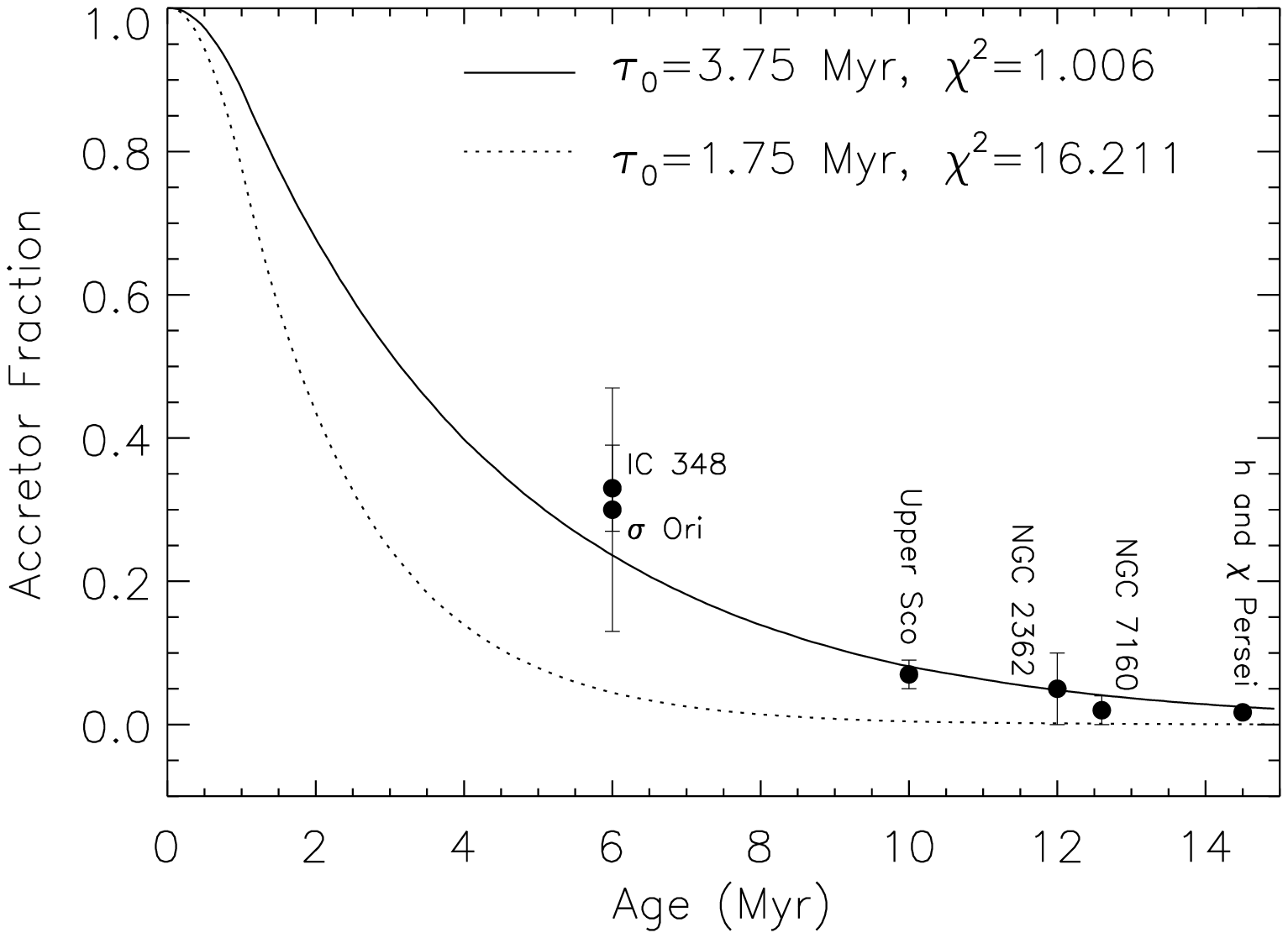}
\caption{Frequency of accreting protoplanetary disks in 1--20 Myr old 
clusters adopting nominal cluster ages (left) and considering revised 
cluster ages from \citet{Bell2013} (right).  The solid lines depict the 
predicted accretor frequency vs. time for the best-fit curve to the data. 
The dotted line in the right panel shows the predicted curve for a 1.75 $Myr$ 
lifetime to compare with the best-fit value of 3.75 $Myr$.
}
\label{accfreqevo}
\end{figure}

Our results for circumstellar gas evolution mimic the general trends found 
in the above section for dust (Figure \ref{accfreqevo}).  The best-fit accretion timescale 
 adopting nominal cluster ages is 1.75 $Myr$ ($\chi^{2}_{\nu}$ $\sim$ 1.4), slightly smaller than our nominal 
dust timescale and marginally consistent with 
previous analyses \citep{Fedele2010}.  The predicted gas disk frequency 
as a function of time well matches the combined sample 
with the possible exception of $\rho$ Oph where it overpredicts the accretion 
frequency and slightly older (t $>$ 5 $Myr$) regions like $\eta$ Cha and 25 Ori, where 
it underpredicts the accretion frequency.

As with dust, the revised cluster ages for many clusters significantly extend the 
gas disk lifetime.  Our best-fit value is $\tau_{o}$ = 3.75 $Myr$, more than
 a factor of two longer than the timescale we obtain from nominal cluster ages.
Taken literally, adopting revised cluster ages yields gas disk timescales slightly shorter 
than warm dust lifetimes, in slight contrast to previous results finding similar lifetimes for 
gas and dust \citep{Fedele2010,CurrieKenyon2009}.  Although sample-specific biases in how disk dust and gas may 
instead be responsible for these differing gas and dust lifetimes, the more basic result is robust: 
the revised cluster ages would more than double the gas \textit{and} warm dust lifetimes.

\subsection{Transitional Disk Evolution}
\begin{figure}
\epsscale{1.1}
\centering
\plottwo{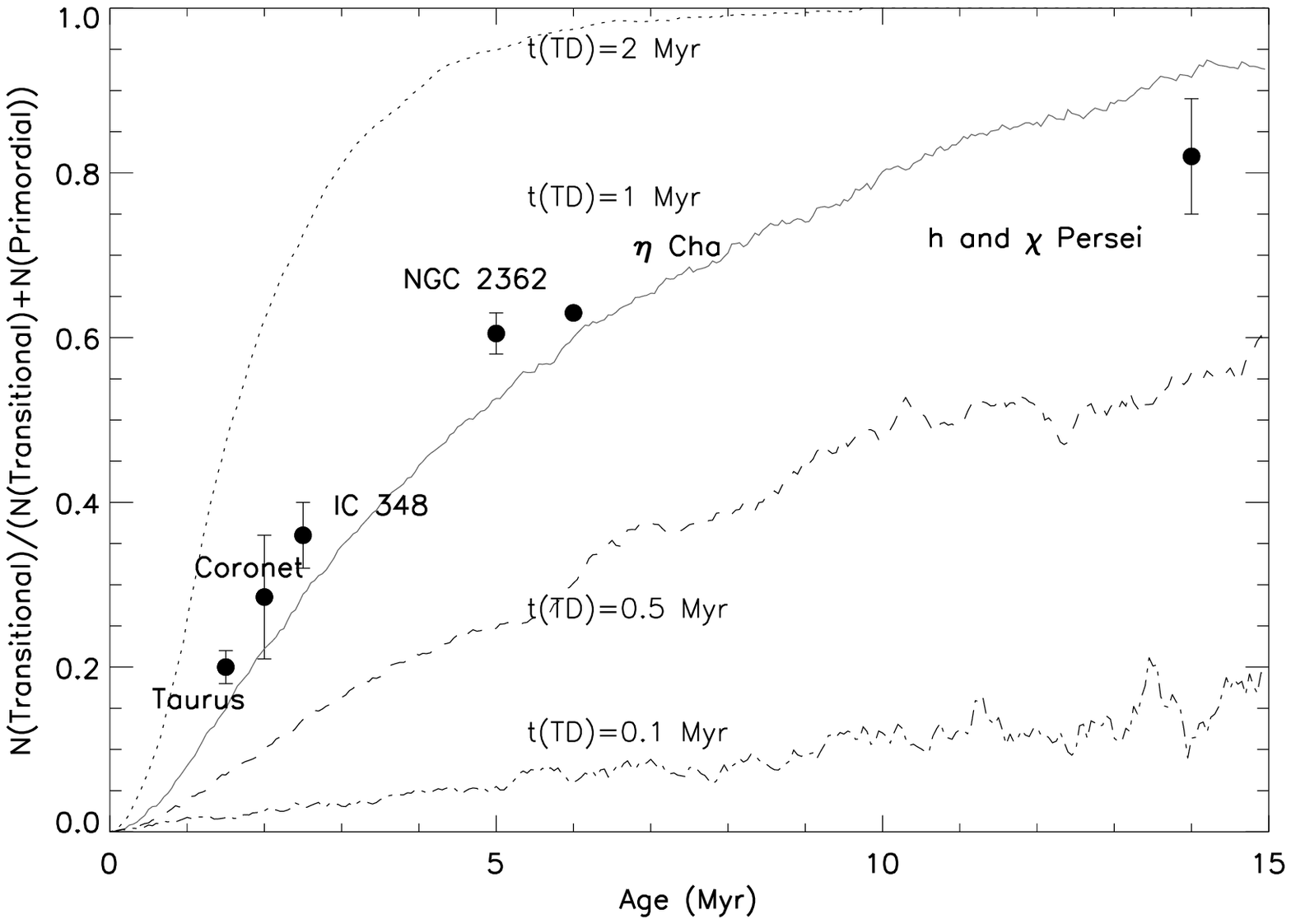}{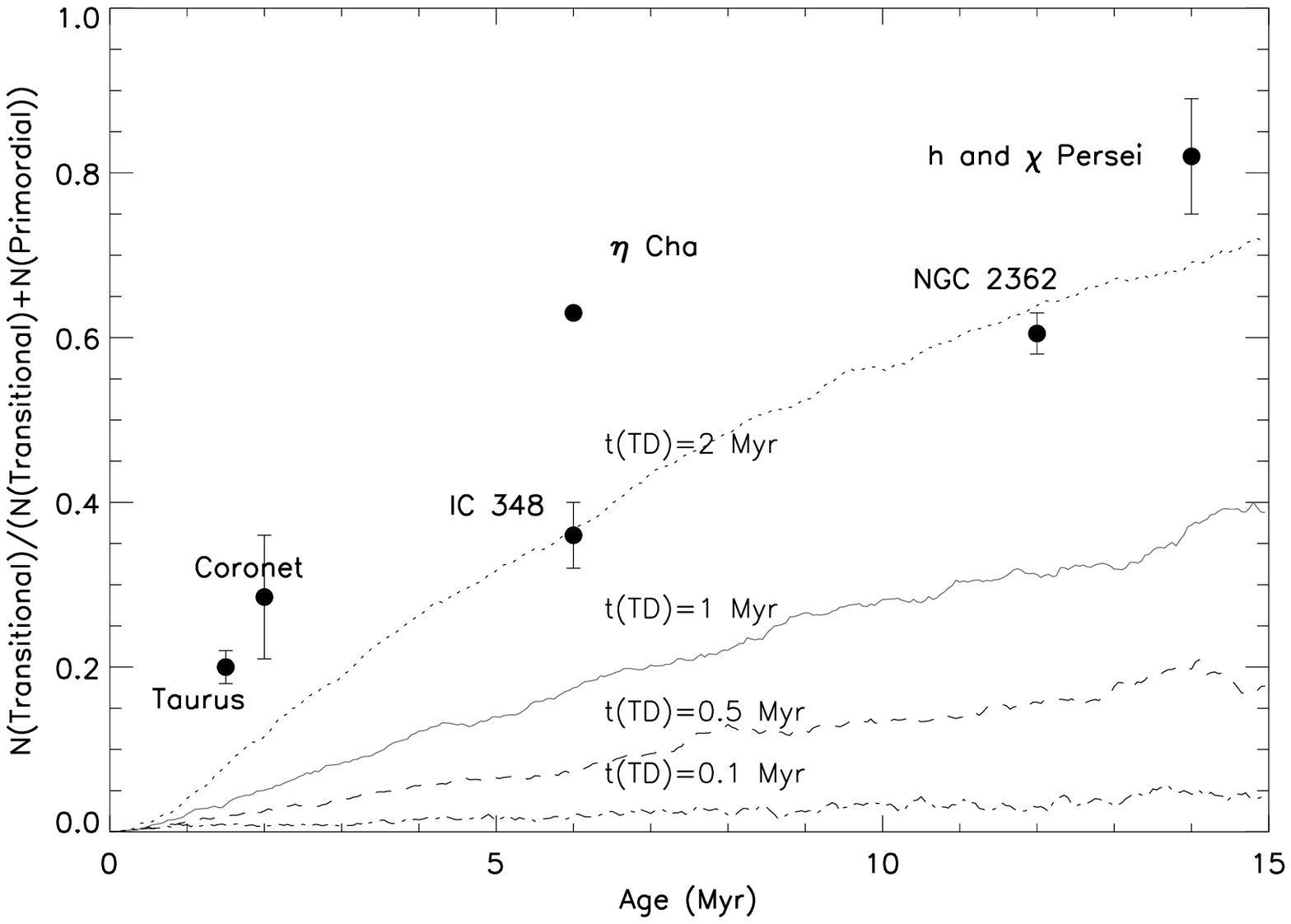}
\caption{The relative frequency of transitional disks vs time in Taurus, the Coronet 
Cluster, IC 348, NGC 2369, $\eta$ Cha as in \citet{CurrieSiciliaAguilar2011} updated 
to include h and $\chi$ Persei and to consider nominal (left) and revised (right) 
cluster ages from \citet{Bell2013}.   For each case, we overplot predicted 
transitional disk frequencies from our parametric model, adopting a typical 
protoplanetary disk lifetime of 2.5 Myr (left) and 6 Myr (right).  In both cases, the 
duration of the transitional disk phase is significantly longer than 0.1 $Myr$ and 
more comparable to $\sim$ 1 Myr.  }
\label{tranfreqevo}
\end{figure}

Figure \ref{tranfreqevo} compares transitional disk frequencies derived for Taurus, the Coronet Cluster, 
IC 348, NGC 2362, and $\eta$ Cha from \citet{CurrieSiciliaAguilar2011} with 
those derived here for our limited h and $\chi$ Persei sample.  Lines denote 
predicted transitional disk frequencies from our Monte Carlo simulations for a range of bulk transitional disk durations ranging 
between 0.1 $Myr$ and 2 $Myr$ assuming a protoplanetary disk lifetime of 2.5 $Myr$ for nominal cluster ages (left) 
and 6 $Myr$ for revised ages (right).  For h and $\chi$ Persei, the error bars correspond to upper and lower 
limits for the transitional disk frequency determined from whether stars with weak excesses are considered to be homologously depleted 
transitional disks or warm debris disks.  For other clusters, the error bars likewise reflect ambiguities in interpreting the properties 
of the most weakly-emitting disks \citep[see discussion in ][]{CurrieSiciliaAguilar2011}.  

Despite uncertainties in inferring the properties 
of many disks, our analysis clearly points to a transitional disk phase that lasts $\approx$ 1 $Myr$ when 
averaged over entire cluster samples and all possible pathways: transitional disks with holes/gaps depleted 
in dust that clear from the inside-out and those that deplete homologously.
For revised ages for IC 348, NGC 2362 and other clusters, this trend is even stronger.  
For all cases, the transitional disk phase 
lasts on average $\approx$ 1/4--1/3 of the total protoplanetary disk lifetime.  Our results are consistent with recent 
detailed studies of transitional disk frequencies \citep{CurrieSiciliaAguilar2011,CurrieLada2009}.

Many sources in the oldest clusters -- i.e. NGC 2362, h and $\chi$ Per -- identified as having homologously depleted transitional disks may instead have warm debris disks.
Nevertheless, considering primordial disks and transitional disks with inner holes alone for these clusters 
still point to an extended transitional disk duration.  For NGC 2362 and h and $\chi$ Persei, we would still derive 
transitional disk fractions of 5/9 and 6/8, respectively (cf. Currie \& Sicilia-Aguilar 2011 for NGC 2362; this work for h and $\chi$ Per), which 
are consistent with a 1 $Myr$-long average transitional disk phase.   

\subsection{Frequency of Dust around 4--6 $M_{\odot}$ stars}
\begin{figure}
\centering
\plotone{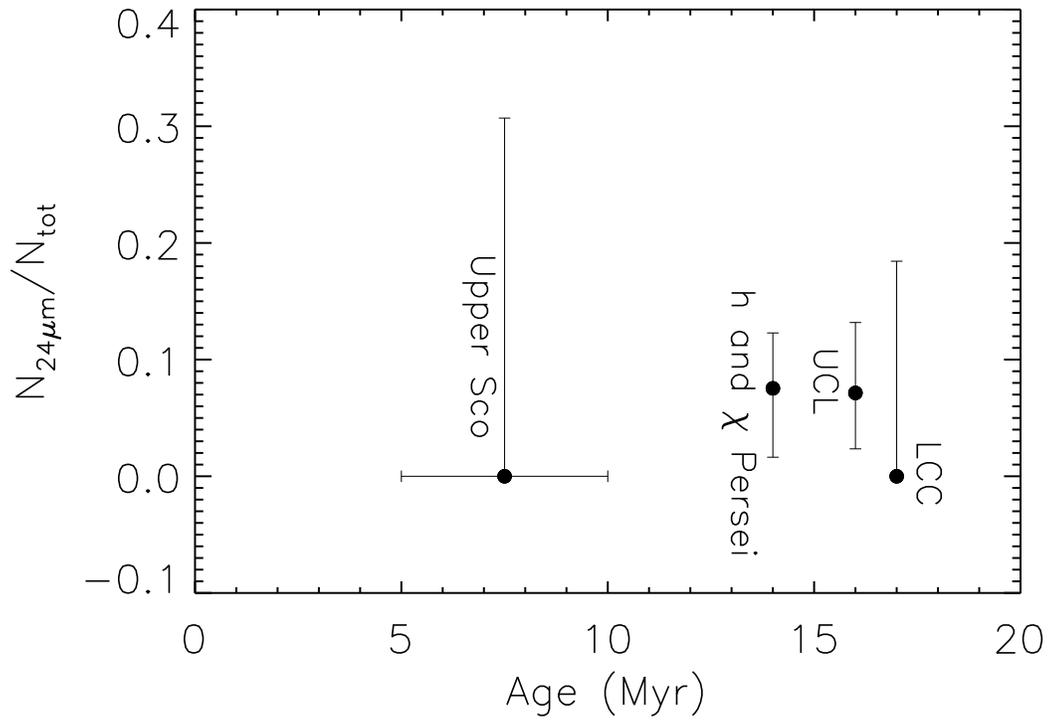}
%{24micfreq_4to6.ps} 
\caption{Frequency of 24 $\mu$m excess emission around 4--6 $M_{\odot}$ stars h and $\chi$ Persei and the 
the three Sco-Cen subgroups: Upper Scorpius, Lower Centaurus Crux, and Upper Centaurus Lupus.} 
\label{freq24exc_massive}
\end{figure}

Although the relatively shallower depth of MIPS compared to IRAC for stellar photospheres means we have far weaker constraints 
on the presence of cooler dust producing 24 $\mu$m excess emission, the MIPS data are 
deep enough that we can compare frequencies for massive stars to those for similar stars 
in other clusters.  
Figure \ref{freq24exc_massive} compares the frequency of dust around 4--6 $M_{\odot}$ stars 
to those for the Sco-Cen subgroups: Upper Scorpius (5--10 $Myr$), Upper Centaurus Lupus (16 $Myr$) 
and Lower Centaurus Crux (17 $Myr$).  To derive frequencies for Sco-Cen we focus on 
B2--B6 stars and simply compare the number flagged as excess sources in \citet{Carpenter2009b} 
and \citet{Chen2012}, trimming out Be stars and post-main sequence objects. 

All samples exhibit a very low frequency of 24 $\mu$m excess from disks 
around 4--6 $M_{\odot}$ stars, ranging 
between zero and $\sim$ 7\% (0/4 for Upper Sco, 0/8 for LCC, and 1/14 for UCL).  
The small sample sizes and small numbers of excess sources preclude us from identifying
any trend in 24 $\mu$m excess frequency vs. time as has been found for 
1.5 $M_{\odot}$ stars \citep{Cpk2008}.   
However, these massive stars exhibit lower MIPS-24 $\mu m$ excess frequencies than do 
1--2.5 $M_{\odot}$ stars  at 10--30 $Myr$ for many clusters \citep[$\sim$ 25--50\%][]{Cpk2008,Forbrich2008,Chen2011,Chen2012}.
%This trend holds true if we consider \textit{all} sources of excess emission, including 
%dust from Be stars (dotted lines): 1/4 for Upper Sco, 
%Thus, at ages covered by these samples (5-20  $Myr$), a peak in the 24 $\mu$m excess 
%frequency is evident at 1.5-2 $M_{\odot}$ or typically corresponding to A and F stars.
%\subsection{Evolution of Colder, 24 $\mu$m Dust Luminosity with Time}
%\subsection{Frequency of 24 $\mu$m Excess Emission with Stellar Mass}
%--see new fig. (placeholder is Figure \ref{risefall})

\section{Discussion} \label{sectdiscussion}
\subsection{Summary of Results}
%(describe what we were trying to do, summarize results)\\
We present new constrains on the time evolution of planet forming disks from new, deep IRAC and MIPS photometry of 
$\sim 12,500$ stellar members of the Double Cluster, h and $\chi$ Persei, 
%Within our sample 263 members exhibit evidence for warm circumstellar dust and 152 exhibit evidence for cooler circumstellar dust as probed by excess 8 $\mu$m and 24 $\mu$m emission respectively. 
%In the mid-IR passbands of particular interest our stellar samples are 
deeper at 8 $\mu$m and 24 $\mu$m by $\sim 1.5$ magnitudes compared to previous studies \citep{Currie2007a,Currie2008a}. 
Relative to regions of a similar age, the Double Cluster contains a large sample of stars which is vital for producing 
robust statistical estimates of the frequency of warm dust and gas.

Our study yields the following primary results:

%In this study we perform a deep photometric, infrared study of h and $\chi$ Persei members using $IRAC$ and MIPS on the \emph{Spitzer Space Telescope}. In the bandpasses of particular interest to our study we detect cluster members at $\sim 1.5$ magnitudes dimmer in both [8] and [24] compared to \cite{Currie2007a} (IRAC) and \cite{Currie2008a} (MIPS) (cf. Figs. \ref{numbercounts}, \ref{mag24error}).

%We first identify the IR-excess sources via their reddened colours and proceed to quantify the frequency of warm and cooler dust as a function of [V], [I], spectral type, and/or inferred stellar mass where the three former variables are used as proxies for stellar mass. This is done both including and excluding known Be stars where applicable. 
%We find in general that the frequency of warm dust increases around stars of later types between $\sim 1-5$\% ignoring the initial spike in disk frequency among the visually brightest/earliest type stars. Due to the smaller sample of detected sources at 24 $\mu$m (cf. left Fig. \ref{mag24error}), a trend of cooler dust frequency versus stellar mass is less apparent. However we observe that the disk frequency appears to be $\sim 5$\% among the most massive stars in our sample ($4-6 M_{\odot}$).  

\begin{itemize}
\item The frequency of warm dust, as probed by IRAC $8 \mu$m excess, is on the order of $\sim$ 2--3\% for the sample as a whole, increasing from near zero for mid-B stars to $\approx$ 4--5\% for sub-solar mass stars. \\
%. Across our entire sample of stars that show evidence for warm circumstellar dust, 
%the frequency of such material rises from $\sim 0.5-5$\% as a function of infered spectral type.

\item Although the MIPS 24 $\mu$m data are not sufficiently deep to conduct a similar analysis for all h and $\chi$ Persei members, they provide the first constraints on the disk frequency for the region's most massive planet-bearing stars (4--6 M$_{\odot}$).  About $\sim$ 3--8\% of these massive stars have MIPS-24 $\mu m$ excesses, lower than the MIPS-24 $\mu m$ frequencies found for 1--2.5 $M_{\odot}$ stars in other clusters.  

%\item A similar dust frequency analysis of $24 \mu$m excess sources is conducted. However due to the 
%relatively small sample size of cluster members that show evidence for colder dust, a general trend 
%of the frequency of colder dust as a function of spectral type is not observed. Instead we find that 
%the observed frequency of $24 \mu$m excess sources to be between $\sim 3-8$\% among the most massive 
%stars (4-6 M$_{\odot}$).    

\item The frequency of accretion amongst h and $\chi$ Persei stars is $\sim 2$\%, comparable to the warm dust frequency. Similarly, the accretion frequency may be higher for later type, lower-mass stars. We do not find evidence for a correlation between the presence of warm dust and accreting gas. \\

\item SED modeling of h and $\chi$ Persei stars with IRAC \emph{and} MIPS excesses allows us to investigate the morphology of rare disks with warm dust at 10--15 Myr. Our modeled stars have disks exhibiting a wide range of evolutionary states. In particular, a large fraction of disks appear to have SEDs consistent with transitional disks, indicative of a rather extended transitional disk duration. \\

%We then place our results in the context of global disk evolution over time. % by comparing the frequency of various empirical quantities in h and $\chi$ Persei to that of other clusters from $\sim 0-15$ Myr. 
%Through the use of Monte-Carlo simulations of evolving disks over the same timescale we derive an e-folding time for each of warm dust in disks, accreting gas, and transitional disks after adapting nominal cluster ages first and again with revised cluster ages where applicable. For warm circumstellar dust we derive an e-folding timescale of 2.75 Myr using nominal cluster ages which gets extended to 5.75 Myr if we adopt the revised ages. Similarly the e-folding timescale for accreting gas in disks increases from 1.75 to 3.75 Myr and from $\sim 1$ to $\sim 2$ Myr for the ratio of transitional disks to transitional and primordial disks. 

\item We combine our results for h and $\chi$ Persei to those for other clusters to conduct a global analysis of disk evolution in young clusters. Assuming nominal cluster ages, we derive characteristic e-folding timescales of 2.75, 1.75, and 1 Myr for the warm dust lifetime, gas lifetime, and transitional disk duration. Assuming alternate cluster ages \citep{Bell2013,Pecaut2012}, these timescales increase to 5.75, 3.75, and 1.5 Myr, respectively. Thus, the protoplanetary disk phase may last significantly longer than previously thought. 
\end{itemize}
%The empirically derived frequency of warm dust, gas, and transisitonal disks in h and $\chi$ Persei 
%are compared to other clusters with ages between 0-15 Myr. Assuming nominal cluster ages, characteristic 
%e-folding timescales ($\tau_0$) for each of the aforementioned quantities are determined to be 2.75 Myr, 
%1.75 Myr, and 1 Myr respectively. We then revise our set of cluster ages according to the findings of 
%\cite{Bell2013} and arrive at the increased $\tau_0$s of 5.75 Myr, 3.75 Myr, and $\sim 1.5$ Myr.    

%We also conduct a brief analysis of the frequency of cooler dust around $4-6$ M$_{\odot}$ stars in h and $\chi$ Persei and the Sco-Cen subgroups with ages from $\sim 5-20$ Myr. As a result of the lack of observed trend in the data we do not fit an e-folding timescale for cooler dust.  

\subsection{Comparisons with Earlier Studies}
\subsubsection{Earlier Studies of the Disk Population of h and $\chi$ Persei}
% \citep{Currie2007a,Currie2007b,Currie2007c,Currie2008a}}
This study builds upon analyses of Cycle 1 IRAC and MIPS Spitzer and ground-based data, which comprised the first 
studies of circumstellar disks around stars in the Double Cluster \citep{Currie2007a,Currie2007b,Currie2007c,Currie2008a}.
Over an area comparable to this study, \citet{Currie2007a} identified about 5000 (7000) stars detected in the IRAC 
[8] ([4.5]) filters and located near previously known members of h and $\chi$ Per, used 2MASS all-sky survey data to select 
stars whose near-IR color-magnitude diagram positions were plausibly consistent with those for 2.3 $kpc$ distant stars, 
and studied the frequency of warm IRAC excess emission by comparing photometry derived from 2MASS $K_{s}$ to IRAC flux densities.
\citet{Currie2007b} identified a subsample of stars with IRAC excesses 
in multiple bands, added in preliminary MIPS photometry, and compared the stars' 
SEDs to model predictions for warm debris disks whose dust 
is indicative of active terrestrial planet formation.  
\citet{Currie2007c} identified about two dozen stars whose color-magnitude diagram positions were consistent 
with values expected for h and $\chi$ Persei members and whose $H_{\alpha}$ equivalent widths were consistent with 
estimates for accreting objects.  
Finally, \citet{Currie2008a} present a full analysis of MIPS data for h and $\chi$ 
Persei, modeled systems with strong MIPS-excess emission, and compared MIPS 24 $\mu$m excesses around BAF stars from 
multiple regions to investigate the time evolution of 24 $\mu$m debris emission.

The data analyzed in our study differ from previous data in several important respects.  
Our IRAC data are significantly deeper, with completion limits 
roughly two magnitudes deeper in [8]; at the completion limits for the Cycle 1 data, our photometric errors are 
$\approx$ 3 times smaller.  We achieve similar increases in survey depth at MIPS 24 $\mu$m.  While \citet{Currie2007c} 
draw from preliminary analysis of 5000 stars with optical spectroscopy, our study incorporates the full sample.
These previous studies either focused on all 2MASS-detected objects in the field or relied on very preliminary photometric 
cuts to focus on candidate members.  Our study uses the candidate membership list from \citet{Currie2010a} based on 
extremely deep optical photometry and optical spectroscopy to remove background stars from our analysis.

Our results modify some previous conclusions about h and $\chi$ Persei's disk population.
% in minor ways. 
\citet{Currie2007a} found that few ($\sim$ 2--3 \%) stars have excesses at multiple IRAC bands, although the 
frequency of such excesses rose towards
%the frequency for which rose towards 
fainter stars likely representative of later, lower-mass stars.  We derive similar frequencies at 8 $\mu$m near our 
survey depth limit.  For brighter stars probed by \citet{Currie2007a}, our frequencies are lower likely because our criterion 
for differentiating between an IR excess source and a bare stellar photosphere is more conservative and based on the 
observed dispersion of IRAC colors as a function of magnitude.  While our accretor frequency agrees with 
rough estimates previously reported by \citet{Currie2007c}, we fail to confirm the trend between 
IRAC excess and $H_{\alpha}$ equivalent width indicated by their preliminary analysis.

\citet{Currie2007b} and \citet{Currie2008a} identify 
stars with weak IRAC excesses as candidate terrestrial planet-forming systems.  A full analysis of the optical spectroscopy 
sample and the IRAC data together demonstrates that h and $\chi$ Persei contains a mixed disk population: sources with evidence 
for gas but no dust, dust but no evidence for gas, and those with evidence for both.  It is likely that 
many (most?) of our weak IRAC excess sources are terrestrial planet forming systems, since such systems are found 
around more nearby, better studied stars \citep{Rhee2007,Lisse2008}.  However, to be conservative we remain 
agnostic as to the state of these systems: they could either be warm debris disks or homologously depleted transitional disks.

Finally, we did not reinvestigate the ``rise and fall" of 24 $\mu$m emission from debris disks around BAF stars.  Although a 
full analysis of these targets within the context of debris disk evolution will be left for a future paper, studies 
published subsequent to \citet{Currie2008a} support a more limited version of our previous results.  For example, 
\citet{Carpenter2009a} find evidence for a peak in debris emission around F stars at $\sim$ 10--15 $Myr$ but do not find a 
similar trend for earlier or later-type stars.  Focusing only 
on planet mass (by accounting for the different mapping between spectral type and mass vs. age), \citet{Chen2011} find 
evidence for an increase in debris emission at 15--20 $Myr$ (the age of the older Sco-Cen subgroups)
 around 1.5 $M_{\odot}$ stars.  But this behavior is not seen so clearly for solar-mass stars nor for 2--2.5 $M_{\odot}$ stars.
  While \citet{Currie2008a} and later \citet{KenyonBromley2008} interpret 
a peak in debris emission within the context of oligarchic growth in icy planetesimal swarms, \citet{Chen2011} argue that 
the debris luminosities for solar-mass stars are \textit{higher} than predicted in debris disk evolution models.  
The distribution of debris disk temperatures as a function of stellar mass and age also may hint at key effects not 
explicitly considered in the \citeauthor{KenyonBromley2008} models such as planetesimal belt sculpting by 
massive planets \citep[e.g.][]{Chen2014}.

\subsubsection{The Lifetime of Protoplanetary Disks and Timescale for Gas Giant Planet Formation}
This work provides a quantitative, updated comparison to earlier estimates of the lifetime of warm dust and gas 
in protoplanetary disks and assessment of how stellar age uncertainties affect derived lifetimes.  
From ground-based $L^\prime$ data of many clusters, \citet{Haisch2001} compare disk frequencies to a 
simple linear fit and estimate that half of all protoplanetary disks disappear by $\sim$ 3 $Myr$ 
and nearly all disappear by 6 $Myr$.  Based on deeper Spitzer/IRAC data able to detect the stellar photosphere 
of subsolar mass stars and at wavelengths where disk excess is more easily identifiable \citep[see discussion in][]{Ercolano2009}, 
\citet{Hernandez2007} estimate a typical lifetime of $\approx$ 3 $Myr$ as well.
%, although they note that 
%many regions nominally age-dated to older than 3--6 $Myr$ have stars with warm protoplanetary disk dust.  
%\citet{CurrieSiciliaAguilar2011} and 
\citet{Fedele2010} match frequencies of gas and dust to an exponential decay function; 
\citet{CurrieSiciliaAguilar2011} match the dust frequency to Monte Carlo simulation predictions with 
an exponential decay probability distribution function as in this work.  In all cases, the results are similar: 
a protoplanetary disk gas and dust lifetime of $\approx$ 2.5--3 $Myr$ for most stars \citep[see also][]{Mamajek2009,Jayawardhana2006}, 
slightly shorter (longer) for early-type higher mass (mid M) stars \citep[see][]{Carpenter2006,CurrieKenyon2009}.

Assuming nominal cluster ages, we derive fairly similar results, although we, like \citet{Fedele2010},
formally favor a slightly longer e-folding timescale for dust than gas ($\tau_{dust}$ $\sim$ 2.75 $Myr$ and 
$\tau_{gas}$ $\sim$ 1.75 $Myr$).  This may be due to some mixing 
of disk populations: e.g. some of the warm dust may be from terrestrial zone debris emission, not protoplanetary 
disk dust, so that our protoplanetary disk frequency is overestimated.  Alternatively, 
surveys may be failing to identify some stars whose weak optical emission lines and veiling 
from accretion are best identified from high signal-to-noise, high resolution echelle spectra 
\citep[see][]{Dahm2008}.

Our results support and quantify \citet{Bell2013}'s assertion that alternate ages derived for many 
young clusters would significantly lengthen the inferred lifetimes for protoplanetary disk dust.  
As expected, the gas disk lifetime is also longer.  \citet{Bell2013} quote a ``half-life" of disks 
of 5--6 $Myr$ and report a 10--20\% disk fraction at 10--12 $Myr$.
  Assuming that the disk lifetime decays as an exponential function and including 
slightly different samples (i.e. Upper Scorpius instead of the more distant $\lambda$ Ori), our Monte Carlo 
analysis yields largely similar results for dust: $\tau_{dust}$ $\sim$ 5.75 $Myr$ and a dust fraction of 
$\approx$ 10--20\% at 10--12 $Myr$ and then $<$ 5\% at later ages.   While the gas disk lifetime we derive is shorter ($\approx$ 3.75 $Myr$) 
it is still more than twice as long as the one we derive assuming nominal cluster ages.

At present, it is not yet clear which set of ages is correct \citep{Soderblom2014}, although older ages derived 
for pre-main sequence stars in Upper Scorpius are more consistent with ages derived for 
the region's post-main sequence population \citep[see ][]{Pecaut2012}\footnote{While every age derivation is model dependent, arguably the least model-dependent age-dating method (Li depletion) likewise favors systematically older ages for regions previously age-dated to 10 $Myr$ \citep[e.g.][]{Binks2014}.}.  Thus, these different estimates should be 
understood to bracket possible values.
While the stars we focus on are predominately solar to slightly subsolar in mass, the disk lifetime may be 
even longer for the latest stars and highest-mass brown dwarfs: for $\approx$ 10 $Myr$-old Upper Scorpius, 
\citet{Luhman2012} still find that 1/4 of all mid M stars show evidence for warm circumstellar dust.

A revised, more lengthy characteristic protoplanetary disk lifetime would have important 
consequences for planet formation, especially for gas giants.  For instance, wide separation 
super-jovian mass planets discovered through direct imaging \citep{Marois2008} are
extremely difficult to form via standard core accretion models \citep{Pollack1996}, 
yet at least some of them have properties (e.g. mass ratio/separation; C/O abundances) 
suggesting that they may have formed this way \citep{Currie2011,Konopacky2013}.
Modifications to core accretion models 
%focused on rapidly accreting dynamically colder planetesimals 
\citep{Lambrechts2012,KenyonBromley2009}
are required to account for these cores within the nominal, short protoplanetary disk lifetime.
With the exception of ROXs 42B \citep{Currie2014}, the host stars for these planet-mass companions 
are older than either the nominal or revised protoplanetary disk lifetimes.  
Thus, a longer disk lifetime relaxes these constraints placed on planet formation by core accretion 
as well.

\subsubsection{Duration of the Transitional Disk Phase}
%Structures of Protoplanetary Disks and the Duration of Disk Clearing}
%While the youngest clusters by themselves provide a poor constraint on the 
%lifetime of the transitional disk phase, the predicted relative sizes of the transitional disk population 
%at older ages are very different for $\tau_{TD}$ $\approx$ 0.1 $Myr$ to 2 $Myr$ 
%\citep[e.g.][,\S 6.4 here]{CurrieSiciliaAguilar2011}.  Thus, the disk population of populous 
%10--15 $Myr$ regions like h and $\chi$ Persei can be used to investigate transitional disk durations.
 Our analyses support previous results 
\citep[][]{CurrieLada2009,CurrieSiciliaAguilar2011} 
%using 
%a combination 
%IR colors and sophisticated SED modeling to find
finding
that the transitional disk phase lasts longer 
than proposed in most early studies
%analyzing $IRAS$ satellite and 
%ground-based data 
\citep{Skrutskie1990,Wolk1995}, 
%\footnote{Note here that \citet{Skrutskie1990} 
%clearly consider the possibility of a longer transitional disk duration.  Explicitly, their claim 
%regarding a short transitional disk phase focuses on comparing the ratio of disks with 
%$\Delta$K $<$ 0.2, $\Delta$N $<$ 1.2 to disks with stronger emission.  However, if disks with 
%no near-IR excess but strong ($\Delta$N $\ge$ 1.2) mid-IR excess are ``regarded as candidate transition 
%cases, then our transition time estimate \textbf{must be increased to $t$ $\sim$ 1 $Myr$}" (emph. added).  
%Several objects with such properties are widely considered to be transitional disks, including GM Aur.}: 
closer to 1 $Myr$ than 0.1 $Myr$.
%, or 
%about 1/4--1/3 of the protoplanetary disk lifetime.  
In contrast, a few recent studies which use
empirical breaks in IR 
colors alone arrive at somewhat shorter durations \citep{Luhman2010,Luhman2012}.
The major drawback with these other recent studies is that IR colors alone do not accurately distinguish between 
normal protoplanetary disks and disks that are becoming optically thin, since 
the color space between the two populations overlaps \citep{Merin2010,CurrieSiciliaAguilar2011}  
\footnote{For example, many ``gapped" transitional disks (e.g. LkCa 15, UX Tau, MWC 758) 
have ``normal" IR colors but large, $\gtrsim$ 10 $AU$-wide cavities indicative of disk clearing 
and likely opened by infant planets
\citep{Espaillat2010,Espaillat2012,Andrews2011,DodsonRobinson2011}.
%\footnote{Broadly speaking,
%\citet{Luhman2010,Luhman2012} consider 
%Note that \citet{Luhman2010} treat
%such gapped disks, often referred to as so-called ``pre-transitional" disks, are from the standpoint of 
%their analysis, normal pro }.  
Since these disks have experienced significant structural evolution, they cannot be considered `primordial' 
as claimed in \citet{Luhman2012}, since disks are not born with 50 $AU$-scale gaps in large dust.
Homologously depleted transitional disks 
likewise overlap in some IR color space with normal protoplanetary disks.}.  Thus, a criterion for identifying 
transitional disks based on perceived breaks in IR colors will underestimate this population's size
 and in turn underestimate the transitional disk duration.  

To alleviate these shortcomings, \citet{CurrieSiciliaAguilar2011} adopt the SED modeling-driven criteria 
described in their \S 4.2 and briefly summarized in \S 5.2 in this work.
While \citet{Luhman2012} rightly note that theoretical models are imperfect and have degeneracies, 
\citet{CurrieSiciliaAguilar2011} explicitly addressed these issues by treating their classifications 
probabilistically.  For instance, the intrinsic distribution of their modeling grid in disk mass is heavily 
biased towards high masses.
Their SED modeling identifies many transitional disks inspite of this strong prior in favor of classifying a disk 
as primordial\footnote{\citet{Luhman2012} also cite \citet{Espaillat2012} as highlighting ZZ Tau as a candidate transitional disk that 
could also be fit by a primordial disk model with significant dust settling.
However, the \citeauthor{Espaillat2012} model is optically thick only over tiny, $\sim$ $AU$ scales, while the modeled disk size 
is $\approx$ 100 $AU$.  Visual inspection of their Figure 13 indicates 
that their the model overpredicts emission at $K$-band and the IRAC bands, which
comes hot dust located at these inner disk regions.  Our own model comparisons suggest that ZZ Tau has only a clear broadband excess at 
$\lambda$ $\ge$ 4.5 $\mu m$ \citep{CurrieSiciliaAguilar2011}.  Thus, their optical depth in inner disk regions is likely overestimated.
Finally, they require a significantly lower disk mass than typically reported for optically thick disks 
(5 $\times$ 10$^{-4}$ $M_{\odot}$), which is precisely the point made in \citet{CurrieSiciliaAguilar2011}.}.  

The key addition used in \citet{CurrieSiciliaAguilar2011}'s classifications -- a disk mass limit 
separating some transitional disks from primordial disks -- is justified.  
Our ignorance of the dust opacity and assumed (not measured) dust-to-gas ratios for many disks precludes 
an absolute calibration of disk masses \citep{CurrieSiciliaAguilar2011}.
  However, since disk masses for both primordial disks and (candidate) transitional disks
were derived self-consistently, we can identify whether or not homologously depleted transitional disks have 
\textit{relatively} lower masses.  
Although \citet{CurrieSiciliaAguilar2011}'s adopted limit (0.1\%$\times$$M_{\star}$) was typically just above the detection 
limit for most of the submm data we considered \citep{Andrews2005}, 
%this limit is irrelevant for 
%their conclusions: 
nearly all of the optically-thick disks in \citet{Andrews2005} were detected, 
and \citet{CurrieSiciliaAguilar2011} modeled \textit{all} of them anyway.  Candidate homologously depleted disk 
masses can then be compared to the mass distribution for primordial disks.  
%$Thus, they were able to determine whether candidate homologously depleted disks 
%identified from IR data have 
%estimated masses below this limit.  
\footnote{Subsequent, deeper (sub)mm data do not change this result. \citet{Andrews2013} find that 
their sample, which includes both optically thick disks and those that are optically thin in the near-to-mid IR
 (e.g. V819 Tau, V410 X-ray 6, FP Tau, ZZ Tau, have submm-inferred masses of 0.2--0.6 \%.  The lower-quartile of their 
$M_{disk}$/$M_{\odot}$ distribution does extend below 0.1\%, but the masses are estimated from the submm alone and this 
trend is most likely driven by submm non-detections.   Similar trends in \citet{Andrews2005} (``full sample" and their Table 2 
sample) were not recovered by 
\citet{CurrieSiciliaAguilar2011} who estimated disk masses from from modeling \textit{all} data, 
not estimating masses based solely on the submm flux.
SED modeling-derived masses for the \citet{Andrews2013} will 
be higher and will further widen the gap between \citet{CurrieSiciliaAguilar2011}'s adopted 
limit and the mass range for normal, primordial disks.  
Data for older clusters reinforce this point.  \citet{Luhman2012} argue based on $K_{s}$-[24] colors that 
Taurus and 10 $Myr$-old Upper Sco have similar transitional disk populations.  
However, masses for Upper Sco disks are \textit{systematically much lower} than those for Taurus \citep{Mathews2013}.  They likely provide 
evidence for homologous depletion of disk material for most of the population\citep{Mathews2013}.  Our own independent analysis 
of Upper Sco targets shows evidence for a much larger transitional disk population.}  

In summary, detailed modeling of very young nearby clusters like Taurus, coarser modeling of populous older regions like 
NGC 2362 and h and $\chi$ Persei, and statistical analyses of multiple clusters support a more lengthy 
transitional disk duration of $\approx$ 1 $Myr$.  Averaged over entire clusters and over all possible disk clearing 
mechanisms, transitional disks occupy about 1/4 to 1/3 of the total disk lifetime.  As shown in this paper,
 the upwards-revised ages for clusters 
in \citet{Bell2013} and \citet{Pecaut2012} only serve to strengthen the argument in favor of a more extended transitional 
disk duration.

%As different durations for the transitional disk phase
%- classic results.  WW95, Skrutskie (notes this).  Standard repeated often short timescale.  based on 
%Taurus etc.  CL09 showed that this does not work, CSi11 follow.  CSi11 obj with weak near IR not just dust settling.
%colcol diagrams not sufficient.  w/e LM12 says.  reply.  Now even clearer.  1/3 of ppd lifetime.  so if ppd lifetime
%increases then td duration increases. 

%-Hernandez et al. 2007, 2008; Fedele et al.2010\\
%-Currie \& Lada 2009; Luhman et al 2010; Currie \& SiciliaAguilar 2011; Luhman et al. 2012; Espaillat et al. 2012\\
\subsection{Future Work}
The Double Cluster, h and $\chi$ Persei, provides an excellent laboratory with which to study stellar evolution from the 
pre-main sequence to post-main sequence \citep{Currie2010a,Bell2013,Slesnick2002} and circumstellar disks 
at an age probing the protoplanetary to debris disk transition \citep[this work;][]{Currie2007a,Currie2008a}.  
\citet{Currie2010a} highlighted several key areas of future research on h and $\chi$ Per's stellar population, and here 
we list future avenues of research to study its disk population:
\begin{itemize}
\item \textbf{Disk Evolution as a Function of Environment} -- As noted in \citet{Currie2010a}, the coverage area for optical spectra is substantially 
larger than that for the optical photometry we combine it with to establish a list of members, meaning that there could be many 
more stars associated with the halo population of h and $\chi$ Persei \citep{Currie2007a} that we have yet to identify.  
A wide-field optical photometric survey would identify additional candidate members of h and $\chi$ Persei.  Furthermore,
because our Cycle 1 Spitzer data (and, to a lesser extent, the data we present here) were obtained over a similarly large area, 
we could compare the frequency of warm dust/gas between stars in the low density regions to those in the cluster-dominated regions 
to better understand how disk evolution depends on the circumstellar environment.  Recent studies indicate that disk lifetimes 
may be signficantly longer for extremely tenuous associations \citep{Fang2012}, or much shorter in extremely dense globular cluster-like 
environments \citep{Thompson2013}.  The Double Cluster allows us to study the environmental dependence of the frequency of warm dust in disks
within a single coeval region.
\item \textbf{The Disk Properties of Subsolar-Mass Stars} -- Our current spectroscopic sample does not well probe 
the h and $\chi$ Persei disk later than $\sim$ mid G \citep[see Fig. 6 in ][]{Currie2010a}.  
In addition to identifying low-mass (candidate) members, optical spectroscopy 
such as with MMT/Hectospec (and later Binospec) or LBT/MODS could measure accretion diagnostics (e.g. $H_{\alpha}$) that will allow us to 
better characterize the properties of the faintest stars with IRAC and MIPS excesses.  In principle, we could see whether 
all such stars show evidence for accretion or, like the stars studied here, include both stars showing evidence for gas (protoplanetary disks) 
and those lacking evidence for gas (candidate debris disk-hosting stars).
\item \textbf{Observations of the h and $\chi$ Persei Disk Population with \textit{JWST}} --
Finally, the \textit{James Webb Space Telescope} (JWST) should provide a significant advance in our understanding 
of the Double Cluster's disk population.  In particular, while Spitzer/IRAC data are highly sensitive to 
photospheric emission/weak excesses around subsolar mass stars, our sensitivity further into the infrared 
is far poorer, as MIPS is only sensitive to the photospheres of early/mid B stars (see \S 2.5).  

In contrast, 
MIRI with \textit{JWST} has a 50$\times$ smaller beam size and a predicted 5-$\sigma$, 2500 $s$ sensitivity of 28 $\mu$Jy at 25.5 $\mu$m, or roughly 
two magnitudes fainter than our typical MIPS-24 $\mu$m limits, capable of detecting the photosphere 
of early A stars in the Double Cluster.  Sensitivity gains at slightly shorter wavelengths 
(e.g. 20 $\mu$m) are even more substantial, allowing us to study disks around even lower-mass stars.
The far superior mid-IR sensitivity of JWST/MIRI then means we will better be able to put a study of 
h and $\chi$ Per's disk population on a more comparable footing to other less populous but much closer regions 
(e.g. Sco-Cen; $\sigma$ and $\lambda$ Orionis).  
%MIRI includes multiple filters in between the wavelengths 
%currently covered by IRAC and MIPS, and the relative brightnesses of disks in these filters 
%gives insightMIRI will 
%- look for disks aroundstars in the halo population\\
%- spectroscopy of lower-mass stars\\
%- compare with x-ray data\\
%-JWST will solve everything else\\
\end{itemize}
\acknowledgements We thank the anonymous referee for suggestions that improved the quality of this paper.  
John Carpenter and Eric Mamajek also provided very helpful early draft comments and other suggestions that improved the 
quality of this paper.  We also thank Andras Gaspar, Todd Thompson, and Cameron Bell for additional helpful conversations.  
T.C. is supported by a McLean Postdoctoral Fellowship.
{}
\appendix
\subsection{Expected Levels of Extragalactic Contamination of h and $\chi$ Persei Members}
From our line of sight, h and $\chi$ Persei an extremely populous cluster projected 
out of the galactic plane by $\sim$ 200pc, and the Spitzer data reaches completeness 
limits of $\sim$ 51 $\mu Jy$ at [8], making extragalactic contamination of at least 
one cluster member very likely.  To assess the probable frequency of contamination 
as a function of wavelength and magnitude, we use the IRAC number counts derived 
for the Bootes field from \citet{Fazio2004} and the MIPS-24 $\mu$m counts derived 
from \citet{Papovich2004}.  In \citet{Fazio2004}, the number counts are differential, 
estimating the number of galaxies per magnitude in half-magnitude bins\footnote{Thus the differential
number counts at m[IRAC] = 15.5 are defined from galaxies with m[IRAC] = 15.25--15.75.  To compute the 
number of galaxies expected with m[IRAC]=15.25--15.75, we must multiply by the bin width of 0.5 magnitudes.}.

We estimate the frequency of galactic contamination using the following 
 relation: 
\begin{equation} \label{eq:contam}
N_{contamination} = \pi \times r_{m}^{2} \times N_{g} \times N_{star}/A,
\end{equation} 
where $r_{m}$ is the matching radius (1\arcsec{}), $N_{g}$ is the number of galaxies in a square degree, 
and $N_{star}$ is the number of members in our survey area, A.  Here, $N_{star}$ is the 
subset of $\sim$ 13,960 probable cluster members located within the IRAC/MIPS fields: $\approx$ 12,000
for IRAC and $\approx$ 9600 for MIPS.  
The survey area covers $\sim$ 0.5 square degrees in the IRAC 8 $\mu$m filter 
and 0.4 square degrees in MIPS-24.

From the IRAC 8 $\mu$m differential number counts at m(IRAC) = 14, 15, 15.5, and 16 yields,
we then expect the following number of members 
contaminated by galaxies on our IRAC mosaics as a function of magnitude:
 $\sim$ 4, 9, 22, and 31 galaxies with magnitudes of m[8] = 14, 15, 15.5, and 16, respectively. 
  The IRAC color-magnitude diagram for [8] vs. [3.6,4.5]-[8] implies that 
a member must have have [3.6,4.5]-[8] $\gtrsim$ 0.75 to be flagged as an excess source for the 
faintest objects detected at 5-$\sigma$ at [8].  Our faint detection limit is at 
m[8] $\approx$ 15.5, so galaxies as faint as m[8] $\sim$ 16.25 can contaminate a member, making it 
appear as bright as m[8] = 15.5 and giving it a [3.6,4.5]-[8] color $\gtrsim$ 0.75.  Thus, galaxies 
in the m[8] = 16 bin (m[8] = 15.75-16.25) are the faintest ones that could possibly contaminate 
a member's photometry.  Summing from m[8] = 12 to 15.5 (16), we expect 47 (78) members to 
have IRAC photometry contaminated by a galaxy.

The density of galaxies brighter than [24] = 11.25 is expected to be $\sim$ 10$^{7}$ sr$^{-1}$ 
\citep[cf.][]{Papovich2004}.  Over our 
0.4 square-degree MIPS area, we expect $\sim$ 1200 galaxies brighter than this limit.  
   Adopting $N_{star}$ = 9600 and an area of A = 0.4 sq. deg., we predict that 
18 members are contaminated from galaxies brighter than [24] = 11.25.   
As depicted in Figure \ref{iracmipsmem},
the number of excess sources at 8 and 24 $\mu$m is significantly larger than 
the number of expected extragalactic contaminants.

\subsection{Extragalactic Contamination of the h and $\chi$ Persei IRAC/MIPS Excess Population}
Not all galaxies measurably affect the observed mid-IR colors of the stars they contaminate if the 
stars are much brighter than the galaxies in the IRAC bands, as many 
have near zero IRAC color \citep[see Fig. 1 in ][]{Stern2005}.  However, 
broad-line active galactic nuclei (BL-AGN) and PAH-emission galaxies both have 
very red IRAC colors \citep{Gutermuth2008}.   Member contamination by these galaxies 
can mimick the presence of a disk.
Here we compare the distribution of IRAC colors for members to those expected for 
BL-AGN and PAH-emission galaxies using color-color diagrams as in 
\citet{Gutermuth2008,Gutermuth2009}.

Active star-forming galaxies (e.g. broad-line AGN) have very red [5.8]-[8] colors due to strong PAH emission.  
\citet{Gutermuth2008,Gutermuth2009} define regions in [4.5]-[5.8]/[5.8]-[8] and [3.6]-[5.8]/[4.5]-[8] diagrams 
where these galaxies reside and removed objects located within as contaminants.  We 
first focus on PAH-emission galaxies, definig the region of likely PAH-emission galaxy contamination as follows:
\begin{displaymath}
[5.8]-[8] \ge 1.5 
\end{displaymath}
\begin{displaymath}
[4.5]-[5.8] \le 3\times([5.8]-[8]-1.5)
\end{displaymath}
\begin{displaymath}
[4.5]-[8] \ge 1.7 
\end{displaymath}
\begin{displaymath}
[3.6]-[5.8] \le (2/1.3)\times([4.5]-[8]-1.3).
\end{displaymath}
Compared to the criteria in \citet{Gutermuth2008,Gutermuth2009}, our adopted colors for PAH-emission 
galaxies are $\sim$ 0.5 and 0.7 magnitudes redder in [5.8]-[8] and [4.5]-[8].  However, they 
cover all of the Bootes-field PAH galaxies shown in \citeauthor{Gutermuth2008} and thus should be 
acceptable for targeting this specific type of contaminant.

To compare our member colors to those expected from AGN contaminants, we simply adopt the 
boundaries listed in \citet{Stern2005}: 
\begin{displaymath}
[5.8]-[8] \ge 0.6
\end{displaymath}
\begin{displaymath}
[3.6]-[4.5] \ge 0.2\times([5.8]-[8]) + 0.18
\end{displaymath}
\begin{displaymath}
[3.6]-[4.5] \ge 2.5\times([5.8]-[8])-3.5.
\end{displaymath}
\citet{Gutermuth2008,Gutermuth2009} do not flag AGN from color-color diagrams because the IRAC 
colors of AGN exhibit very strong overlap with young stellar objects in \textit{color-color diagrams} 
but exhibit much weaker overlap with the \textit{color-magnitude diagram} positions of YSOs.  The situation here is 
reversed.  Using here the color-magnitude diagram criteria they adopt would flag as an AGN 
nearly every single h and $\chi$ Persei member with an IRAC excess.  However, 
by adopting the \citet{Stern2005} color-color criteria we can neatly separate h and $\chi$ Persei 
stars from strong AGN contamination.

\begin{figure}
\centering
\plottwo{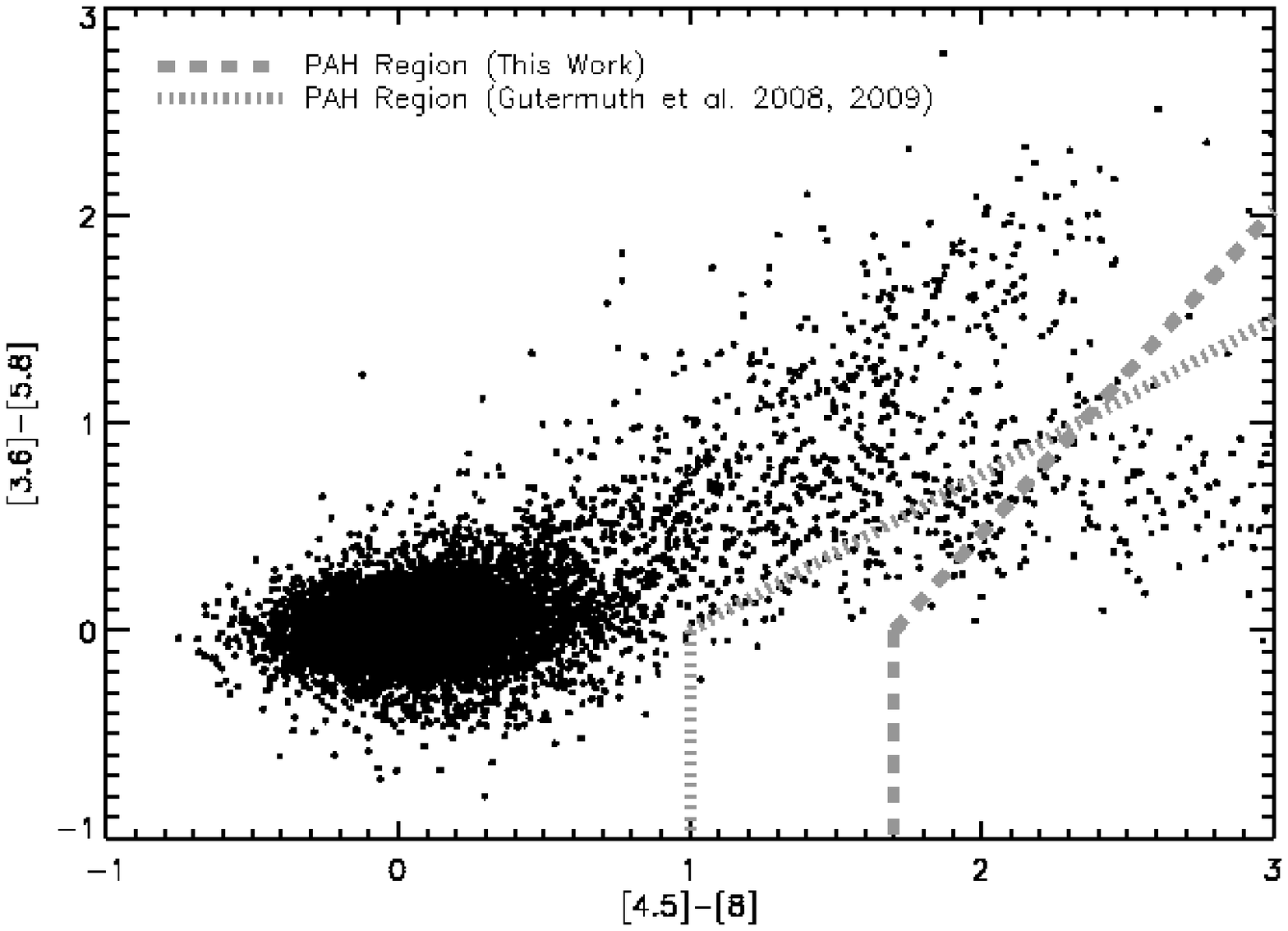}{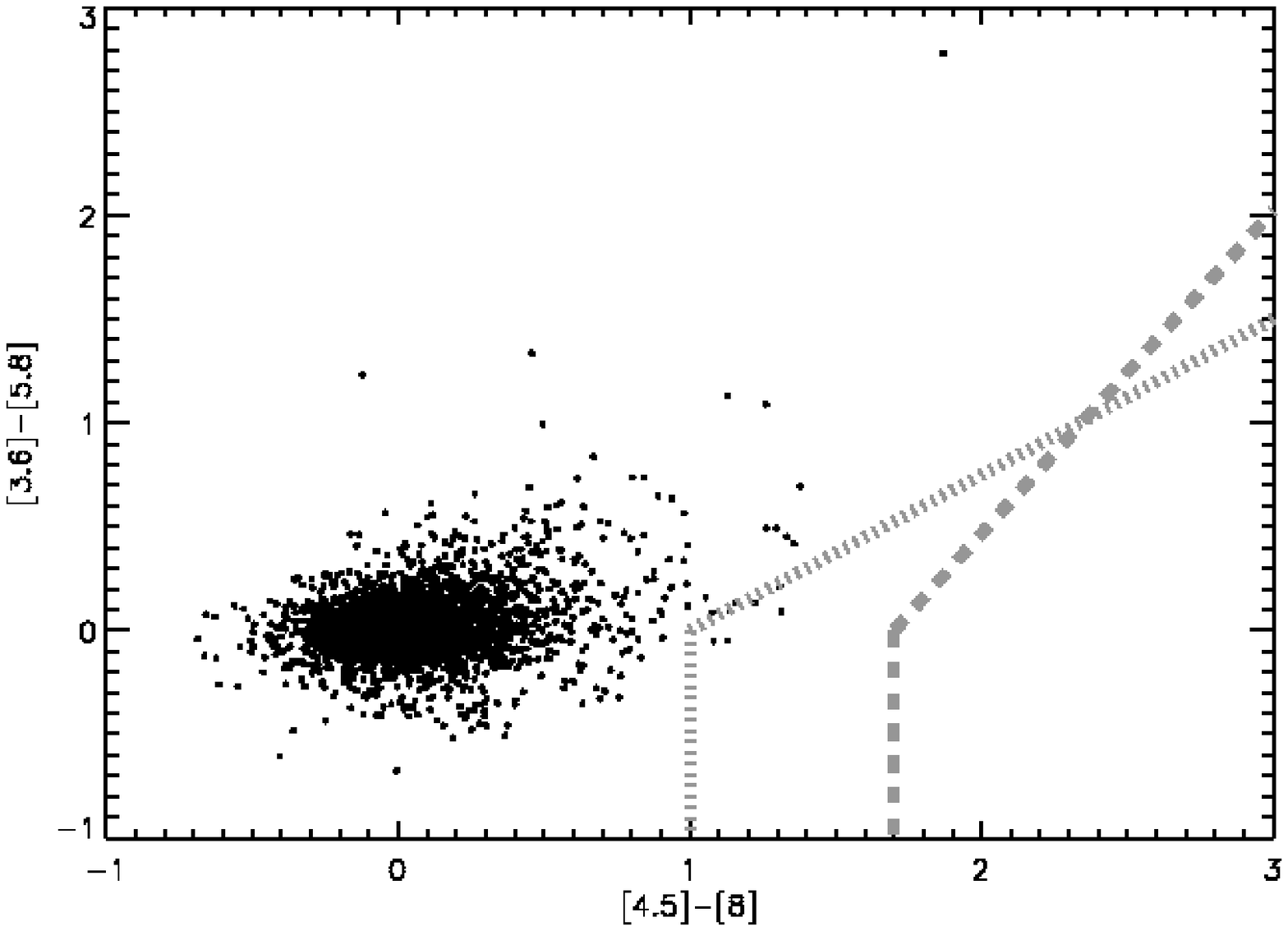}
\plottwo{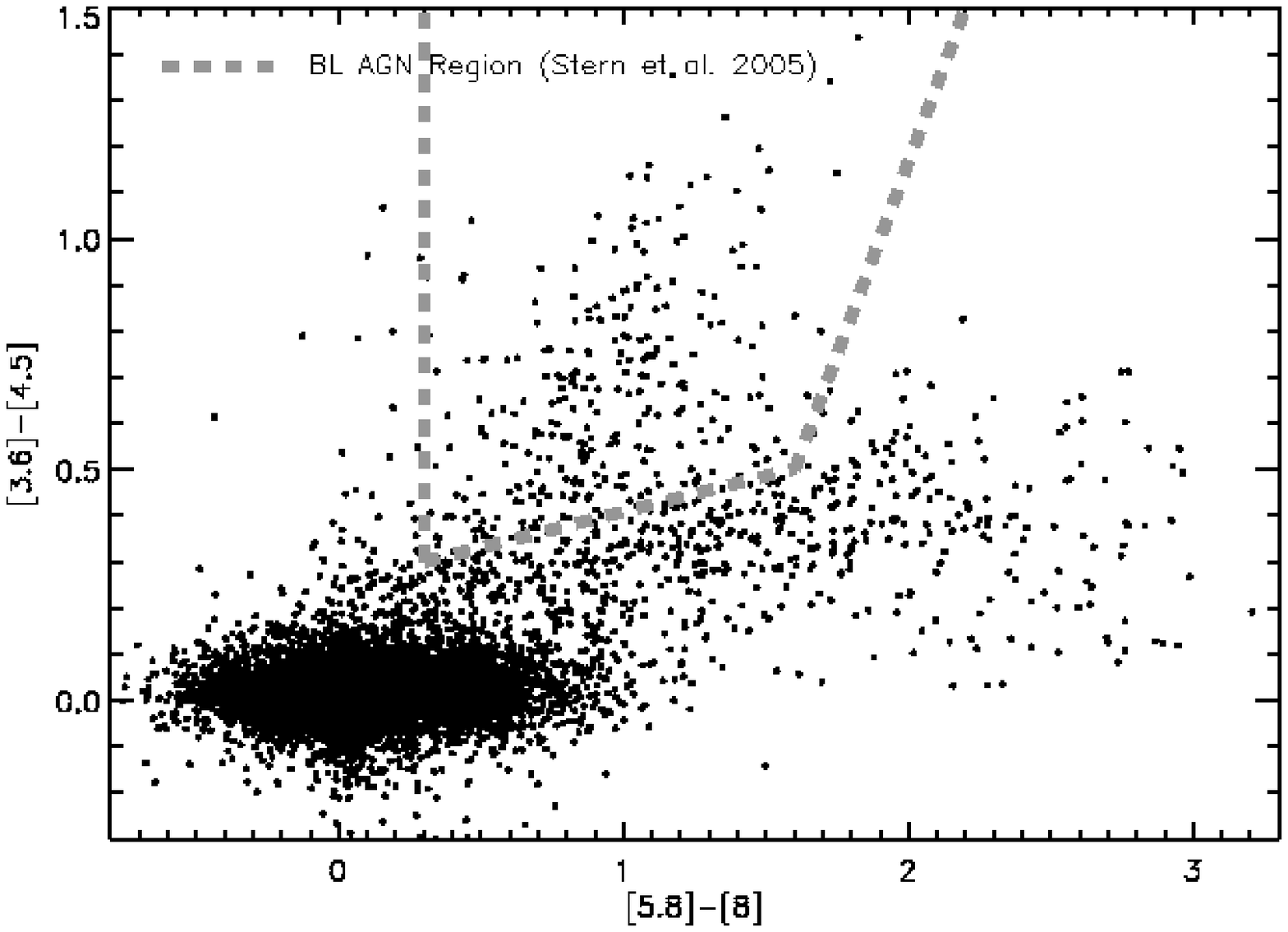}{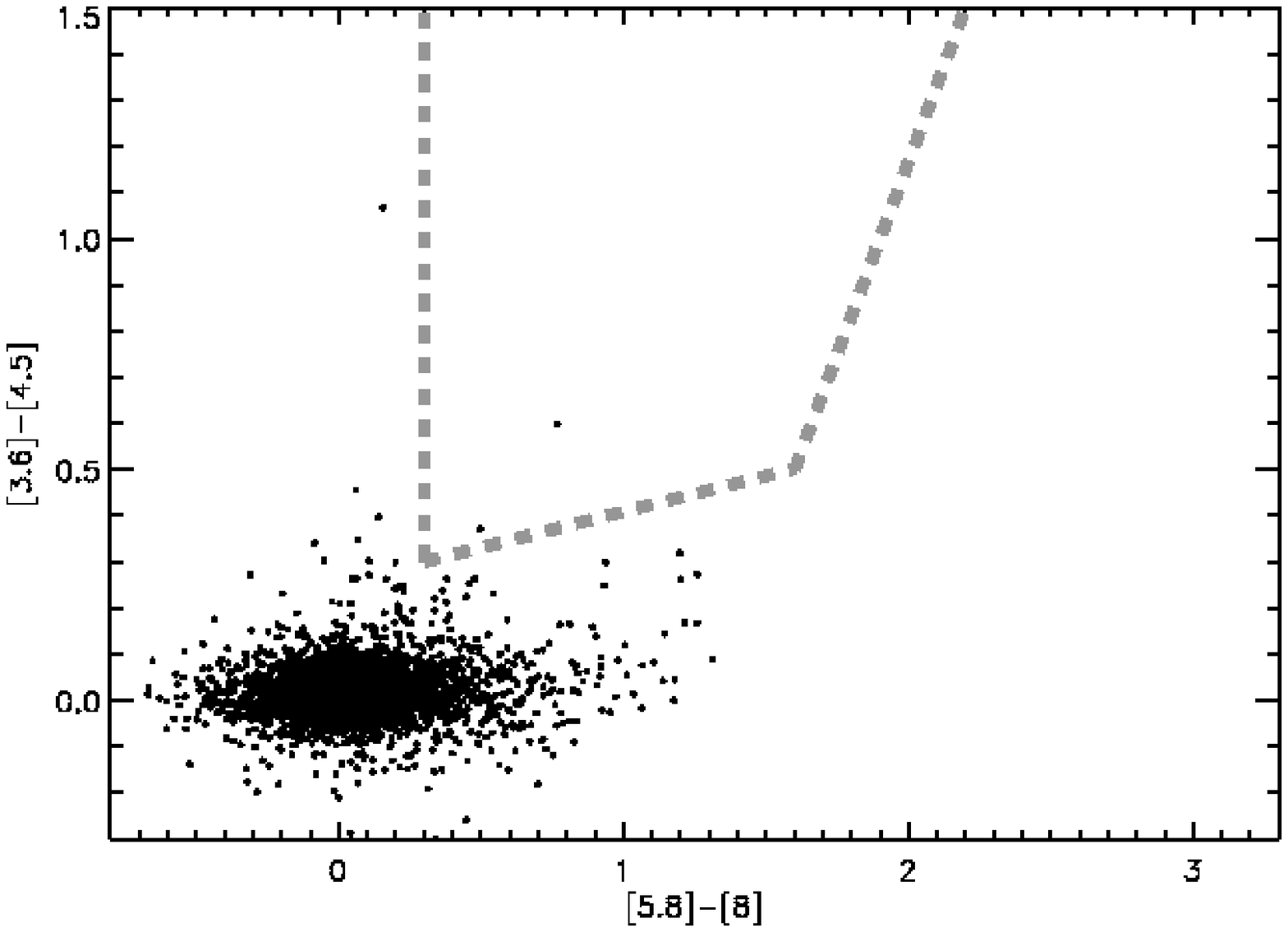}
\caption{[3.6]-[4.5] vs. [5.8]-[8] and [3.6]-[5.8] vs. [4.5]-[8] color-color diagrams for 
all objects detected in our Spitzer data (left panels) and those previously identified as likely cluster members (right panels) 
from \citet{Currie2010a}.  The dotted lines enclose the colors typical of active galaxies \citep{Stern2005, Gutermuth2008}.  
We do not find evidence for significant residual contamination from active galaxies amongst our sample of 
Spitzer-detected members.}
\label{contamcolors}
\end{figure}

Figure \ref{contamcolors} compares the IRAC colors of all objects on the h and $\chi$ Persei field (left panels) 
and cluster members only (right panels) with the predicted colors of PAH-emission galaxies 
and AGN.  Numerous IRAC-detected objects on the field exhibit colors indicative of extragalactic 
contaminants according to our criteria and to \citeauthor{Gutermuth2008}'s.   However, 
contamination is essentially absent when we restrict our analyses to previously identified 
cluster members.  Our criteria fails to find any members strongly contaminated by PAH-emission 
galaxies and only two with colors consistent with AGN contamination.  We likewise 
find no such objects in [4.5]-[5.8] vs. [5.8]-[8] diagrams (not shown), and we would 
fail to find such objects as long as the blue limit in [4.5, 5.8]-[8] is redder than 
1.2.  Adopting \citeauthor{Gutermuth2008}'s criteria, we identify five objects with colors only marginally 
consistent with PAH-emission galaxy contamination.  

Thus, while we cannot prove that no 
cluster member has IRAC excess emission due to extragalactic contamination, we fail to find 
evidence for anything more than a negligible population of \textit{strongly} contaminated members.
The 47--78 members predicted to be contaminated by galaxies then generally are either much 
brighter than the contaminating galaxies themselves or are contaminated by galaxies with intrinsic IRAC colors near zero 
(e.g. not PAH-emission galaxies or BL-AGN).  

\begin{deluxetable}{lllllllllllllllllll}
 \tiny
%\rotate
%\documentstyle[10pt]
%SPMquot"(0pt)
\setlength{\tabcolsep}{0.02in}
%\linewidth{0.1 in}
\tabletypesize{\tiny}
%\tabletypesize{\scriptsize}
\tablecolumns{12}
\tablecaption{List of Candidate/Probable h and $\chi$ Persei Members (Revised from Currie et al. 2010)}
\tiny
\tablehead{{Member}&{Photometry}&{Spectroscopy}&{Membership Type}&{RA}&{DEC}&{Numerical}&{V}&{$\sigma$(V)}&{I}&{$\sigma$(I$_{c}$)}\\
{ID Number}&{Running Number}&{Running Number}&{}&{}&{}&{Spectral Type}}
\startdata
1 & 3 & 174 & 1 & 34.7691 & 57.1355 & 13.0 & 7.1731 & 0.0001 & 6.6876 & 0.0001 \\
%2 & 12 & 0 & 1 & 34.7946 & 57.1306 & -99.0 & 7.7561 & 0.0001 & 7.2290 & 0.0001 \\
2 & 16 & 60 & 3 & 34.6173 & 57.2084 & 11.5 & 7.7866 & 0.0001 & 7.3575 & 0.0001 \\
3 & 17 & 11 & 3 & 34.5962 & 57.0102 & 10.7 & 7.8288 & 0.0001 & 7.4510 & 0.0001 \\
4 & 30 & 8 & 3 & 34.4577 & 57.0904 & 10.5 & 8.0183 & 0.0001 & 7.5970 & 0.0001 \\
5 & 39 & 69 & 3 & 34.6996 & 57.0673 & 11.5 & 8.2215 & 0.0001 & 7.6884 & 0.0001\\
\enddata
\tablecomments{The membership type has the following meaning: Candidate members identified from 
1 = Photomety, 2 = Spectroscopy 3 = Spectroscopy and Photometry.  
A zero for either the photometry or spectroscopy running number means that the source lacks either 
optical photometric or spectroscopic data.  The spectroscopy running number corresponds to the running number from the combined photometry + spectroscopy 
table (Table 3), not the running number from Table 2.  The optical photometry listed here comes from that obtained in 
Currie et al. (2010), which may be saturated for the brighest stars.}
\label{memberlist}
\end{deluxetable}

\begin{deluxetable}{clllllllllllllllllllllll}
% \tiny
%\rotate
%\landscape
%\documentstyle[10pt]
%SPMquot"(0pt)
\setlength{\tabcolsep}{-15pt}
%\linewidth{0.1 in}
\tabletypesize{\tiny}
%\tabletypesize{\scriptsize}
%\tabletypesize{\footnotesize}
\tablecolumns{16}
%\tablecaption{List of Candidate/Probable h and $\chi$ Persei Members (Revised from Currie et al. 2010)}
\tablecaption{Photometry for Candidate/Probable h and $\chi$ Persei Members}
%\tiny
%\tablehead{{Member}&{}&{Spectroscopy}&{Membership Type}&{RA}&{DEC}&{Numerical}&{V}&{$\sigma$(V)}&{I}&{$\sigma$(I$_{c}$)}\\
%{ID Number}}
%\tablehead{{Member}&{RA}&{DEC}&{SpT}&{V}&{$\sigma$(V)}&{$I_{\textrm{c}}$}&{$\sigma$($I_{\textrm{c}}$)}&{J}&{$\sigma$(J)}&{H}&{$\sigma$(H)}&
%{K$_{\textrm{s}}$}&{$\sigma$(K$_{\textrm{s}}$)}&{[3.6]&{$\sigma$([3.6])}&{[4.5]}&{$\sigma$([4.5])}&
%{[5.8]}&{$\sigma$([5.8])}&{[8]}&{$\sigma$([8])}&{[24]}&{$\sigma$([24])}}
\tablehead{{ID}&{RA}&{DEC}&{SpT}&{V}&{$\sigma$(V)}&{[3.6]}&{$\sigma$([3.6])}&{[4.5]}&{$\sigma$([4.5])}&{[5.8]}&{$\sigma$([5.8])}&{[8]}&{$\sigma$([8])}
&{[24]}&{$\sigma$([24])}}
\startdata
  2 & 34.6173 & 57.2084 & 11.5 & 7.7866 & 0.0001 & 7.480 & 0.002 & 7.475 & 0.002 & 7.427 & 0.002 & 7.415 & 0.001 & 7.339 & 0.047 \\
  3 & 34.5962 & 57.0102 & 10.7 & 7.8288 & 0.0001 & 7.974 & 0.003 & 7.974 & 0.003 & 7.975 & 0.002 & 7.967 & 0.001 & 7.861 & 0.038 \\
  4 & 34.4577 & 57.0904 & 10.5 & 8.0183 & 0.0001 & 8.260 & 0.002 & 8.248 & 0.002 & 8.203 & 0.002 & 8.210 & 0.001 & 8.187 & 0.032 \\
  5 & 34.6996 & 57.0673 & 11.5 & 8.2215 & 0.0001 & 7.979 & 0.002 & 7.881 & 0.002 & 7.717 & 0.001 & 7.578 & 0.001 & 6.659 & 0.031 \\
  7 & 34.7002 & 57.2855 & 12.0 & 8.3787 & 0.0001 & 8.569 & 0.002 & 8.583 & 0.002 & 8.570 & 0.002 & 8.552 & 0.002 & 8.505 & 0.048 \\

%2&34.6173 & 57.2084 & 11.5 & 7.7866 & 0.0001 & 7.3575 & 0.0001 & 7.5100 & 0.0250 & 7.4970 & 0.0330 & 7.4160 & 0.0200 & 7.4802 & 0.0024 & 7.4745 & 0.0022 & 7.4269 & 0.0016 & 7.4146 & 0.0011 & 7.3395 & 0.0468\\
 %3&34.5962 & 57.0102 & 10.7 & 7.8288 & 0.0001 & 7.4510 & 0.0001 & 7.9130 & 0.0190 & 7.9030 & 0.0530 & 7.8740 & 0.0180 & 7.9736 & 0.0027 & 7.9743 & 0.0028 & 7.9746 & 0.0017 & 7.9674 & 0.0012 & 7.8610 & 0.0380\\
 %4&34.4577 & 57.0904 & 10.5 & 8.0183 & 0.0001 & 7.5970 & 0.0001 & 8.2760 & 0.0540 & 8.2430 & 0.0400 & 8.1770 & 0.0240 & 8.2597 & 0.0017 & 8.2475 & 0.0022 & 8.2034 & 0.0015 & 8.2103 & 0.0011 & 8.1870 & 0.0320\\
 %5&34.6996 & 57.0673 & 11.5 & 8.2215 & 0.0001 & 7.6884 & 0.0001 & 8.2980 & 0.0430 & 8.1880 & 0.0450 & 8.0950 & 0.0240 & 7.9786 & 0.0021 & 7.8809 & 0.0022 & 7.7169 & 0.0014 & 7.5777 & 0.0010 & 6.6590 & 0.0310\\
 %7&34.7002 & 57.2855 & 12.0 & 8.3787 & 0.0001 & 7.8966 & 0.0001 & 8.5960 & 0.0200 & 8.4970 & 0.0260 & 8.5370 & 0.0230 & 8.5686 & 0.0016 & 8.5832 & 0.0023 & 8.5700 & 0.0018 & 8.5515 & 0.0017 & 8.5045 & 0.0481\\
\enddata
\tablecomments{}
\label{spitzerphot}
\end{deluxetable}

\begin{deluxetable}{lllllllllllllllllll}
 \tiny
%\rotate
%\documentstyle[10pt]
%SPMquot"(0pt)
\setlength{\tabcolsep}{0.02in}
%\linewidth{0.1 in}
\tabletypesize{\tiny}
%\tabletypesize{\scriptsize}
\tablecolumns{7}
\tablecaption{Mapping Between Observed and Intrinsic Properties for h and $\chi$ Persei Stars}
\tablehead{{Stellar Mass (M$_{\odot}$)}&{$T_{eff}$ (K)}&{SpT (Cu10a)}&{SpT (P13)}&{$V_{obs}$}}
\tiny					
\startdata
6.0&    19950&  12.5/12& 12.25& 11.8\\
5.0&    16900&  14/13& 14& 12.8\\
4.0&    14790&  15.5&   16& 13.3\\
3.5&    13400&  17&     17.5& 14.2\\
3.0&    12100&  18&     18& 14.3\\
2.5&    10900&  19&     19& 14.6\\  
2.0&    9300&   21.5&   21& 15.2\\
1.5&    7210&   30&     30  &  16.2\\ 
\\
1.4&	6402&	35&	35.5&		16.5\\
1.3&	5750&	45&	42.5&		17.4\\
1.2&	5218&	50&	50.5&		18.3\\
1.1&	4794&	53&	53&		19.0\\
1&	4470&	54.5&	55&		19.7\\
0.95&	4326&	55&	55.5&		20.1\\
0.9&	4185&	56&	56&		20.4\\
0.8&	3961&	57.5&	58&		21.1\\
0.7&	3811&	60.5&	60.5&		21.6\\
0.6&	3694&	61.5&	61&		22.1\\
0.5&	3693&	61.5&	61&		22.6\\
0.4&	3452&	63&	62.5&		23.1\\
\enddata
\tablecomments{For 1.5--6 $M_{\odot}$ stars we use \citet{PallaStahler1999} to map between intrinsic and observable properties.  
For lower-mass stars, we use \citet{Baraffe1998}.  For comparison with our spectral type/$T_{eff}$ conversions, we list the 
spectral types from \citet{Pecaut2013}, which show excellent agreement for nearly every entry.}
\label{obsint}
\end{deluxetable}

\begin{deluxetable}{lllllllllllllllllll}
 \tiny
%\rotate
%\documentstyle[10pt]
%SPMquot"(0pt)
\setlength{\tabcolsep}{0.02in}
%\linewidth{0.1 in}
\tabletypesize{\tiny}
%\tabletypesize{\scriptsize}
\tablecolumns{7}
\tablecaption{Candidate Accreting Stars}
\tiny
\tablehead{{RA}&{DEC}&{ST}&{$EW(H_{\alpha})$}&{[3.6]-[8]}&{[4.5]-[8]}}
%\tablehead{{Running}&{RA}&{DEC}&{J}&{ST}&{$\sigma$(ST)}&{$EW(H_{\alpha})$}&{$EW(H_{\beta}$)}&{[3.6]}&{[8]}}
\startdata
%1 & 2 & x&35.4850 & 57.1471 & 29.00 & 3.00 & x&13.3940 & 0.0450 & 13.2400 & 0.0380 & 13.0114 & 0.1614 \\
%Accreting Stars Previously Reported in Currie et al. (2007c)\\
35.8788&56.9368&43.4& -5.60&-99.000&  0.010\\
34.8733&57.0952&45.0&-19.30&  0.084&  0.051\\
34.8106&56.8745&62.9& -6.50&  0.114&  0.031\\
34.7858&57.1891&39.8& -3.30&  0.088&  0.037\\
34.7723&57.1504&40.7& -6.50& -0.088& -0.084\\
34.8196&57.4537&40.8& -1.30& -0.178& -0.193\\
34.8565&57.1389&41.3&-12.60&  0.068&  0.054\\
34.8455&57.1830&62.8& -5.30& -0.423& -0.535\\
34.8293&57.2154&47.2& -2.10&  0.519&  0.460\\
35.9386&57.2865&44.6& -2.20& -0.123& -0.115\\
34.8521&57.0314&46.4&-21.30&  0.119& -0.042\\
34.9686&56.9170&42.4& -0.80&-99.000&-99.000\\
36.2416&57.3407&42.7& -4.20& -99.000& -99.000\\
36.1889&57.4141&52.8&-82.90& -99.000& -99.000\\
35.9331&57.3837&54.0&-43.10& -99.000& -99.000\\
36.1516&57.4623&45.5& -6.20& -99.000& -99.000\\
35.7826&56.9684&50.8&-19.10&  0.052&  0.061\\
35.5140&57.1173&42.2&-20.00& -0.049& -0.065\\
35.8746&57.3021&48.7& -8.10&  0.066&  0.064\\
34.9739&57.0482&43.4& -8.30& -0.011&-99.000\\
36.1228&57.3046&42.2&-10.10& -99.000& -99.000\\
35.4319&56.8294&53.3&-17.20&  0.058&  0.057\\
35.0511&57.1230&43.5& -0.50&  0.272&  0.255\\
34.8044&57.2098&40.0& -7.70&  0.133&  0.093\\
34.7157&57.1875&43.2& -7.00&-99.000&-99.000\\
35.0980&57.0256&42.6& -7.00& -0.027& -0.024\\
34.7179&57.0764&47.0&-16.10& -0.129& -0.180\\
34.6226&57.5886&64.5& -6.80&-99.000&-99.000\\
35.0717&57.1476&44.2& -0.70& -0.045& -0.081\\
34.8229&57.1354&38.2& -1.40&  0.040&  0.016\\
34.9248&57.3163&45.6& -7.40&  0.161&  0.176\\
35.5106&57.0000&45.1&-11.80& -0.023& -0.027\\
34.8816&57.0840&38.9&  0.70&  0.198&  0.242\\
34.3993&56.8156&51.7&-25.00& -99.000& -99.000\\
34.5443&57.0386&45.1&  0.80&-99.000&-99.000\\
34.8499&57.0401&44.4& -0.50&  0.063&  0.122\\
35.1955&56.8353&41.8& -8.30& -0.039& -0.059\\
35.9421&57.2128&18.8&  7.10&  0.183&  0.173\\
34.8515&57.0719&52.6& -5.30&  0.500&  0.512\\
34.7243&57.1288&36.6& -6.20&-99.000&-99.000\\
34.6675&57.1491&43.0&-11.10&  0.159&  0.180\\
35.9057&56.7800&39.5& -1.20& -99.000& -99.000\\
35.5901&57.0668&44.1&-13.74&  0.184&  0.159\\
35.6647&57.2619&40.2& -4.50&  0.011&-99.000\\
34.4040&57.0196&40.5& -1.80& -0.103& -0.107\\
34.4798&56.9342&42.7& -6.46&  0.101&  0.079\\
35.4435&57.0456&44.2& -5.75&  0.033&  0.017\\
35.5172&57.3804&39.0& -3.40&  0.191&  0.145\\
35.0596&56.7393&40.9&-12.40&-99.000&-99.000\\
35.2182&57.2180&46.0& -8.50& -0.071& -0.026\\
35.8477&57.0570&40.2& -3.20&  0.240&  0.196\\
34.5238&57.3162&41.6&-15.42&-99.000& -0.015\\
34.5224&57.0245&42.5&-10.15&  0.041&  0.012\\
35.1531&57.0935&42.4& -8.98& -0.011& -0.029\\
35.8870&57.1159&45.0& -8.60&  0.147&  0.083\\
34.2805&57.1869&44.1& -9.79&  0.063& -0.085\\
35.6276&57.2429&47.3&-12.93&  0.133&  0.098\\
35.7026&57.2102&48.0& -7.90&  0.297&  0.258\\
35.4154&57.2804&44.6&-12.74& -0.044& -0.072\\
35.6574&57.2178&39.8& -2.86& -0.150& -0.177\\
35.5286&57.1292&50.7&-13.07&  0.140&  0.124\\
35.3857&57.0089&40.4& -5.00& -0.057& -0.111\\
35.4009&57.1858&42.2&-23.00& -0.023&-99.000\\
35.4913&57.0680&50.0&-12.96&  0.149&  0.104\\
34.7779&56.8716&50.6& -8.80&  0.205&  0.150\\
35.3994&57.0897&51.1&-29.93&  0.060&  0.052\\
34.4520&57.3922&51.6&-19.40&-99.000&-99.000\\
34.6519&57.0209&46.0&-11.59&  0.126&  0.179\\
35.3151&57.0772&52.5&-26.40&  0.017&-99.000\\
34.4112&57.4461&53.9&-59.50&-99.000&-99.000\\
34.5489&57.0786&57.4&-47.32&  0.678&  0.431\\
34.6662&57.1706&25.0&  0.00&  0.078&  0.056\\
\enddata
\tablecomments{The numerical spectral type has the following formalism: 10=B0, 11=B1 ... 68=M8.  
The \textbf{Source} column refers to the source of the spectroscopic
data: Hectospec (1), Hydra (2), or FAST (3).  \textbf{Archive Number} refers to the 
numerical suffix as stored in the spectral archive at the Center for Astrophysics' \textit{Telescope Data Center}, as 
described in \citet{Currie2010a}.  The equivalent widths of $H_{\alpha}$ and $H_{\beta}$ are listed in units of 
angstroms.  Spitzer/IRAC photometry at [3.6] and [8] are from this paper.}
%\tablecomments{The photometry flags have the following meanings: -1 = unsaturated, uncrowded, -2 = crowded, PSF fitting used, -9 = likely saturated.}
\label{speccatalog}
\end{deluxetable}

%\documentclass{article}
%\usepackage[showframe=true]{geometry}
%\usepackage{changepage}
%\usepackage{amsmath}
%\usepackage[landscape]{geometry}%[top=2in, bottom=1.5in, left=0.25in, right=0.5in]{geometry}
%\def\mic{$\mu$m }

%\begin{document}

%\makebox[\textwidth]{
\begin{deluxetable}{llllllllllll}
\tiny
\setlength{\tabcolsep}{0.01in}
\tablecolumns{11}
\tablecaption{Properties of Modeled Disk-Bearing h and $\chi$ Persei Stars}
%\centering
%\begin{adjustwidth}{-2cm}{}
%\tablecaption{Properties of h and $\chi$ Persei Stars with Excesses at IRAC and MIPS}
\tablehead{{$\alpha$} & {$\delta$} & {SpT} & {E(B-V)} & {SpT est.} & {E(B-V) est.} & {[3.6]-[8]}
 & {[4.5]-[8]} & {[3.6]-[24]} & {[4.5]-[24]} & {R$_{disk}$/R$_{sub}$}&{Disk Type}}
\startdata
      34.4833    &      57.1675    &       A7.4    &     0.572    &       -
 &     -    &     0.133    &     0.116   &   2.077  &    2.060 & 6 & TDHD/WDD
    \\
      34.5675    &      57.0963    &       B8.5    &     0.673    &       -
    &     -    &     0.618    &     0.433  &    1.758   &   1.573 & 400 &TDHD/WDD (TDIH)
    \\
      34.6778    &      57.0202    &       -    &     -    &       M2
    &     0.650    &      1.126    &      1.081   &   4.056  &    4.012 & 20 & TDIH
    \\
      34.7561    &      57.0997    &       -    &     -    &       K0
    &     0.650    &     0.250    &     0.225   &   1.900   &   1.875 & 1& TDHD/WDD
    \\
      34.7786    &      57.0364    &       G5.2    &     0.848    &       -
    &     -    &     0.412    &     0.244   &   3.514  &    3.345 & 1& TDHD/WDD
    \\
      34.7824    &      57.2346    &       F8.3    &     0.582    &       -
    &     -    &      1.239    &      1.157   &   4.232   &   4.150 & 10 & TDHD/WDD
    \\
      34.9145    &      57.4071    &       F5.1    &     0.605    &       -
    &     -    &     0.709    &     0.665   &   2.684  &    2.639 & 1& TDHD/WDD
    \\
      34.9447    &      57.1925    &       G1.1    &     0.675    &       -
    &     -    &      1.004    &     0.929   &   5.230   &   5.154 & 40 & TDIH
    \\
      35.0298    &      57.1013    &       -    &     -    &       M4
    &     0.700    &     0.998    &     0.905   &   4.129  &    4.035 & 20 & TDIH
    \\
      35.0501    &      57.0758    &       -    &     -    &       M1
    &     0.700    &      1.859    &      1.261   &   4.742   &   4.144 & 8& PD
    \\
      35.1028    &      56.8800    &       -    &     -    &       K0
    &     0.500    &     0.368    &     0.447   &   2.203   &   2.282 & 1& TDHD/WDD
    \\
      35.3153    &      57.2595    &       -    &     -    &       A5
    &     0.300    &     0.199    &     0.178   &   1.564   &   1.544 & 30 & TDHD/WDD (TDIH)
    \\
      35.4073    &      57.0433    &       -    &     -    &       K7
    &     0.520    &     0.592    &     0.474   &   4.805   &   4.686 & 300 & TDIH
    \\
      35.4106    &      56.9030    &       F5.9    &     0.474    &       -
    &     -    &      1.391    &      1.224   &   4.706   &   4.539 & 20 & TDHD/WDD (TDIH)
    \\
      35.6905    &      56.9388    &       -    &     -    &       M5
    &     0.520    &     0.948    &     0.803   &   5.266   &   5.121 & 15& TDIH
    \\
      35.6914    &      57.2379    &       -    &     -    &       K7
    &     0.520    &     0.874    &     0.822   &   3.816   &   3.764 & 1& TDHD/WDD
    \\
      35.6923    &      56.9258    &       -    &     -    &       M0
    &     0.560    &     0.751    &     0.620   &   4.518   &   4.386 & 30 & TDIH
    \\
      35.8256    &      56.8723    &       -    &     -    &       M5
    &     0.700    &     -99.    &      1.533   &  -99.   &   5.167 & 30 & PD (TDIH)
\enddata
%\end{tabular}
%\end{adjustwidth}
\tablecomments{Observed properties of 8 $\mu m$ and 24 $\mu m$ excess sources in h and $\chi$ Persei. Dashes indicate that a value was not applicable to that specific column. The final entry does not have a 3.6 $\mu m$ detection, this is indictated with a value of -99. for its computed colors involving [3.6].   R$_{disk}$/R$_{sub}$ denotes the disk inner radius in units of the disk sublimation radius derived from fitting the source SED with the \citeauthor{Robitaille2006} radiative transfer model grid.  Sources with inner holes have a best-estimated inner disk radius of R$_{disk}$ $>$ 10$\times$ R$_{sub}$.  The disk classification (right-most column) is determined by comparing source SEDs with the fiducial disk model: PD = primordial disk, TDIH = a transitional disk with an inner hole, and TDHD/WDD = a homologously depleted transitional disk or a warm debris disk.  If SED model fitting from the \citet{Robitaille2006} grid favors a different classification, we put that alternate classification in parentheses.  \label{irexcesstable}}
\label{diskevostate}
\end{deluxetable}

%\end{document}

%\documentclass{article}
%\usepackage{pdflscape}
%\begin{document}

%\begin{landscape}
\begin{deluxetable}{lllllll}
\tiny
\setlength{\tabcolsep}{0.02in}
\tablecolumns{6}
\tablecaption{Frequencies of Warm Circumstellar Dust and Gas from Spitzer Surveys}
%\hline
\tablehead{
{Name} & {Nominal Age (Myr)} & {Revised Age (Myr)} & {Disk Fraction} & {Accretion Fraction} & {References} 
}
%\hline
\startdata
%\\
NGC 1333 & 1 & - &	$0.83 \pm 0.11$ &-& 1 \\
$\rho$ Oph & 1 & - &- &$0.50 \pm 0.16$ & 2\\
Taurus & 1.5 & - & $0.63 \pm 0.09$ &$0.59 \pm 0.09$ & 2,3 \\	
NGC 7129	& 1.5 & - &	$0.54 \pm 0.14$ & -& 4 \\
NGC 2068/71 & 2 & - &-& $0.61 \pm 0.09$ & 5\\
Cha I & 2 & - &-& $0.44 \pm 0.08$ & 2\\
IC 348 & 2.5 & $6$ & $0.47 \pm 0.12$ &$0.33 \pm 0.06$&	2,6 \\
$\sigma$ Orionis & 3 & $6^{+7.4}_{-1.3}$ & $0.36 \pm 0.04$ &$0.30 \pm 0.05$ & 7,8 \\
NGC 6231 & 3 & -&-&$0.15 \pm 0.05$ & 8\\
Tr37 & 3.5 &- &$0.48 \pm 0.05 $ &-& 9 \\
NGC 2362 & 5 &$12^{+3.3}_{-4.1}$	&$0.20 \pm 0.03$ &$0.05 \pm 0.05$& 10,11 \\
Upper Sco & 5 & $10$ &$0.19 \pm 0.05$ &$0.07 \pm 0.02$& 2,12 \\
OB1b & 5 & -&$0.17 \pm	0.04$ & -& 13 \\
$\eta$ Cha & 6 &- &$0.40 \pm 0.14$ &$0.27 \pm 0.19$& 14, 15 \\
%OB1a & 7.5 & $0.074 \pm	0.034$ & 13 \\
NGC 6531 & 7.5 & - & -&$0.08 \pm 0.05$ & 8\\
TWA & 8 & - &-& $0.06 \pm 0.06$ & 15\\
25 Orionis & 8.5 &- &$0.06 \pm 0.023$ &$0.06 \pm 0.02$& 7,16 \\
NGC 2169 & 9 & - &-& $0.0^{+0.03}$ & 17\\
NGC 7160 & 11.8 &$12.6^{+1.3}_{-2.1}$ &$0.04 \pm 0.04$ &$0.02 \pm 0.02$& 9 \\
$\beta$ Pic MG & 12 & - & -&$0.0^{+0.13}$ & 15\\
h and $\chi$ Persei & 14 & & $0.022 \pm 0.002$ &$0.017 \pm 0.003$& This work \\
%h and $\chi$ Persei & 14 & $14.5^{+2.2}_{-1.7}$&$0.022 \pm 0.002$ &$0.017 \pm 0.003$& This work \\
Tuc-Hor & 27 & -&- & $0.0^{+0.08}$ & 15\\
NGC 6664 & 46 & - &-& $0.0^{+0.04}$ & 8\\
\enddata
\tablecomments{Nominal ages, revised ages from \cite{Bell2013} with empirical disk and 
accretor fractions for the clusters in Fig. \ref{diskfreqevo}.  References: 1) \citet{Gutermuth2008}, 
2) \citet{Mohanty2005}, 3) \citet{Hernandez2007}, 4) \citet{Gutermuth2005}, 5) \citet{Flaherty2008}, 
6) \citet{Lada2006}, 7) \citet{Hernandez2007b}, 8) \citet{Fedele2010}, 9) \citet{SiciliaAguilar2006}, 
10)\citet{DahmHillenbrand2007}, 11) \citet{CurrieLada2009}, 12) \citet{Carpenter2006}, 13) \citet{Hernandez2005}, 
14) \citet{Megeath2005}, 15) \citet{Jayawardhana2006}, 16) \citet{Briceno2007}, 17) \citet{Jeffries2007}, 
18) this work} 
\label{discfreqtable}
\end{deluxetable}
%\end{landscape}

%\end{document}


\clearpage

\clearpage
\clearpage

\clearpage

%\begin{figure}
%%\plottwo{excvagetestf.ps}{excvageavg.ps}
%\caption{Placeholder figure for rise and fall of debris disks.}
%\label{risefall}
%\end{figure}

\clearpage

\begin{figure}
\centering
\epsscale{0.8}
\plottwo{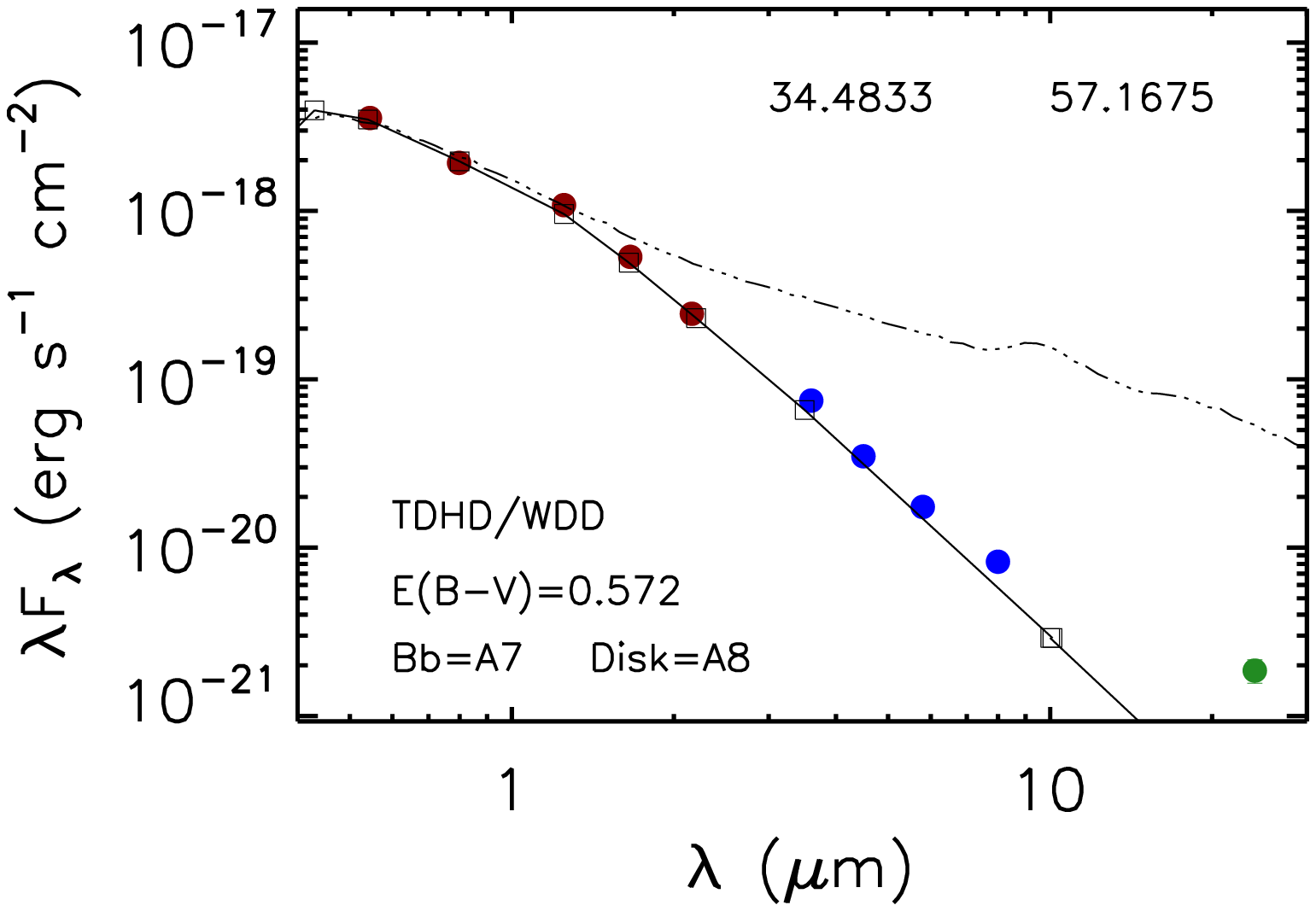}{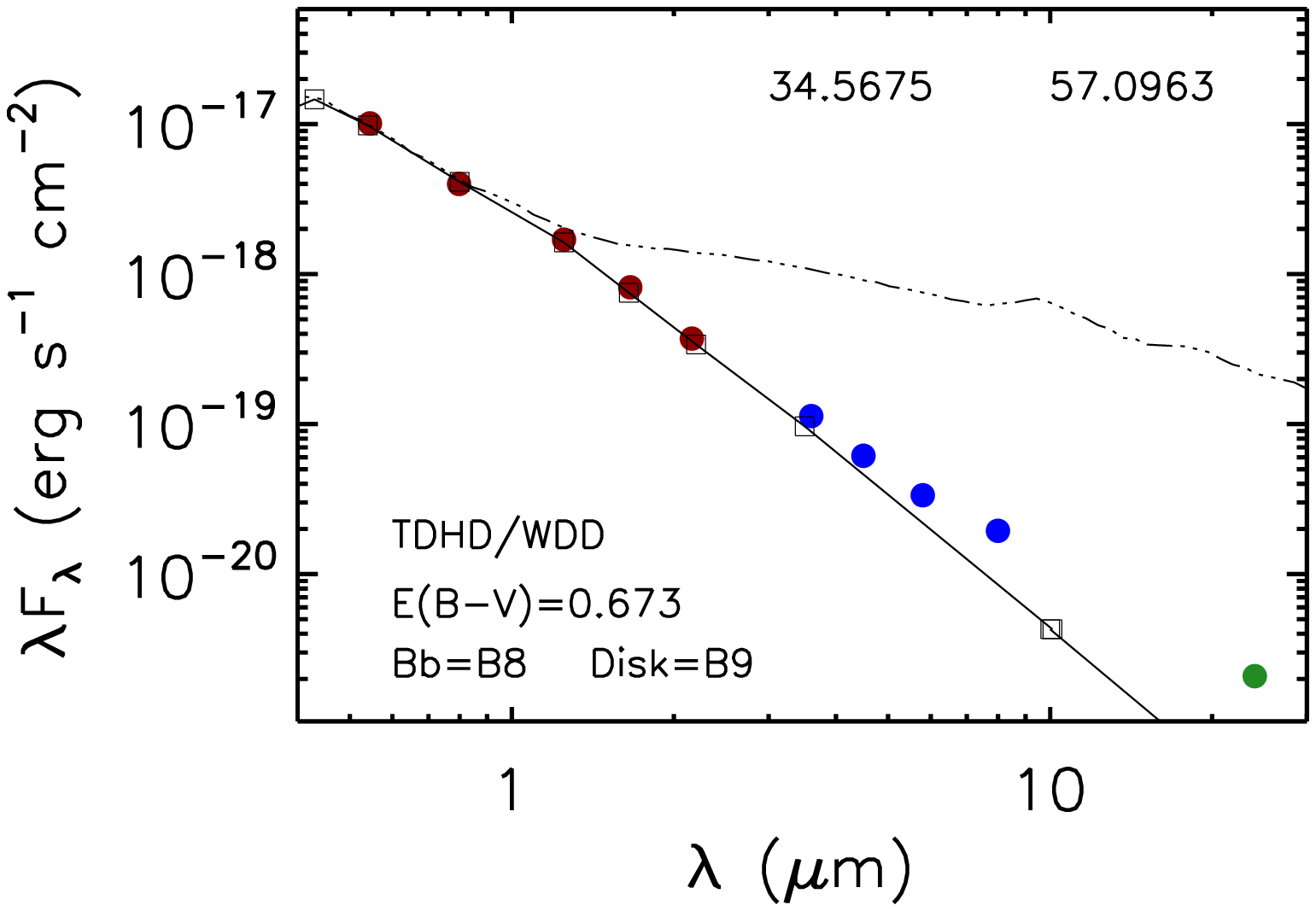}
\plottwo{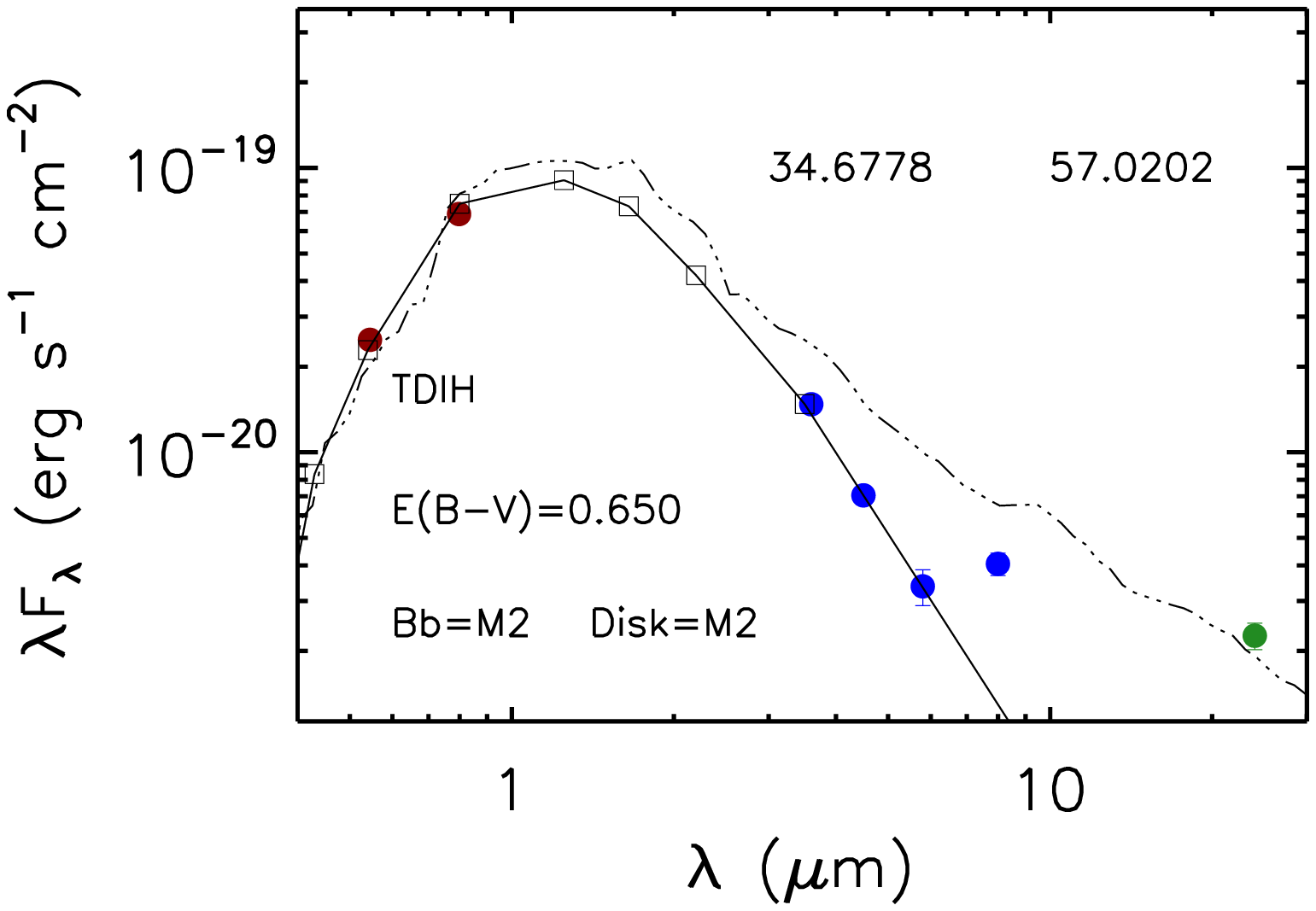}{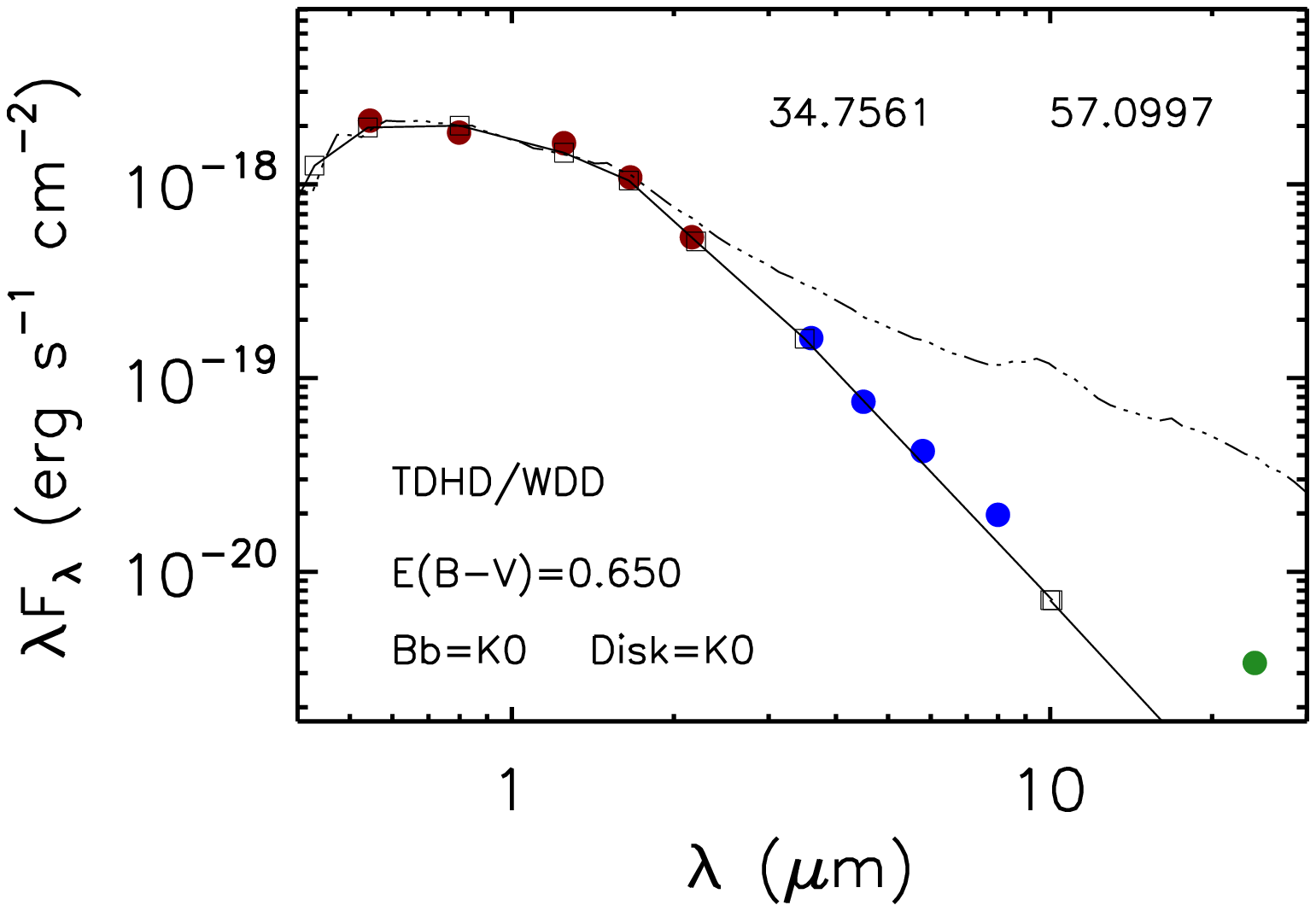}
\plottwo{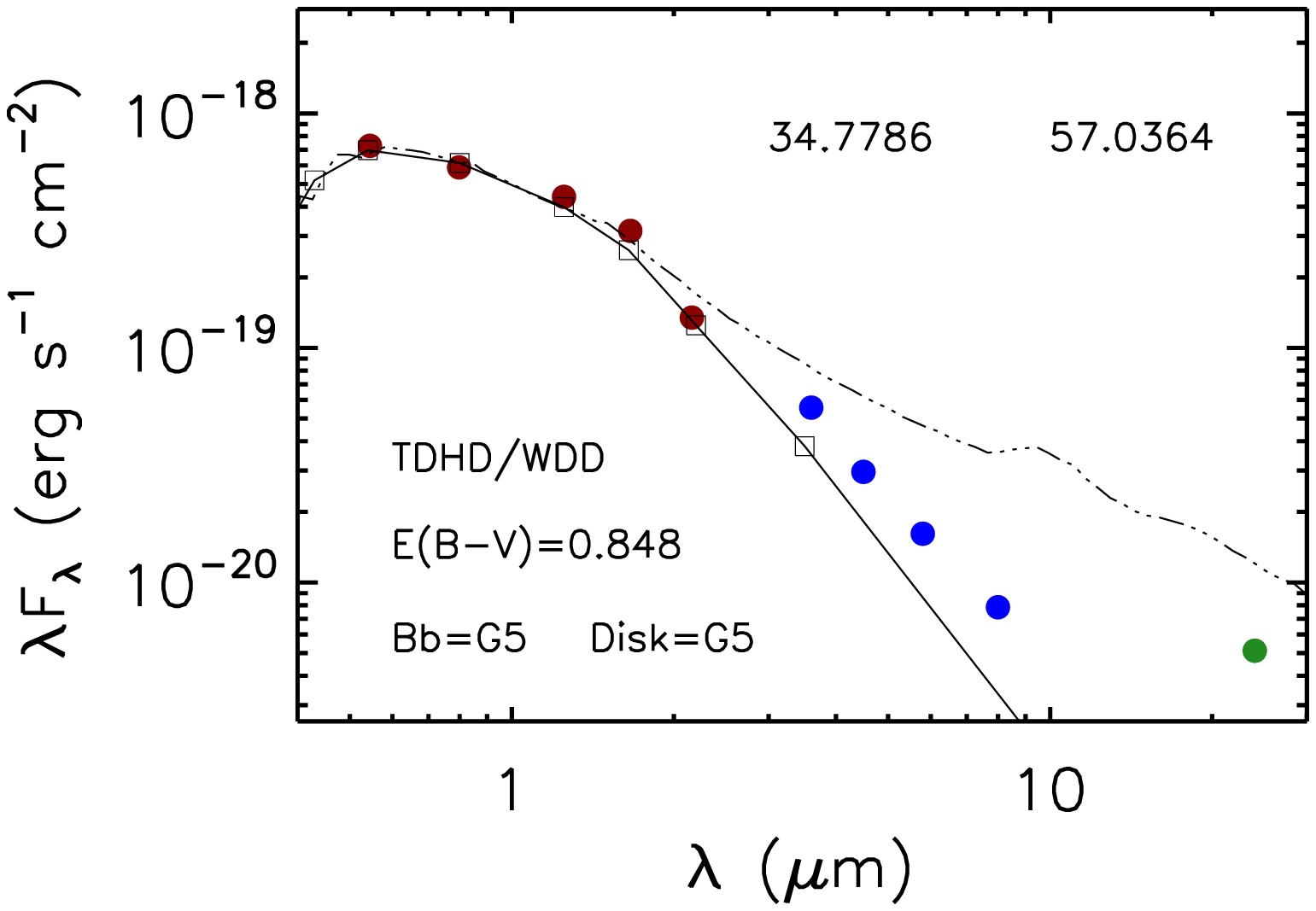}{3sed_dim_sources_with8and24excess5.ps}
\plottwo{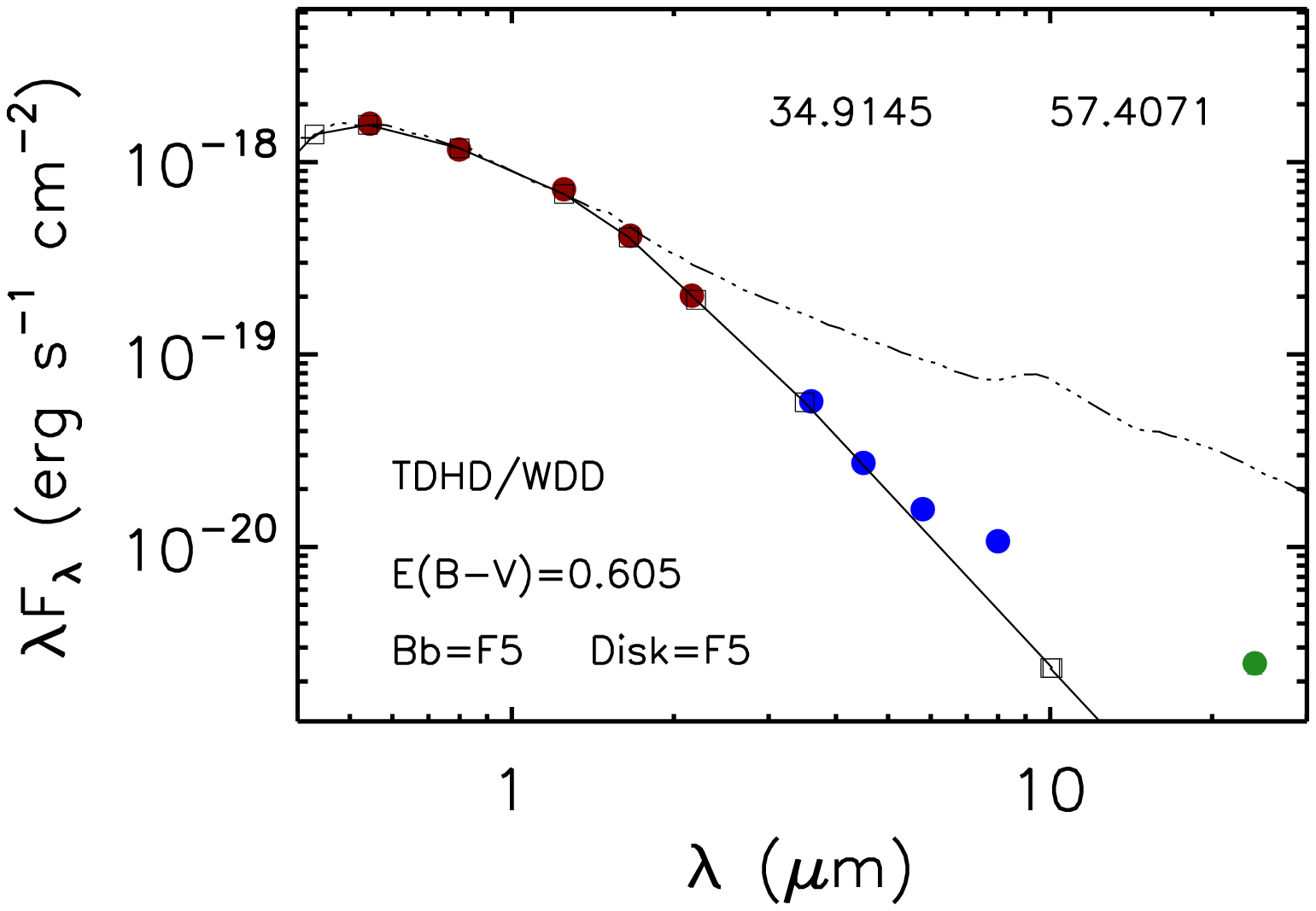}{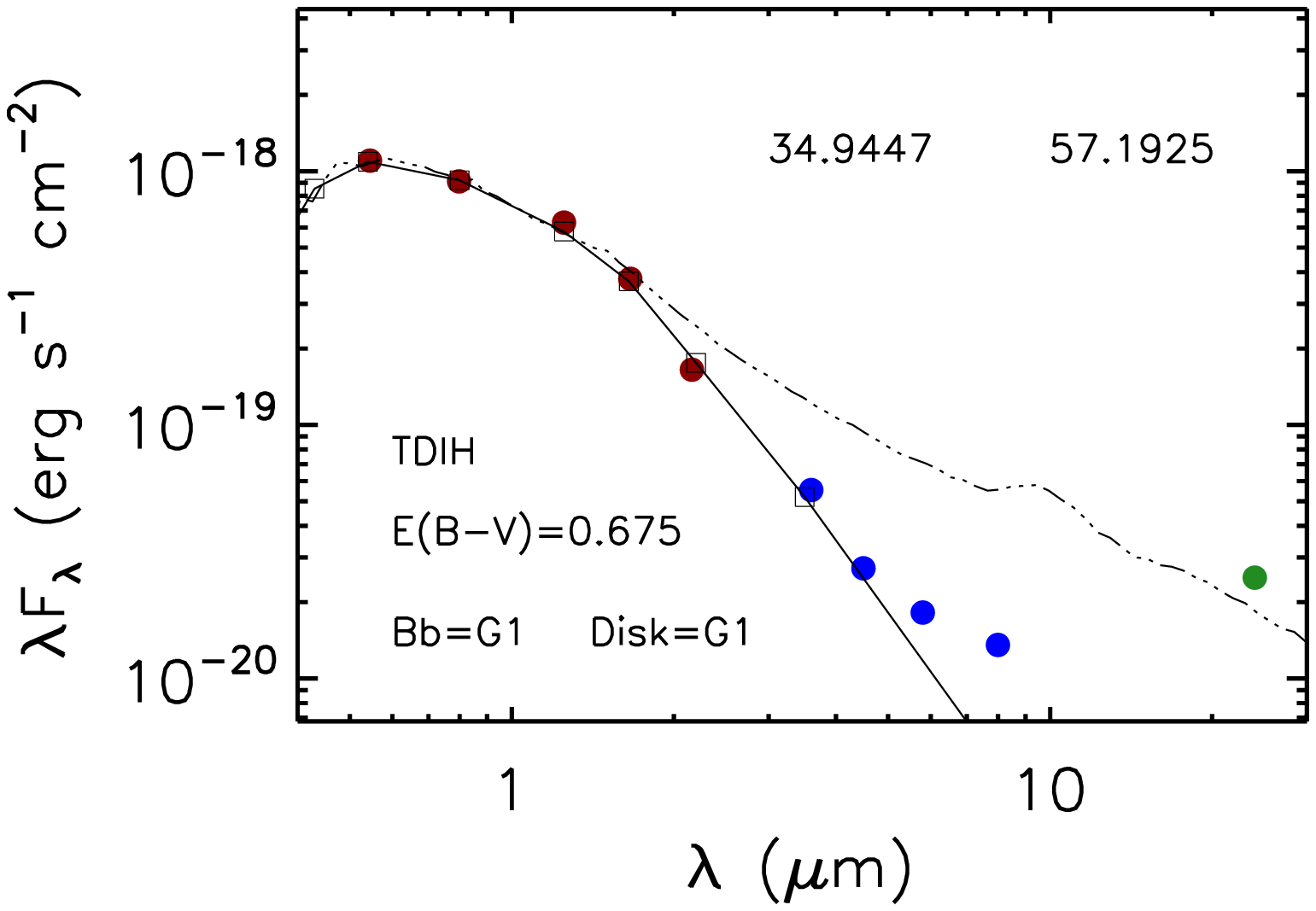}
\plottwo{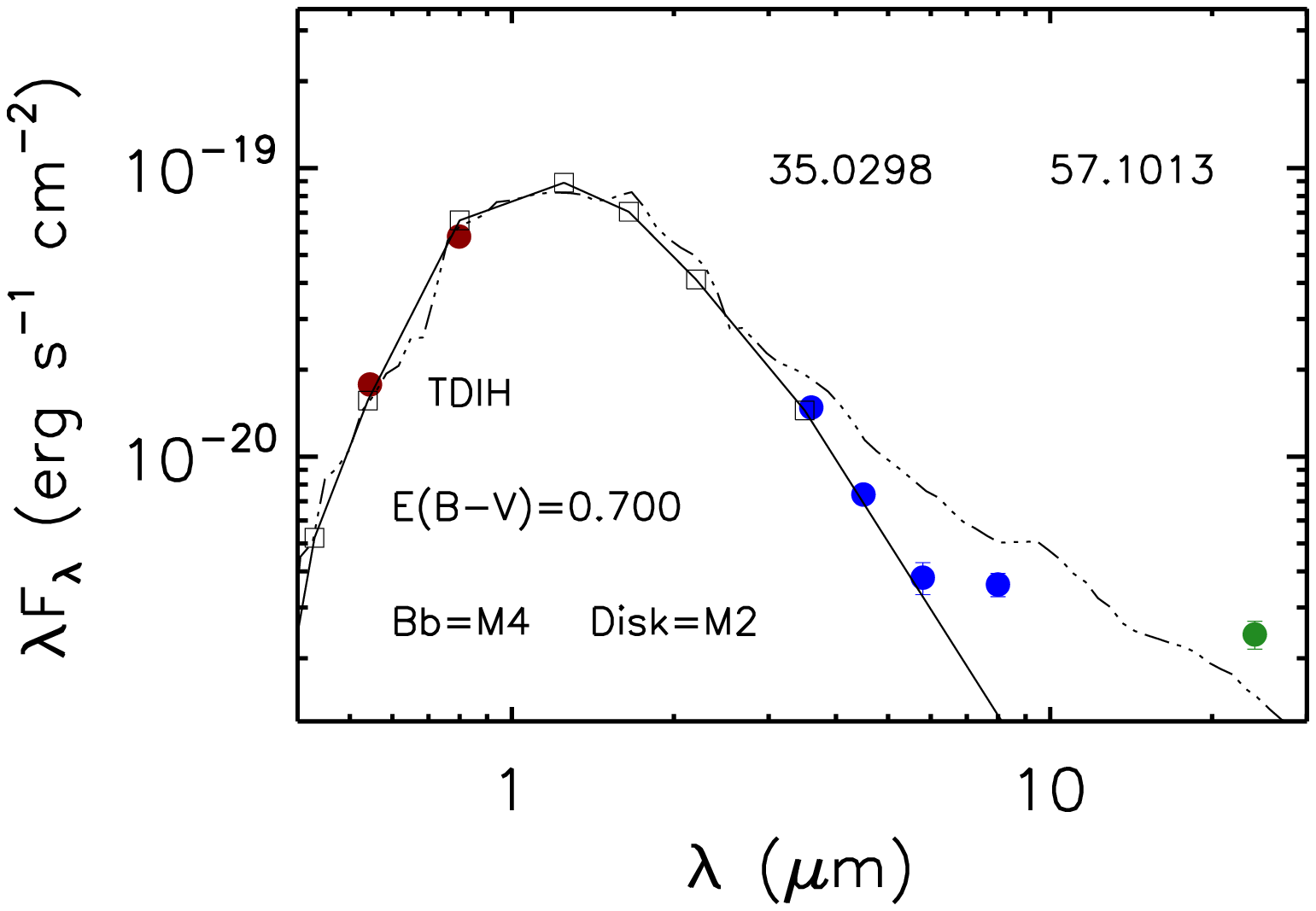}{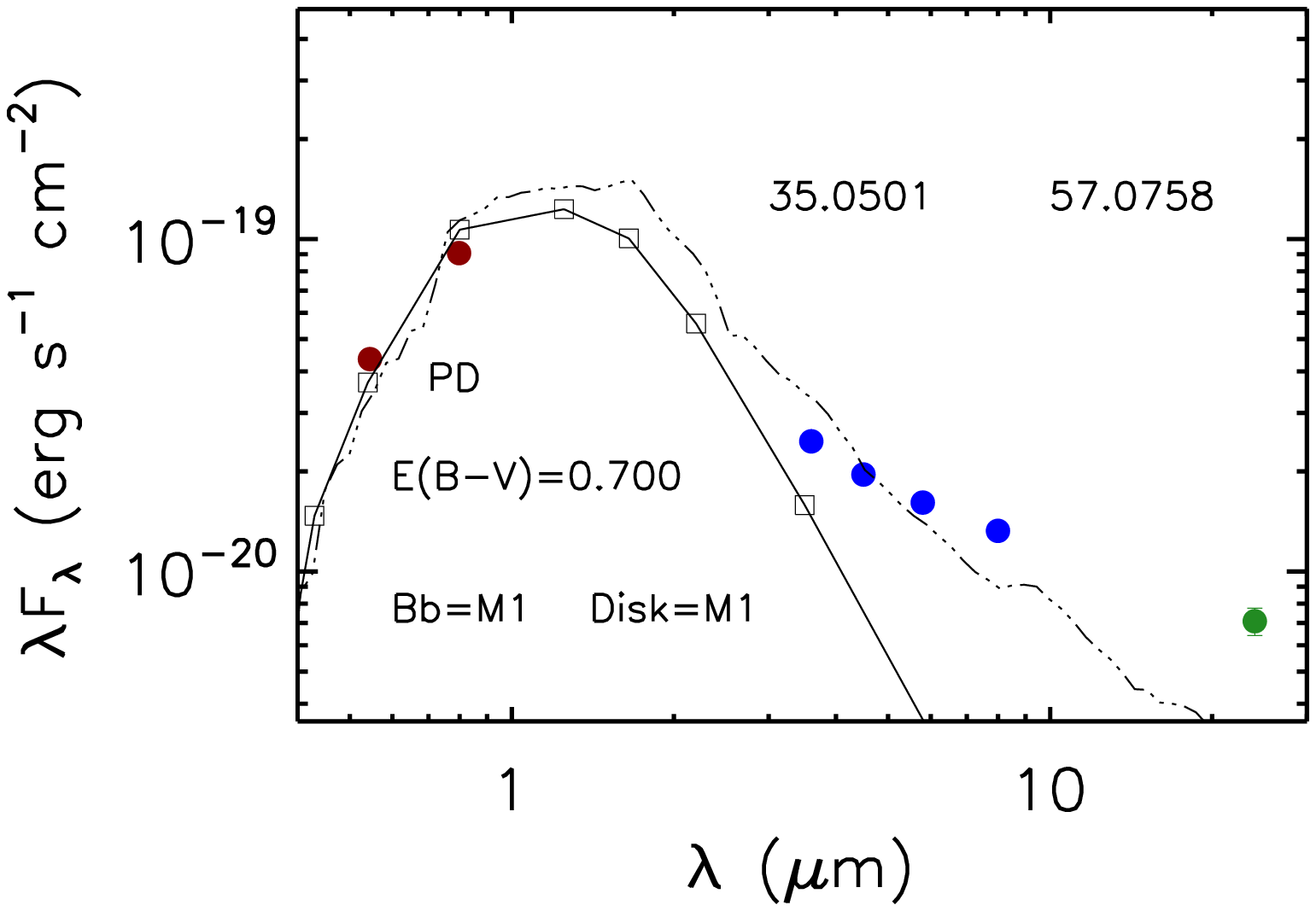}
\caption{Atlas of h and $\chi$ Persei stars with clear excesses in IRAC \textit{and} MIPS bandpasses compared to 
photometric predictions for a bare stellar photosphere (solid line) and an optically-thick, flat reprocessing disk (dashed line).  
}
\label{atlasseds}
\end{figure}

\begin{figure}
\epsscale{0.8}
\centering
\plottwo{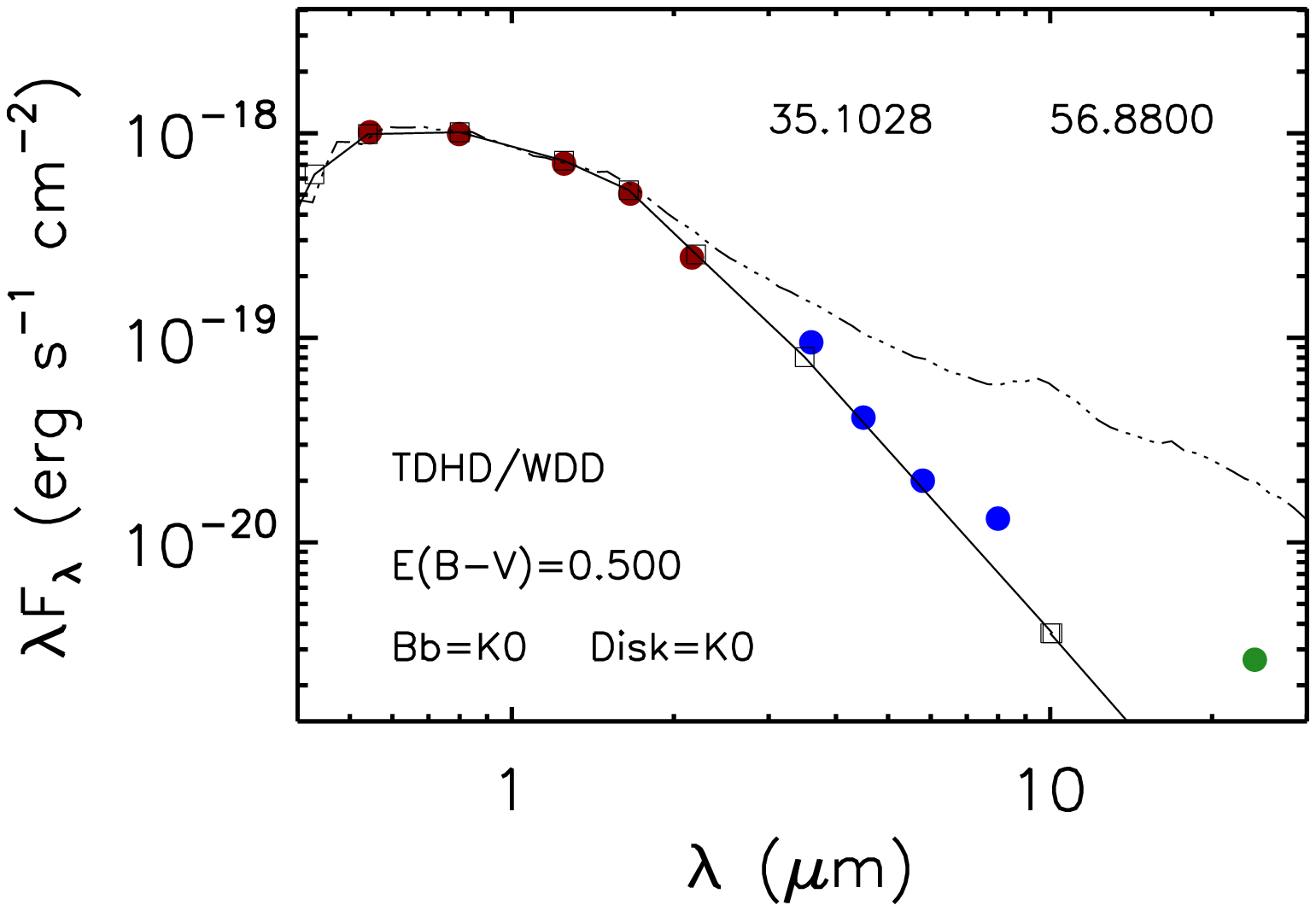}{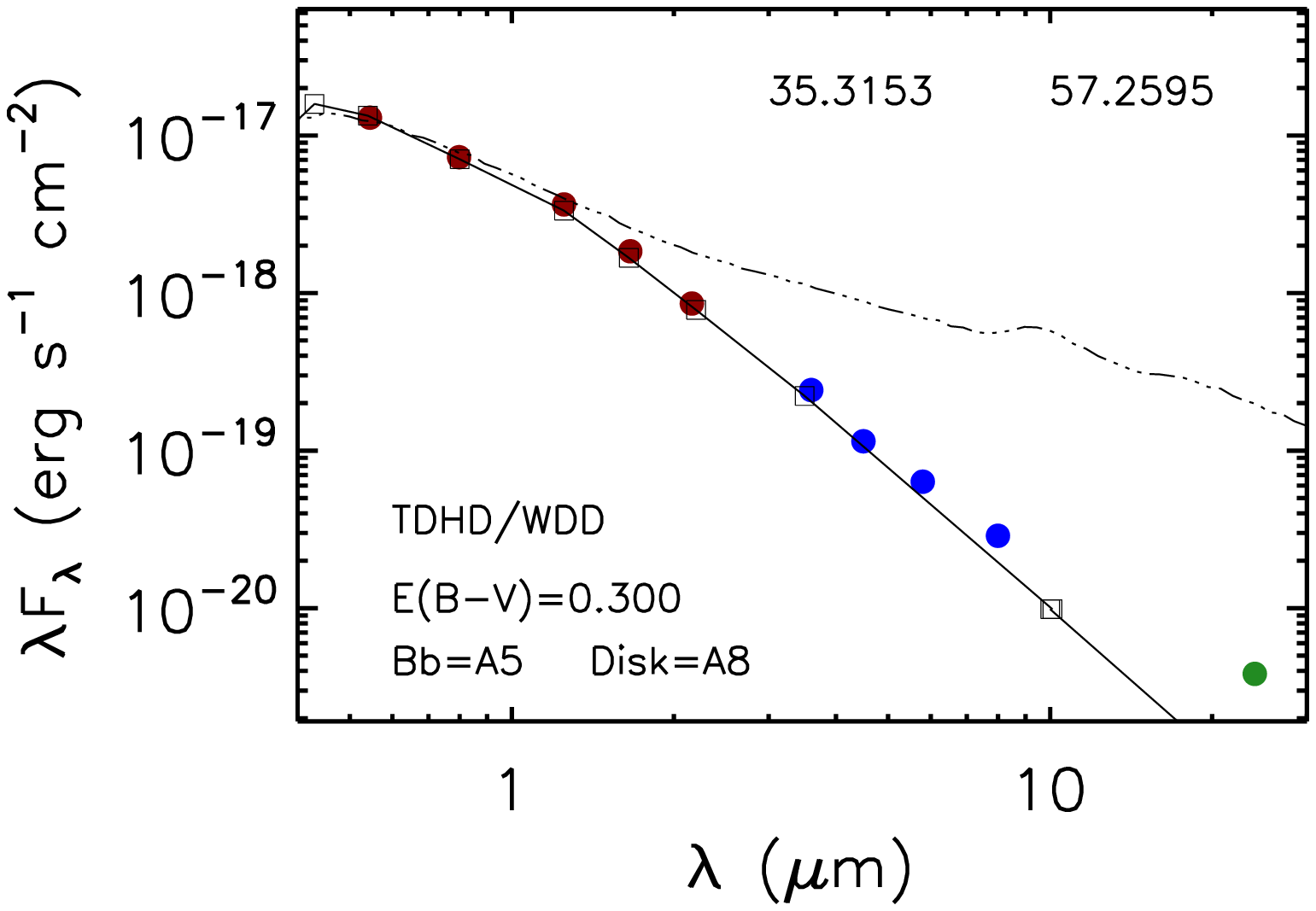}
\plottwo{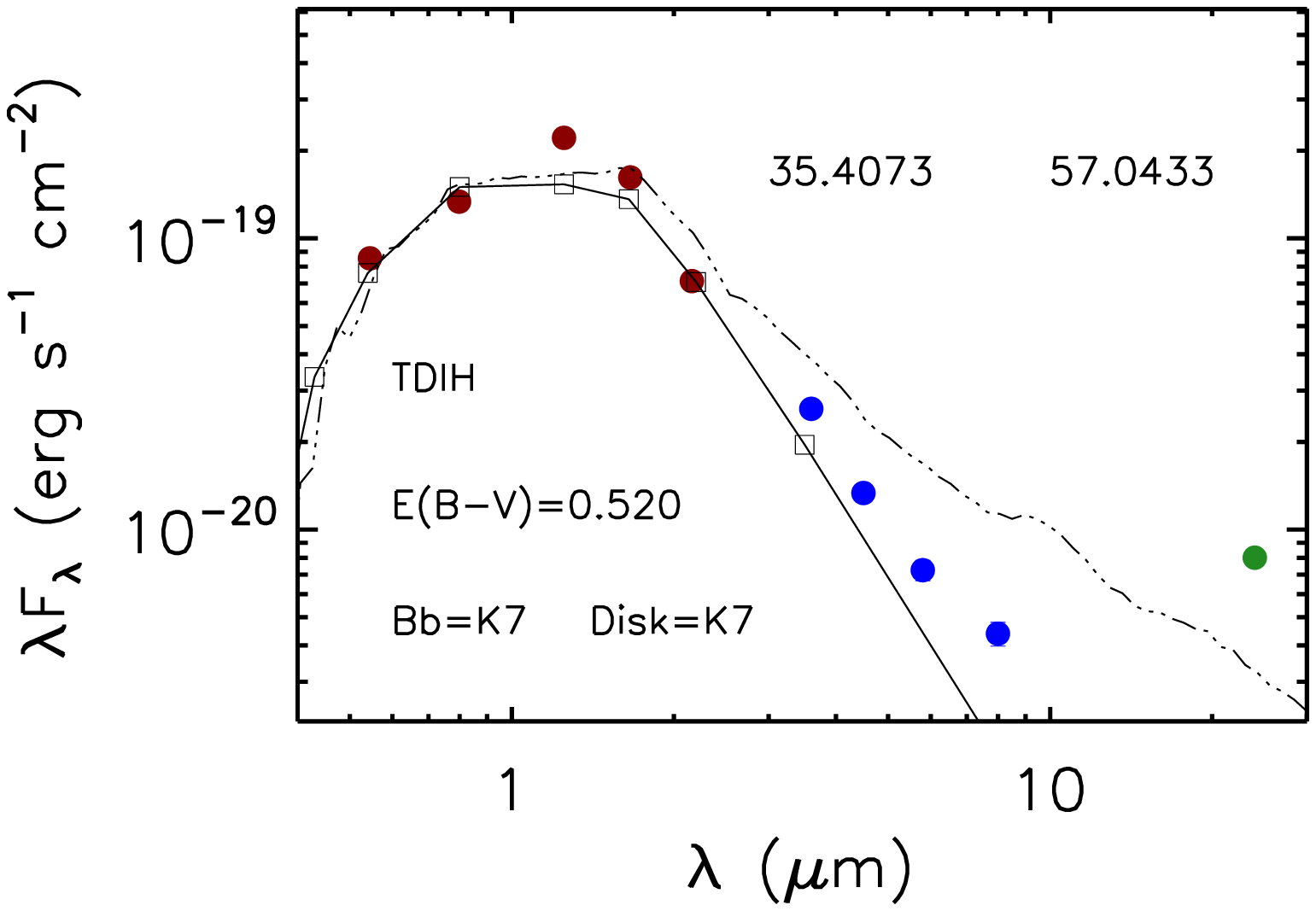}{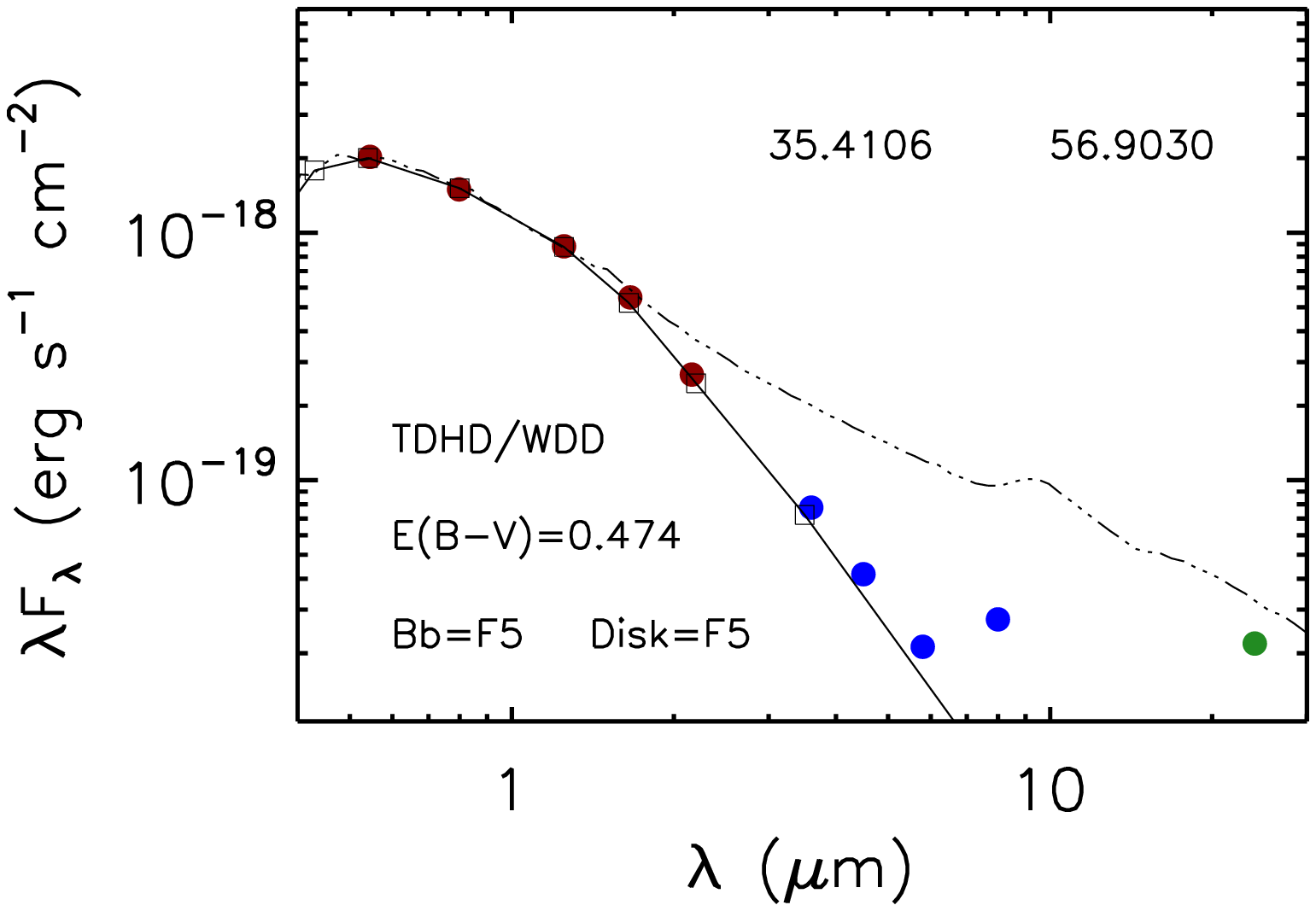}
\plottwo{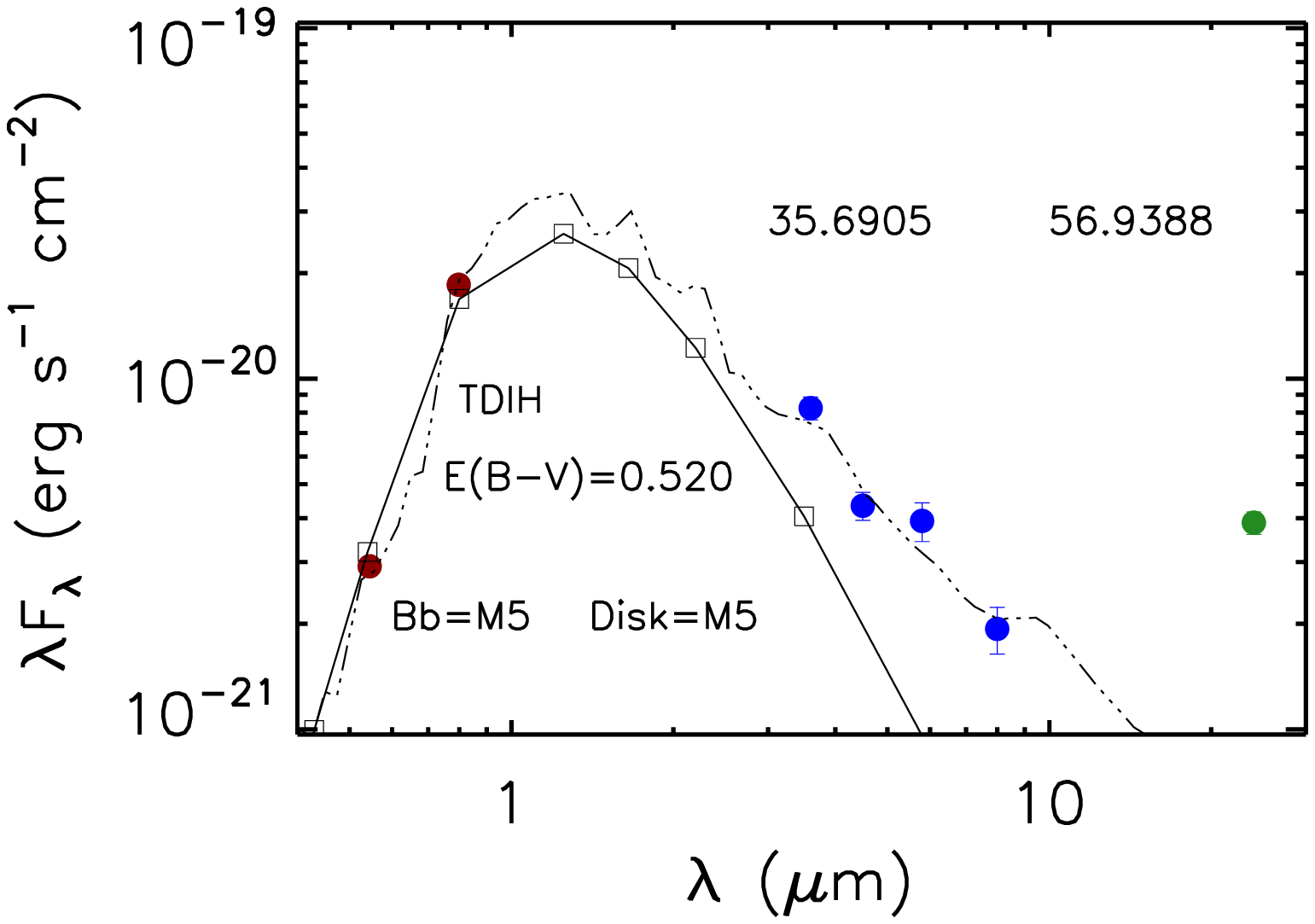}{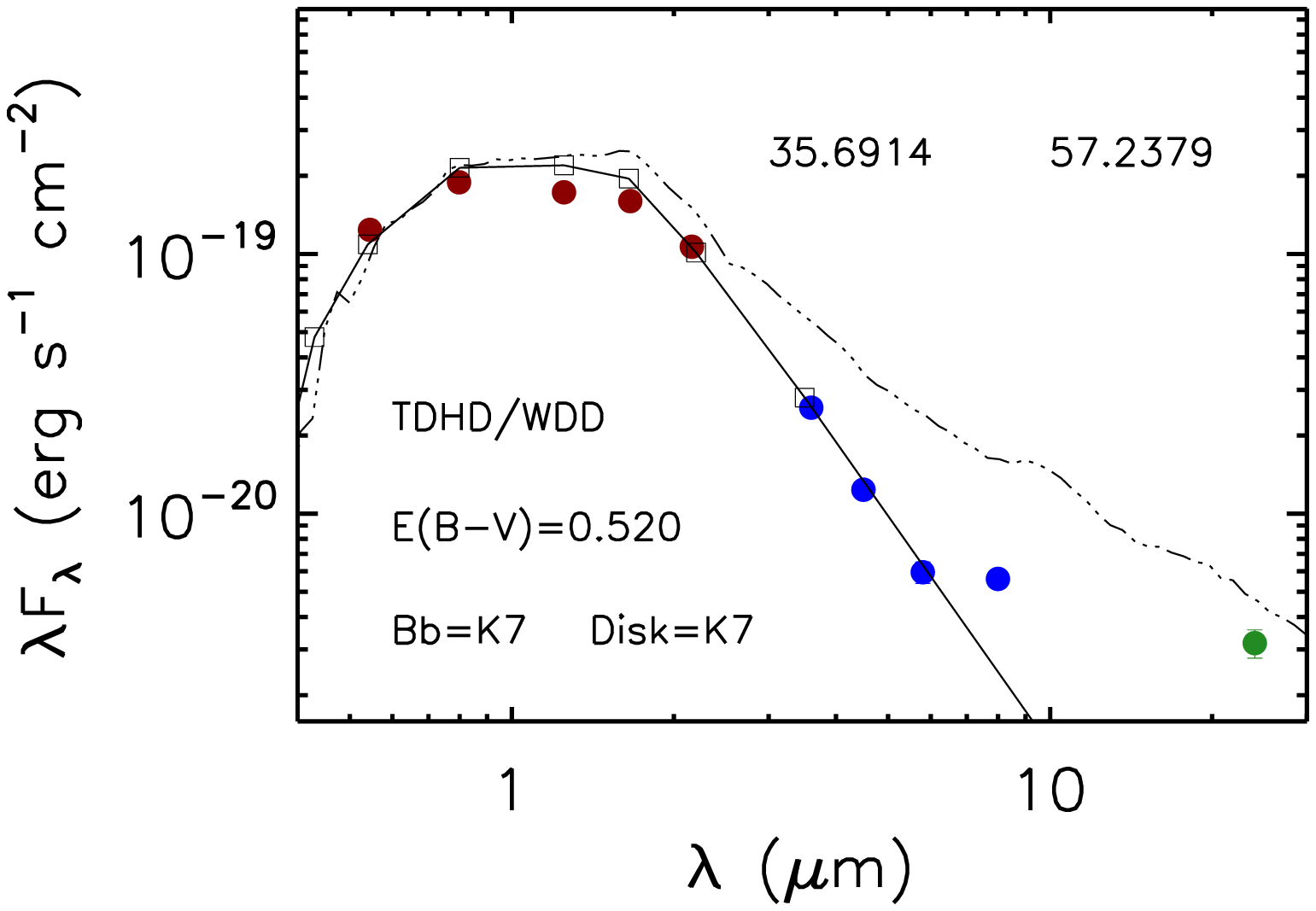}
\plottwo{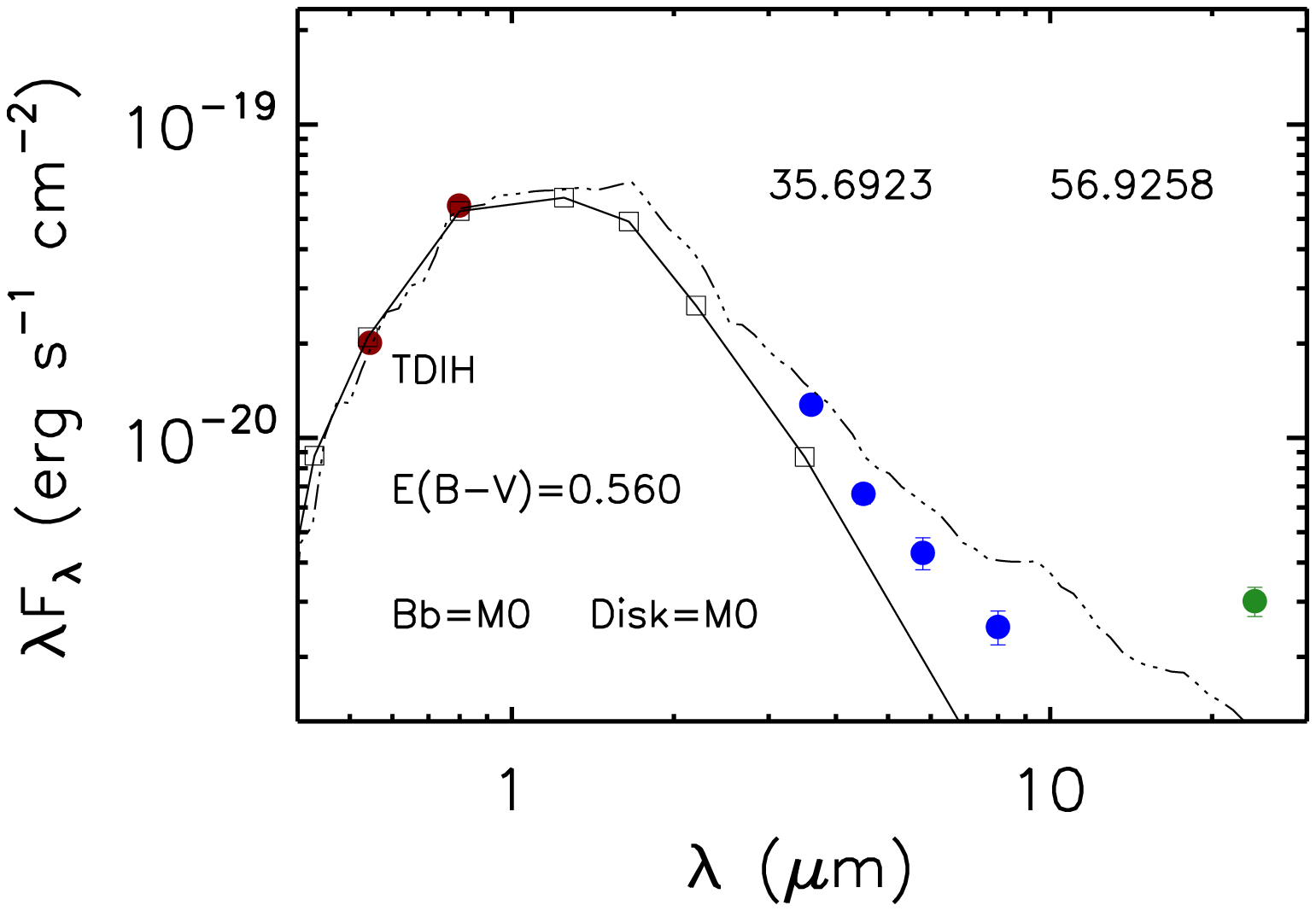}{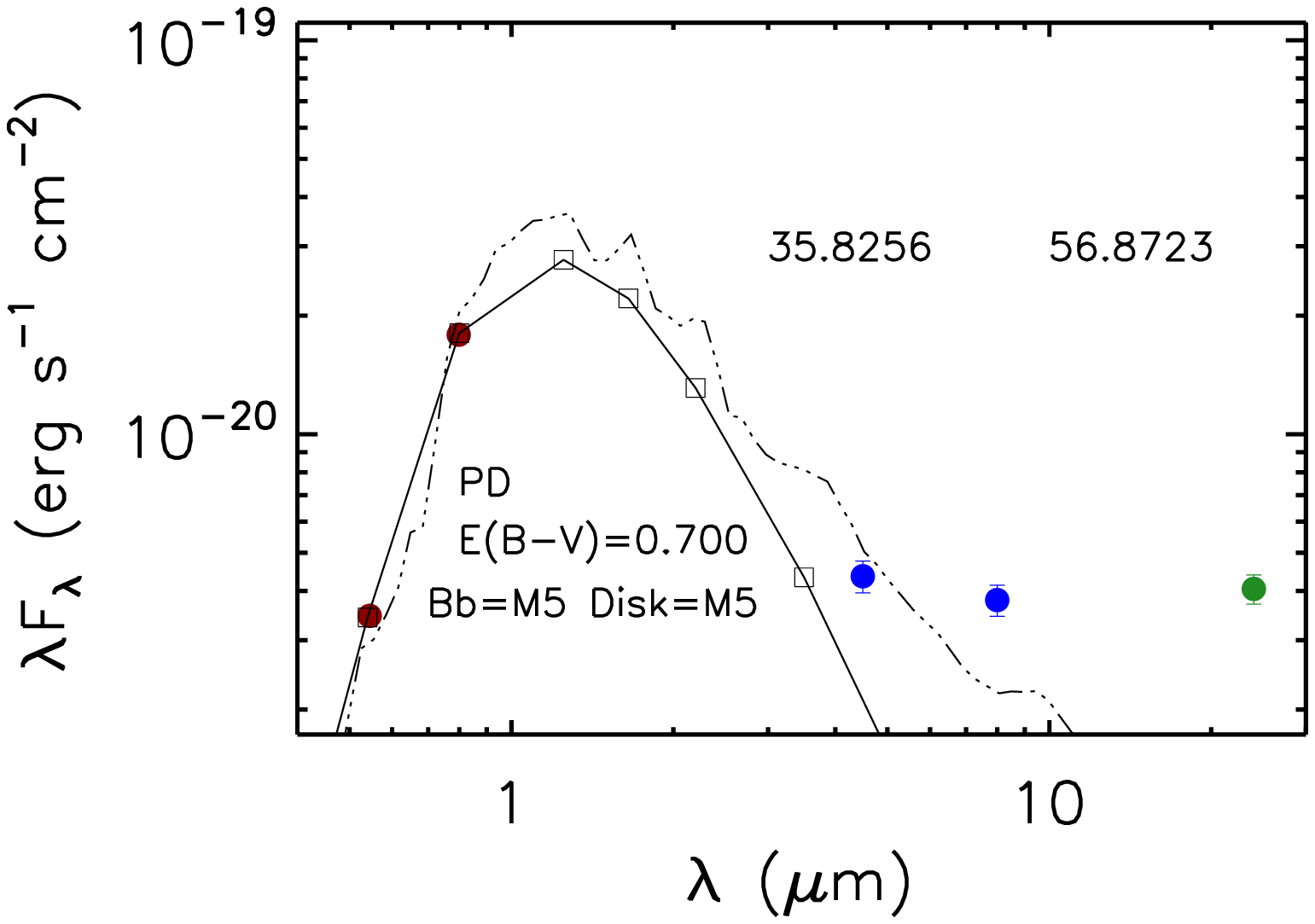}
\caption{Figure \ref{atlasseds} continued.}
\label{atlassedss}
\end{figure}

\end{document}